\documentclass[a4paper,fleqn,usenatbib,useAMS]{mnras}

\usepackage{graphicx}        % Including figure files
\usepackage{amsmath}        % Advanced maths commands
\usepackage{amssymb}        % Extra maths symbols
\usepackage{multicol}        % Multi-column entries in tables
\usepackage{bm}                % Bold maths symbols, including upright Greek
\usepackage{pdflscape}        % Landscape pages
\usepackage{xcolor}
\usepackage{threeparttable}  
\usepackage[T1]{fontenc}
\usepackage{ae,aecompl}

   \title[The study of unclassified B{[}e{]}  stars and candidates]{The study of unclassified B[e] stars and candidates in the Galaxy and Magellanic Clouds\thanks{This study was based on observations with the MPG 2.2-m telescope at the European Southern Observatory (La Silla, Chile) under the agreements ESO-Observat\'{o}rio Nacional/MCTIC and MPI-Observat\'{o}rio Nacional/MCTIC, Prog. IDs.: 075.D-0177(A), 080.A-9200(A), 082.A-9209(A), 096.A-9024(A), 096.A-9027(A), 096.A-9030(A), 097.A-9022(A), and 097.A-9024(A). We also used public data from the ESO archive.}}

\author[Condori et al.]{%
C. A.~H.~Condori,$^{1}$\thanks{E-mail: cesar@on.br, borges@on.br}
M.~Borges Fernandes,$^{1}$
M.~Kraus,$^{2}$
D. Panoglou,$^{1}$ \and
C. A.~Guerrero$^{3}$
\\
$^{1}$Observat\'{o}rio Nacional, Rua General Jos\'{e} Cristino 77, CEP: 20921-400, S\~{a}o Crist\'{o}v\~{a}o, Rio de Janeiro, Brasil\\
$^{2}$        Astronomick\'{y} \'{u}stav, Akademie v\v{e}d \v{C}esk\'e republiky, Fri\v{c}ova 298, 251\,65 Ond\v{r}ejov, Czech Republic\\
$^{3}$ Observatorio Astron\'omico Nacional, Universidad Nacional Aut\'onoma de Mexico, Apartado Postal 877, C.P. 22800, \\Ensenada B.C., Mexico
}

\date{Received ...; accepted ...}

\pubyear{2019}

\begin{document}
\label{firstpage}
\pagerange{\pageref{firstpage}--\pageref{lastpage}}
\maketitle

\begin{abstract}
We investigated 12 unclassified B[e] stars or candidates, 8 from the Galaxy,  2 from the Large Magellanic Cloud (LMC) and 2 from the Small Magellanic Cloud (SMC). Based on the analysis of high-resolution spectroscopic (FEROS) and photometric data, we confirmed the presence of the B[e] phenomenon for all objects of our sample, except for one (IRAS 07455-3143). We derived their effective temperature, spectral type, luminosity class, interstellar extinction and, using the distances from Gaia DR2, we obtained their bolometric magnitude, luminosity and radius.  Modeling of the forbidden lines present in the FEROS spectra revealed information about the kinematics and geometry of the circumstellar medium of these objects. In addition, we analyzed the light curves of four stars, finding their most probable periods. The evolutionary stage of 11 stars of our sample is suggested from their position on the HR diagram, taking into account evolutionary tracks of stars with solar, LMC and SMC metallicities. As results, we identified B and B[e] supergiants, B[e] stars probably at the main sequence or close to its end, post-AGB and HAeB[e] candidates, and A[e] stars in the main sequence or in the pre-main sequence. However, our most remarkable results are the identification of the third A[e] supergiant (ARDB\,54, the first one in the LMC), and of an ``LBV impostor" in the SMC (LHA 115-N82).
\end{abstract}


%%%%%%%%%%%%%%%%%%%%%%%%%%%%%%%%%%%%%%%%%%%%%%%%%%

%%%%%%%%%%%%%%%%% BODY OF PAPER %%%%%%%%%%%%%%%%%%

% The MNRAS class isn't designed to include a table of contents, but for this document one is useful.
% I therefore have to do some kludging to make it work without masses of blank space.
%\begingroup
%\let\clearpage\relax
%\tableofcontents
%\endgroup
%\newpage

\section{Introduction}

\label{sec:introduction}
The nomenclature ``B[e] stars'' was first used by \citet{conti-1976} to designate B-type stars that present forbidden emission lines in the optical spectrum. Later, \citet{Lamers-1998} suggested the expression ``stars with the B[e] phenomenon'' to describe these objects. This phenomenon was revised by \citet{Zickgraf-1999},   who associated it to the presence in the optical spectrum of B-type stars 
with: (i) intense Balmer emission lines, and (ii) permitted and forbidden emission lines of neutral and singly ionized metals, such as O\,{\sc i} and Fe\,{\sc ii}. In addition, these stars also present strong excess in the near-IR and mid-IR, due to circumstellar (CS) dust. 

However, these spectral characteristics are associated to the circumstellar medium and not to the object itself. \citet{Lamers-1998} noted a great heterogeneity among these objects, suggesting the existence of four classes of stars with the B[e] phenomenon, based on their evolutionary stage: pre-main sequence intermediate-mass stars, or Herbig Ae/B[e] or simply HAeB[e]; massive supergiant stars, or B[e] supergiants or sgB[e]; compact planetary nebulae, or cPNB[e]; and symbiotic stars, or SymB[e].  Thus, an important question that needs to be answered is how such different objects can have similar spectroscopic features. A possible answer is linked to the presence of a complex circumstellar environment, composed of a disk, as confirmed by polarimetric \citep{Magalhaes-1992} and interferometric measurements \citep{Domiciano-de-Souza-2011, Borges-Fernandes-2011} or by rings \citep{Kraus-2016}. The effect of binarity cannot be discarded either.

On the other hand, there is a  large number of objects whose evolutionary stage is still unknown or poorly known,  due to the absence of reliable stellar parameters, also including distance and interstellar extinction. This group of objects is usually called as simply unclassified B[e] stars or unclB[e] \citep{Lamers-1998}. \citet{Miroshnichenko_2007} proposed a new group of stars associated to the B[e] phenomenon,  called as FS CMa stars, which is mainly formed by unclB[e] objects that would be close to or still on the main sequence in binary systems with mass exchange.

\begin{table*}
        \caption{Our sample of unclassified B[e] stars and candidates observed with FEROS (our own spectra and also public ones retrieved from the ESO Science Archive Facility). }
        \label{table:objects}
        \centering
        \begin{tabular}{cccccccccc}
                \hline
&Name            & IRAS ID         & R.A.       & Dec.        & Date       & JD                        & t$\sb{\rm exp}$ (s) & N  & {\it S/N}\\
&(1)             & (2)             & (3)        & (4)         & (5)        & (6)                        & (7)                 & (8) &(9) \\ \hline \hline
\multicolumn{10}{c}{\textbf{First Group}}\\\hline
\textbf{Galaxy} &
Hen 3-938       & IRAS 13491-6318 & 13 52 42.8 & -63 32 49.2 & 2005-04-18 & 2453479.3        & 600                 & 1         &15\\
&                &                 &            &             &            &                 & 3600                 & 1 &60\\
&                &                 &            &             & 2016-06-14 & 2457554.1        & 2000                & 2  &40\\
&SS 255          & IRAS 14100-6655 & 14 13 59.0 & -67 09 20.6 & 2016-06-14 & 2457554.1        & 2400                & 2  &10\\
&Hen 2-91        & IRAS 13068-6255 & 13 10 04.8 & -63 11 30.0 & 2016-04-12 & 2457491.3        & 1100                & 1         &6\\
&                &                   &                &              & 2016-08-14 & 2457615.0        & 1800                & 2$^*$         &11\\
&                &                   &                &              & 2016-08-15 & 2457616.0        & 1800                & 1$^*$ &13\\
&                &                   &                &              & 2016-08-16 & 2457617.0        & 1800                & 1$^*$ &15\\
&                &                   &                &              & 2016-08-17 & 2457618.0        & 1800                & 1$^*$ &12\\ \hline
\textbf{SMC}    &
LHA 115-N 82    & $\cdots$        & 01 12 19.7 & -73 51 26.0 & 2008-12-24 & 2454825.6        & 1800                & 1  &60\\
&                &                 &            &             & 2015-07-06 & 2457210.3        & 3000                & 1$^*$ &50 \\ \hline
\textbf{LMC}    &
ARDB 54         & $\cdots$        & 04 54 43.4 & -70 21 27.5 & 2014-11-24 & 2456986.3        & 900                 & 2                 &20\\
&                &                 &            &             & 2015-12-01 & 2457358.1        & 1500                &2        &35 \\
&LHA 120-S 59    & $\cdots$        & 05 45 29.5 & -68 11 45.9 & 2015-12-06 & 2457363.2        & 2400                &2         &40\\ 
&                &                &                &             & 2016-12-04 & 2457727.1        &3400                       &1        &35\\
&                &                &                &             & 2016-12-05 & 2457728.3        &3400                       &1        &42\\ \hline \hline
\multicolumn{10}{c}{\textbf{Second Group}}\\\hline
\textbf{Galaxy} &
TYC 175-3772-1  & IRAS 07080+0605 & 07 10 43.9 & +06 00 07.9 & 2015-12-06 & 2457363.3         & 1500                & 2             &128        \\
&SS 147                & IRAS 07377-2523 & 07 39 48.0 & -25 30 28.2 & 2008-12-20 & 2454821.2        & 1800                & 1             &75\\
&CD-31 5070      & IRAS 07455-3143 & 07 47 29.3 & -31 50 40.3 & 2008-12-20 & 2454821.3        & 1500                & 2             &97        \\
&                &                 &            &             & 2015-12-05 & 2457362.3        & 1200                   & 2                 &110        \\
&                &                 &            &             & 2016-03-13 & 2457461.0        & 1200                & 2 &93 \\
&                &                 &            &             & 2016-04-12 & 2457491.1        & 1100                & 2  &62\\
&V* FX Vel       & IRAS 08307-3748 & 08 32 35.8 & -37 59 01.5 & 2008-12-21 & 2454822.3        & 900                 & 2                 &128\\
&                &                 &            &             & 2015-10-12 & 2457308.3        & 400                 & 2  &132\\
&                &                 &            &             & 2016-03-20 & 2457468.1        & 500                 & 1  &135\\
&                &                 &            &             & 2016-04-12 & 2457491.1        & 400                 & 2  &161\\
&BD+23 3183     & IRAS 17449+2320 & 17 47 03.3 & +23 19 45.3 & 2016-04-12 & 2457491.3        & 500                 & 1  &72\\
&                &                 &            &             &            &                 & 1100                & 1  & 121\\\hline
\textbf{SMC}    &
{[}MA93{]} 1116 & $\cdots$        & 00 59 05.9 & -72 11 27.0 & 2007-10-03 & 2454377.2        & 1800                & 2 &6\\
&                &                 &            &             & 2007-10-04 & 2454378.2        & 1800                & 2 &6\\
\hline        
\end{tabular}
    \begin{tablenotes}
        \item \textbf{Notes 1.} Column information: (1) name of the object; (2) IRAS identifier; (3) and (4) right ascension and declination from epoch 2000 obtained from CDS; (5) date of observation; (6) Julian Date (JD) of the start of the first exposure; (7) exposure time of each spectrum in seconds; (8) number of spectra at each observation (the public spectra obtained from ESO Science Archive Facility have an asterisk);  (9) signal-to-noise ({\it S/N}) around 5500~\AA\, of each spectrum.
        \item  \textbf{Notes 2.} Our sample is divided in 2 groups (see the text).
    \end{tablenotes}
\end{table*}

%==================================================================================
\begin{table*}
        \caption{Photometric data of our sample collected from the literature.}
        \label{table:objectsphotometry}
        \centering
        \begin{tabular}{ccccccc}
                \hline
&Name            & Year             & U                & B                & V                & References \\
&(1)             & (2)              & (3)              & (4)              & (5)              &    (6)    \\ \hline \hline
\multicolumn{7}{c}{\textbf{First Group}} \\ \hline
\textbf{Galaxy} & 
Hen 3-938       & 1990-1998          & 15.33            & 15.03            & 13.50            & 1     \\
&                & 1990-1995          & 15.02            & 14.92            & 13.40            & 2     \\
&                & 2009-10-01                   & $\cdots$                  & 14.78$\pm$0.04   & 13.40            & 3\\
&SS 255          & 2004-11-04       & $\cdots$         & 15.16            & 14.83            & 4  \\
&Hen 2-91       & 1980-03-01          & $\cdots$         & 15.30            & $\cdots$                & 5  \\
&                & 1989-1993                   & $\cdots$                  & 15.20                         & 14.38                        & 6\\ \hline
\textbf{SMC}    &
LHA 115-N 82    & 1989-07-5,7,8    & 14.24$\pm$0.02   & 14.37$\pm$0.02   & 14.25$\pm$0.02   &     7     \\
&                & 1995-11-13 to 23 & 14.78$\pm$0.05   & 14.75$\pm$0.03   & 14.75$\pm$0.03   &     8     \\
&                & 1999-01-08       & 14.43$\pm$0.01   & 14.35$\pm$0.01   & 14.24$\pm$0.01   &     9     \\ \hline
\textbf{LMC}    &
ARDB 54         & 1968             & 12.79            & 12.93            & 12.71            &     10$^*$     \\
&                                & 1995-11-13 to 23 & 12.81$\pm$0.01   & 13.02$\pm$0.01   & 12.77$\pm$0.01   &     11     \\
&                                & 1999-01-08       & 12.80            & 12.96            & 12.71$\pm$0.01   &     9     \\
&LHA 120-S 59    & 1991-12-01       & 13.62            & 14.62            & 14.41$\pm$0.03   &    12     \\
&                                & 1995-11-13 to 23 & 13.37$\pm$0.03   & 14.51$\pm$0.03   & 14.02$\pm$0.02   &    11     \\ \hline  \hline
\multicolumn{7}{c}{\textbf{Second Group}} \\ \hline
\textbf{Galaxy} & 
IRAS 07080+0605 & 2007          & 12.30            & 12.31            & 12.15            &     13$^*$     \\
&                & 1991-04-02           & $\cdots$                  & 12.345                         & 12.741                        & 14\\                
&IRAS 07377-2523 & 2007          & $\cdots$         & $\cdots$                  & 12.8             & 13$^*$  \\
&                & 2009-10-01                   & $\cdots$                   & 13.27$\pm$0.06   & 12.90$\pm$0.02        & 3\\
&IRAS 07455-3143 & 1971-1979        & 12.33            & 12.45            & 11.53            &     8     \\
&                & 2009-10-01                   &$\cdots$                  &        12.41$\pm$0.04   & 11.51$\pm$0.02        & 3\\
&                & 1989-1993                   &$\cdots$                  & 12.119                         & 11.51                    & 6\\
&V* FX Vel       & 1994-02-18       & 10.898           & 10.973           & 10.795           &     15     \\
&                & 2009-10-01                   & $\cdots$                  & 10.1$\pm$0.2         & 10.0$\pm$0.1     & 3\\
&                & 1989-1993                   & $\cdots$                  & 9.776                         & 9.724                        & 6\\        
&IRAS 17449+2320 & 2007          & 10.05            & 10.06            & 10.00            &     13$^*$     \\ \hline
\textbf{SMC}    & 
{[}MA93{]} 1116 & 1985-11-26       & 15.64            & 16.18            & 15.91            &     16     \\
&                & 1999-01-08       & 14.91$\pm$0.07   & 15.56$\pm$0.07   & 15.01$\pm$0.07   &     9     \\ 
\hline
\end{tabular}
\begin{tablenotes}
        \item \textbf{Notes.} Column information: (1) name of the object; (2) date of observation; (3)-(5) photometric data in the U, B, and V-bands; (6) references. Based on the literature, each set of data was taken at the same night.
        \item \textbf{References.} (1) \citet[][Pico dos Dias Survey]{Vieira_2003}; (2) \citet{Torres_1995}; (3) \citet[][AAVSO Photometric All Sky Survey (APASS) DR9]{Henden-et-al-2015}; (4)  \citet[][SPM 4.0 Catalog]{Girard-et-al-2011}; (5) \citet[][2MASS Catalog]{Cutri-et-al-2003}; (6) \citet[][The NOMAD-1 Catalog]{Zacharias-et-al-2004}; (7) \citet{Heydari-Malayeri-1990}; (8) \citet{Orsatti_1992}; (9) \citet{Massey_2002}; (10) \citet{Ardeberg_1972}; (11) \citet{Zaritsky_2004}; (12) \citet*{Gummersbach_1995}; (13) \citet{Miroshnichenko_2007}; (14) \citet[][ASCC-2.5 V3]{Kharchenko-2001}; (15) \citet{de_Winter_2001}; (16) \citet*{Massey_1989}. The asterisk means that no information about the exact dates of the photometric observations is provided, thus, we decided to assume the year of the publication.
\end{tablenotes}
\end{table*}

\begin{figure*}
        \centering
        \includegraphics[width=0.99\textwidth]{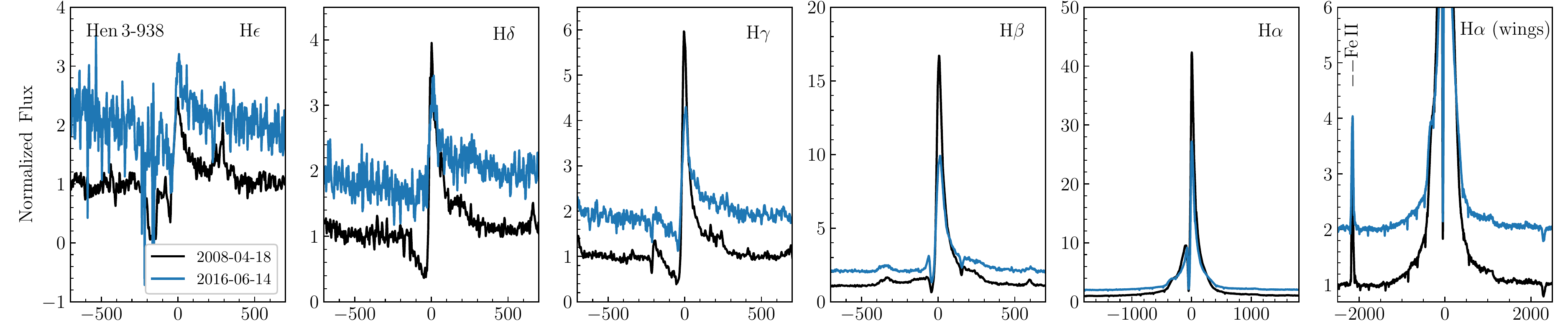}
        \includegraphics[width=0.99\textwidth]{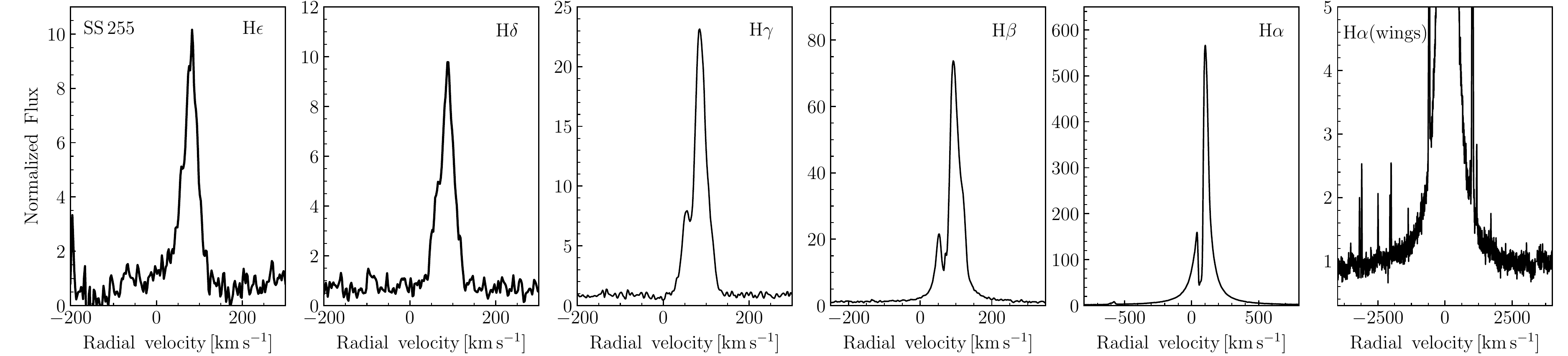}
        \includegraphics[width=0.99\textwidth]{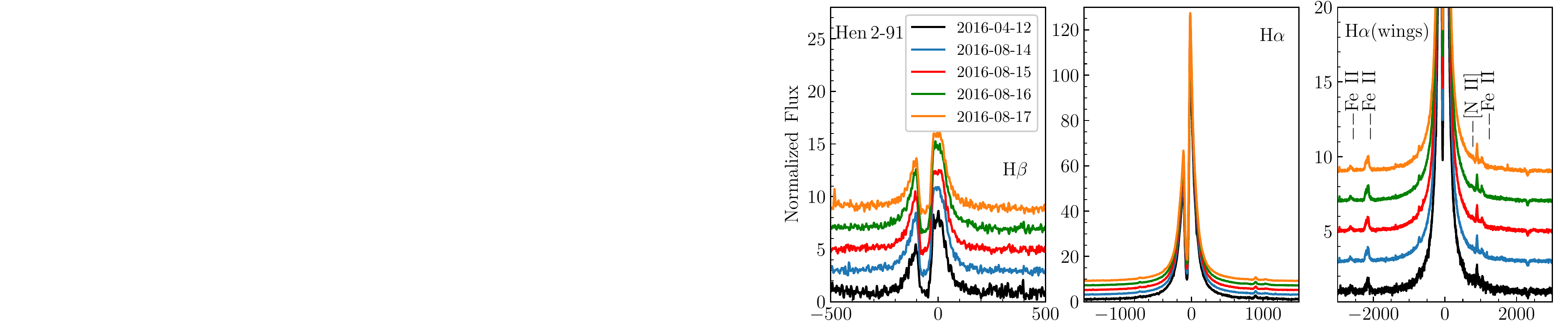}
        \includegraphics[width=0.99\textwidth]{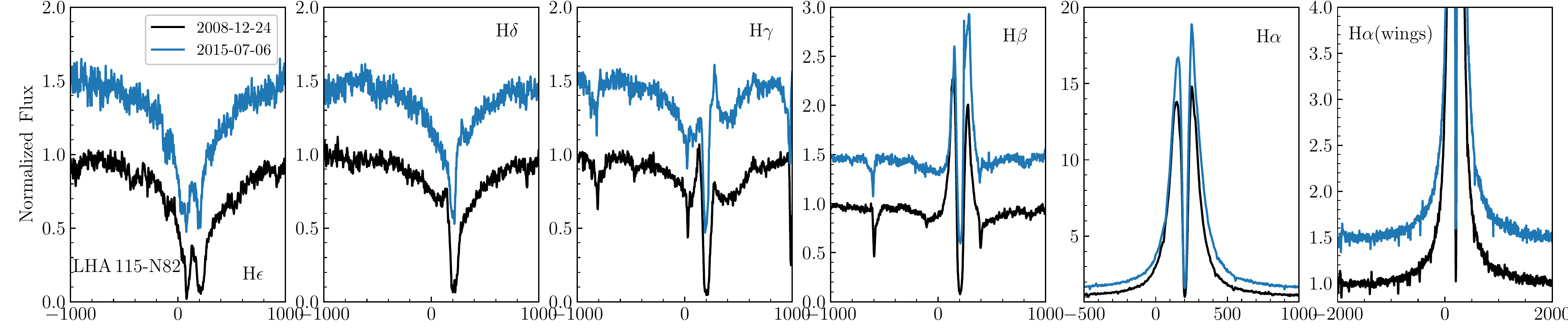}
        \includegraphics[width=0.99\textwidth]{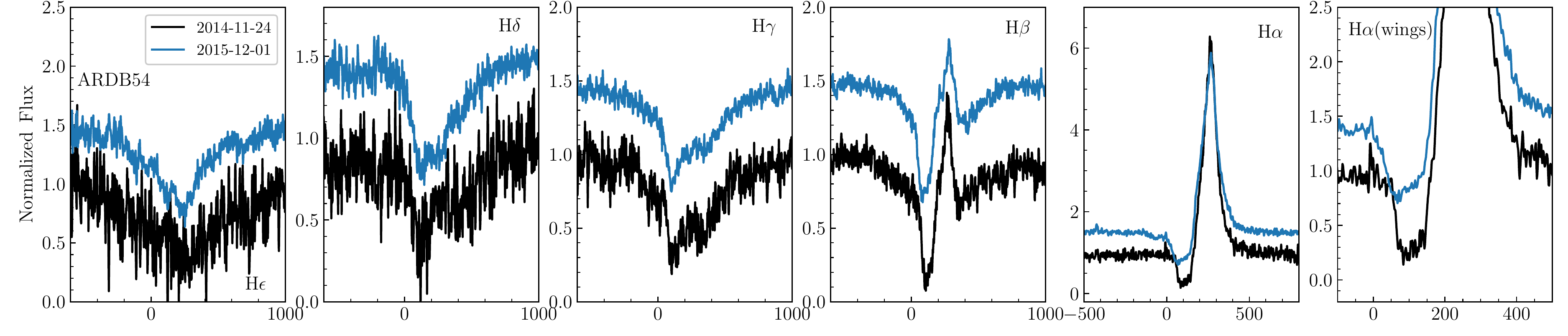}
        \includegraphics[width=0.99\textwidth]{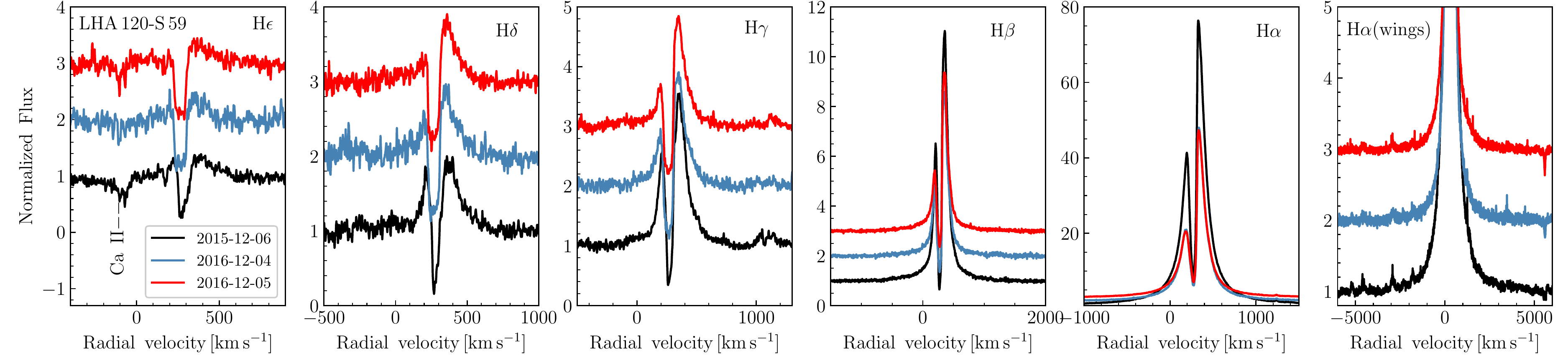}
        \caption{Balmer line profiles observed in the FEROS spectra of our sample (group 1). The first five columns show H$\epsilon$, H$\delta$, H$\gamma$, H$\beta$, and H$\alpha$, respectively. The last column zooms in the H$\alpha$ wings.}
        \label{fig:Balmer-lines}
\end{figure*}

\addtocounter{figure}{-1}
\begin{figure*}
        \centering
        \includegraphics[width=1\textwidth]{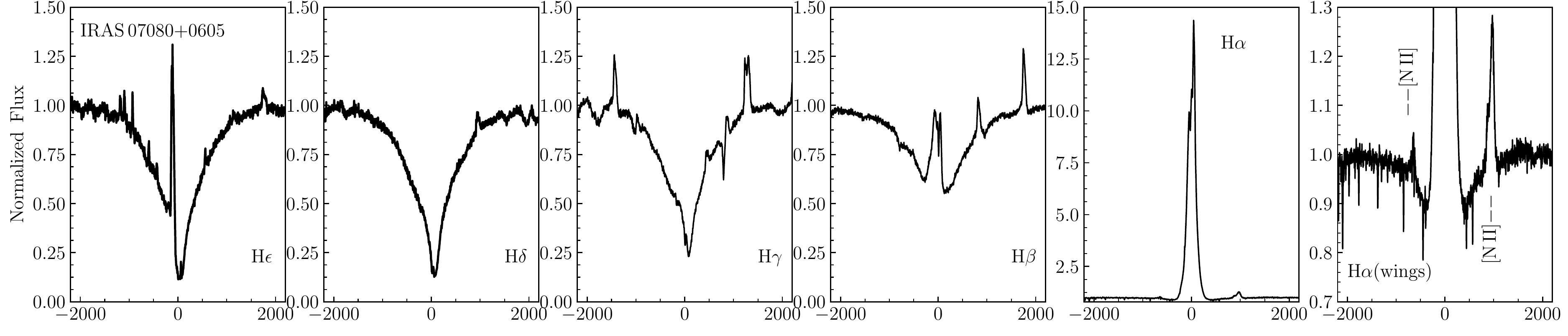}
        \includegraphics[width=1\textwidth]{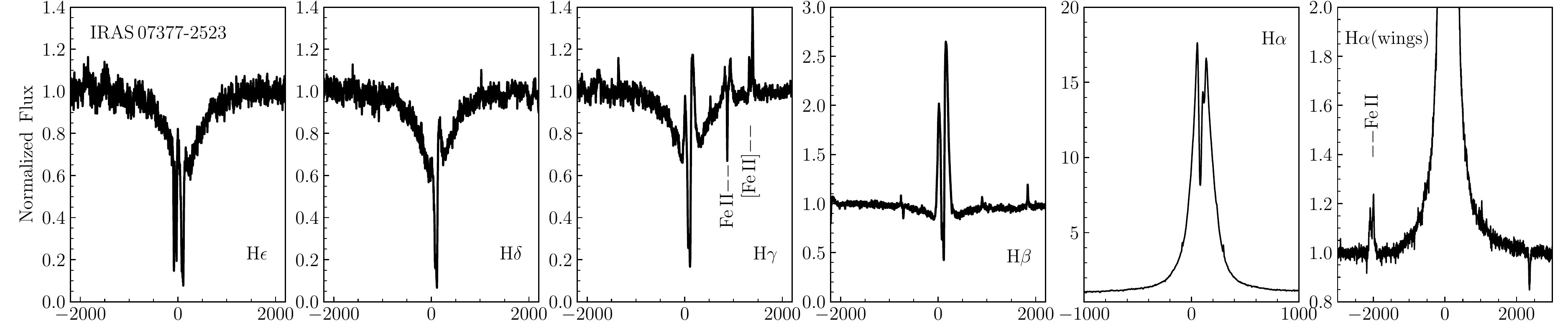}
        \includegraphics[width=1\textwidth]{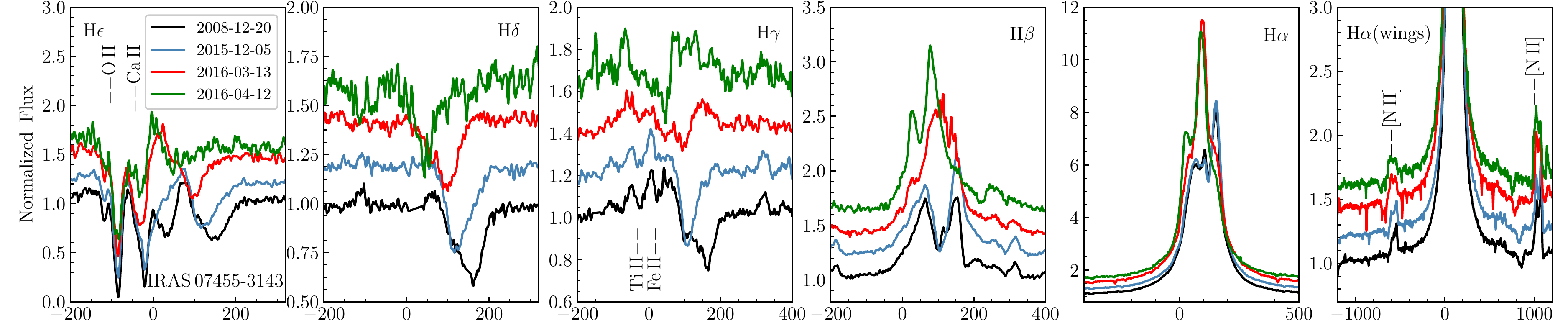}
        \includegraphics[width=1\textwidth]{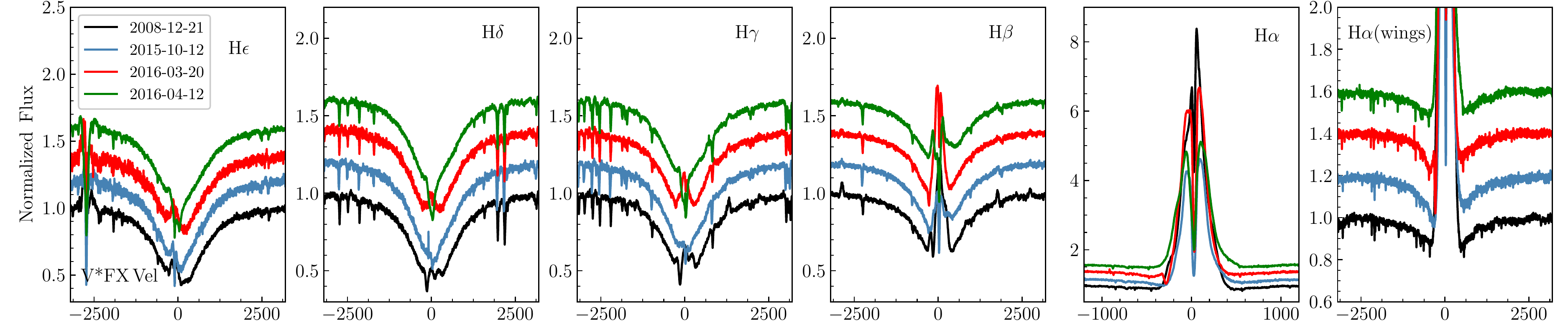}
        \includegraphics[width=1\textwidth]{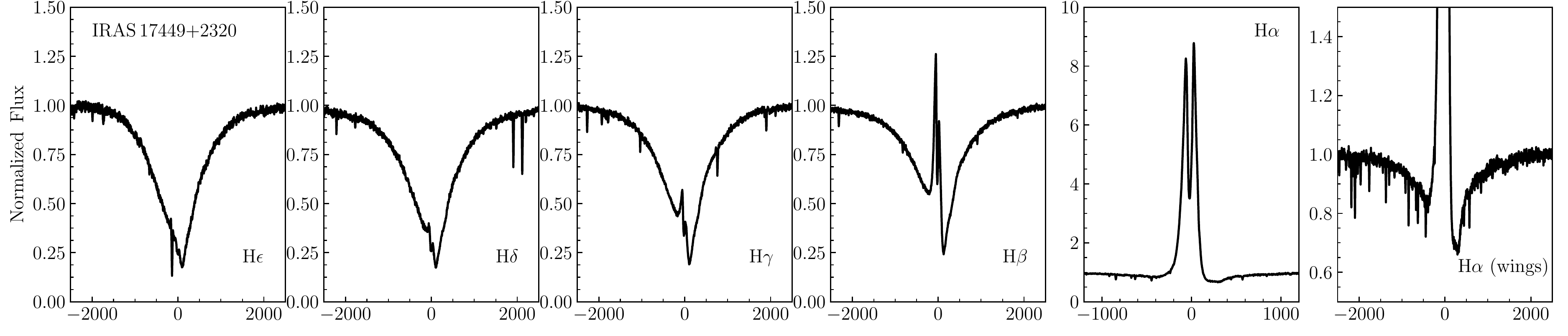}
        \includegraphics[width=1\textwidth]{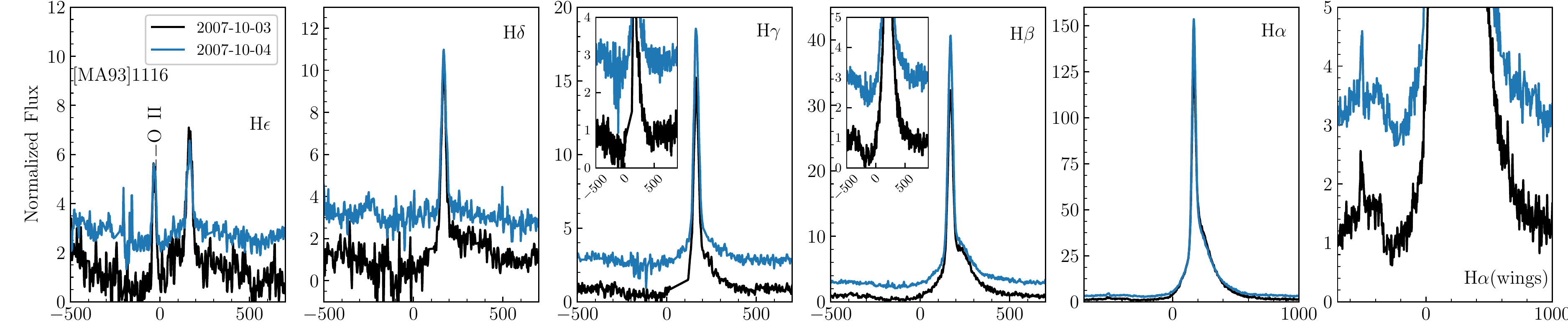}
        \caption{Continued (group2).}
\end{figure*}

Nowadays the number of stars with the B[e] phenomenon is around 150 objects identified in the Galaxy, Large Magellanic Cloud (LMC), Small Magellanic Cloud (SMC), M31, M33, and M81 \citep{Lamers-1998, Miroshnichenko_2007, Kraus-et-al-2014, Levato_2014, Kamath_2014, Miszalski-Mikolajewska-2014a, Kamath-at-al-2017, Humphreys-et-al-2017, 2018MNRAS.480.3706K, 2019AJ....157...22H}. Just a few of them were deeply studied and have their evolutionary stage confirmed. Thus, in this paper we decided to study a sample of 12 unclB[e] stars or candidates to the B[e] phenomenon from the Galaxy, LMC and SMC, through the analysis of photometric and high-resolution spectroscopic data. 

In Sect.~\ref{sec:sample}, we describe our observations and the public data used in this study. 
In Sect.~\ref{sec:Spectral-description}, we present a general description of the main spectral features identified in our sample. In Sect.~\ref{sec:Physical parameters}, we present the methodology used to derive the physical parameters of each star, the kinematics of the circumstellar environment, and the period analysis of light curves of some objects. In Sect.~\ref{sec:Discussion  of the  nature of our objects}, we discuss the possible nature of our objects and in Sect.~\ref{sec:Conclusions}, we summarize our conclusions.

%--------------------------------------------------------------------
%--------------------------------------------------------------------
\section{Our Sample and Observations}
\label{sec:sample}
Our sample is composed of 12 objects: 8 from the Galaxy, 2 from LMC and 2 from SMC, which exhibit or may exhibit the B[e] phenomenon, as seen in Table~\ref{table:objects}. We analysed in a homogeneous way photometric and high-resolution spectroscopic data of these objects. The sample can also be divided in 2 groups: the first group is composed of Hen\,3-938, SS\,255, Hen 2-91, LHA 115-N82, ARDB\,54, and LHA 120-S59, for which the analysis of high-resolution spectra (public or ours) was done for the first time; and the second one is composed of IRAS 07080+0605, IRAS 07377-2523, IRAS 07455-3143, V* FX Vel, IRAS 17449+2320, and [MA93] 1116 that were already studied by different authors using high-resolution spectroscopy, but for which we provide more details about their nature.

%--------------------
\subsection{High-Resolution Spectroscopy}

%----
For our analysis, we obtained high-resolution spectra using the \textit{Fiber-fed Extended Range Optical Spectrograph} \citep[FEROS,][]{Kaufer-et-al-1999}  attached to the 2.2-m ESO-MPI telescope, at La Silla Observatory (Chile). FEROS is a bench-mounted echelle spectrograph, which provides a resolution of 0.03~\AA/pixel (R$\sim$48000) and a spectral coverage from 3600 to 9200~\AA. 

The spectra of our sample were observed in 12 different epochs between 2005 and 2016 (Table\,\ref{table:objects}). The data obtained by us was reduced with the ESO/FEROS pipeline\footnote{\textsf{https://www.eso.org/sci/facilities/lasilla/instruments/feros/tools/ DRS.html}}, except for the spectra taken in 2005, which were reduced using MIDAS routines developed by our group, following standard echelle reduction procedures. The public data obtained from the ESO Science
Archive Facility was reduced by ESO phase 3\footnote{\textsf{http://archive.eso.org/cms/eso-archive-news/feros-pipeline-processed-data-available-through-phase-3.html}}. All spectra were corrected by heliocentric velocity and the {\it S/N} ratio is between 6 and 160 around 5500 \AA. We added up the spectra for stars that did not show variability during the night, in order to increase the {\it S/N}. We used standard \href{http://iraf.noao.edu/}{IRAF}\footnote{IRAF is distributed  by the National
Optical Astronomy Observatories, which are operated  by the Association of Universities for Research in Astronomy, Inc., under cooperative agreement with the National Science Foundation. See \textsf{http://iraf.noao.edu/}} tasks for normalization, cosmic ray removal and equivalent width measurements. We also used the TelFit code\footnote{\textsf{https://pypi.python.org/pypi/TelFit/1.3.2}}  \citep*{Gullikson-et-al-2014} for telluric correction of our FEROS spectra.

%---------------------

\subsection{Photometry}
\label{sec:Photometry}

In addition to the spectroscopic data, we searched for public photometric data, in order to derive the light curve (LC) of some of our objects and identify any possible photometric variation and periodicity. We collected data from \textit{\href{http://vizier.u-strasbg.fr/viz-bin/VizieR}{VizieR}}\footnote{http://vizier.u-strasbg.fr/viz-bin/VizieR}, \textit{\href{http://www.astrouw.edu.pl/asas/}{All Sky Automated Survey}\footnote{http://www.astrouw.edu.pl/asas/}} \citep[ASAS,][]{Pojmanski-2003}, the \textit{\href{http://ogledb.astrouw.edu.pl/~ogle/CVS/}{Optical Gravitational Lensing Experiment}\footnote{http://ogledb.astrouw.edu.pl/$\sim$ogle/CVS/}} \citep[OGLE III,][]{Udalski-Szymanski-2008},  and also from the literature, as can be seen in Table \ref{table:objectsphotometry}.

%---------------------------------------------------
\section{Spectral description}
\label{sec:Spectral-description}

From the high-resolution FEROS spectra, we described the spectral features and derived the radial velocities for all stars of our sample. 

For the identification of the spectral lines, we have used the line lists provided by \citet{Moore_1945}, \citet{Thackeray_1967}, \citet*{Landaberry_2001},  
NIST Atomic Spectra Database Lines Form\footnote{\textsf{http://physics.nist.gov/cgi-bin/AtData/lines/form}} and The Atomic Line List v2.04\footnote{\textsf{http://www.pa.uky.edu/$\sim$peter/atomic/}}. We also used the SpecView\footnote{\textsf{http://www.stsci.edu/institute/software$\_$hardware/specview}} identification tool for 1-D spectral visualization and analysis \citep{Busko_2000, Busko_2002, Busko_2002b, Busko_2012}.

Fig. \ref{fig:Balmer-lines} shows the Balmer lines (from H$\epsilon$ to H$\alpha$) present in the FEROS spectra. We note the different line profiles for each object, whose morphologies can be divided in four groups: (i) broad absorptions, probably of photospheric origin, as those for IRAS 07080+0605, (ii) broad absorptions superimposed with double or triple-peaked emissions of circumstellar origin in IRAS 07080+0605, IRAS 07377-2523 and IRAS 17449+2320; (iii) pure double- or triple-peaked emissions in IRAS 07455+3143, Hen 2-91 and SS 255; and (iv) P-Cygni profiles in Hen 3-938, [MA93] 1116 and ARDB 54. 

As typically seen for stars with the B[e] phenomenon, Fe\,{\sc ii} lines (permitted and forbidden ones) are the most numerous in the spectra of our stars. They also show different line profiles, like single- and double-peaked emission, shell-type, P-Cygni and inverse P-Cygni profiles.   

The [O\,{\sc i}] lines are one of the main defining characteristics of the B[e] phenomenon. These lines display single or double-peaked emission profiles in our sample, except for IRAS 07455+3143, which has no detectable [O\,{\sc i}] emission. Hence for all objects of our sample, except for IRAS 07455-3143, we confirmed the presence of the B[e] phenomenon.

We identified [Ca\,{\sc ii}] lines in some stars of our sample (IRAS 07080+0605, IRAS 07377-2523, IRAS 07455+3143, Hen 3-938, [MA93] 1116, LHA 115 N-82, and ARDB 54). Together with the lines of [O\,{\sc i}], they are excellent tracers for the kinematics of the circumstellar medium, as will be described in Sect.~\ref{subsec:fl}. 

Notably, some of our objects display absorption lines of He\,{\sc i}, Mg\,{\sc ii} and Si\,{\sc ii}, which are used in empirical relations to derive the spectral type of B- and A-type stars (see Sect.~\ref{subsec:Spectral Classification}).

In agreement with the literature, we confirmed the presence of Li\,{\sc i} and Ca\,{\sc i} lines in V* FX Vel and one Ca\,{\sc i} line in IRAS 07455+3143, indicating a possible companion, as will be discussed in Sect.~\ref{subsubsec:FXVel} and ~\ref{subsubsec:IRAS07455}, respectively. 

For some objects, we identified the existence of variability in comparison to the literature. In addition, based on multiple FEROS spectra, a clear variability is seen for four stars of our sample: IRAS 07455+3143, V* FX Vel, LHA 115 N-82, and LHA 120 S-59 (Fig. \ref{fig:Balmer-lines}).

The radial velocities of stars with no noticeable spectral variability were derived from [Fe\,{\sc ii}] and [O\,{\sc i}] lines (see Table~\ref{table:velocities}). These lines were chosen because they have, in general, symmetric emission line profiles. For stars with clear variability, we also measured radial velocities from detected absorption lines of He\,{\sc i}, Mg\,{\sc ii}, Si\,{\sc ii}, Ca\,{\sc i} and Li\,{\sc i} (see Table~\ref{table:velocities2}). The different radial velocities derived from permitted absorption and forbidden emission lines may also indicate binarity.
  
A detailed description of the spectral features present in the FEROS spectra of each star of our sample can be found in Appendix \ref{Apendix:Spectral Descriptions} and Table~\ref{table:atlas}.

%=========================================================
%=========================================================
\section{Physical parameters}
\label{sec:Physical parameters}
%----------------------------------------------------------------------
%----------------------------------------------------------------------
\begin{table}
        \caption{The radial velocity of stars from our sample that do not present sensible spectral variability or for which we have just one spectrum. The radial velocities are the average of values obtained from [O\,{\sc i}] and [Fe\,{\sc ii}] lines.}
        \label{table:velocities}
        \centering
        \begin{tabular}{llccc}
                \hline
&                Name            & Date          & [O\,{\sc i}] and [Fe\,{\sc ii}] \\
&                                                &               & (km~s$^{-1}$)                   \\ \hline\hline
                \textbf{Galaxy} &
                Hen 3-938       & 2005-04-18    & -22$\pm$3                 \\
&                                & 2016-06-14    & -22$\pm$2                                       \\
&                SS 255          & 2016-06-14    & 90$\pm$1                 \\
&                Hen 2-91        & 2016-04-12    & -47$\pm$4                 \\
&                                & 2016-08-14        & -47$\pm$3                        \\
&                                & 2016-08-15        & -46$\pm$2                        \\        
&                                & 2016-08-16        & -47$\pm$2                        \\        
&                IRAS 07080+0605 & 2015-12-06    & 10$\pm$2                  \\
&                IRAS 07377-2523 & 2008-12-20    & 90$\pm$3                  \\
&                 IRAS 17449+2320 & 2016-04-12    & -16$\pm$2                \\ \hline
                \textbf{SMC}    &
                {[}MA93{]} 1116     & 2007-10-03    & 166$\pm$1                \\
&                                & 2007-10-04    & 166$\pm$1                \\        \hline
                \textbf{LMC}    &
                ARDB 54         & 2014-11-24    & 240$\pm$6                 \\
&                                & 2015-12-01    & 235$\pm$4                 \\ \hline
\end{tabular} 
\end{table}

\begin{table*}
\caption{Radial velocities, derived from permitted absorption and forbidden emission lines, for the objects with strong variability.}
\label{table:velocities2}
\centering
\begin{tabular}{llccccccc}
                \hline
&Name            & Date          & [O\,{\sc i}] and [Fe\,{\sc ii}] & He\,{\sc i}   &Mg\,{\sc ii}   & Si\,{\sc ii}  & Ca\,{\sc i}   &  Li\,{\sc i}      \\
&                        &               & (km~s$^{-1}$)                   & (km~s$^{-1}$) & (km~s$^{-1}$) & (km~s$^{-1}$) & (km~s$^{-1}$) & (km~s$^{-1}$)  \\ \hline\hline
\textbf{Galaxy} &
IRAS 07455-3143 & 2008-12-20    & 106$\pm$6   &140$\pm$5        &146$\pm$1  &151$\pm$3  &120$\pm$1  &$\cdots$ \\
&                                & 2015-12-05    & 107$\pm$7   &111$\pm$5        &122$\pm$1  &122$\pm$1  &100$\pm$1  &$\cdots$\\
&                                & 2016-03-13    & 103$\pm$7   &77$\pm$6                &84$\pm$1   &84$\pm$3   &54$\pm$1   & $\cdots$\\
&                                & 2016-04-12    & 97$\pm$8    &37$\pm$7                &42$\pm$1        &44$\pm$5   &22$\pm$1   & $\cdots$\\
                                %------------
&V* FX Vel       & 2008-12-21    & 22$\pm$1            &44$\pm$1           &51$\pm$1        &52$\pm$2  &24$\pm$1        &28$\pm$1\\
&                                & 2015-10-12    & 21$\pm$4    &22$\pm$1     &29$\pm$3        &31$\pm$3  &32$\pm$1        &30$\pm$1\\
&                                & 2016-03-20    & 16$\pm$4    &40$\pm$1     &42$\pm$2        &43$\pm$2  &23$\pm$1        &24$\pm$1\\
&                                & 2016-04-12    & 19$\pm$4    &37$\pm$1     &41$\pm$2        &42$\pm$2  &$\cdots$         &30$\pm$1\\ \hline
\textbf{SMC}    &
LHA 115-N82             & 2008-12-24    & 206$\pm$2   &213$\pm$7        &220$\pm$3        &224$\pm$2 &$\cdots$   &$\cdots$\\ 
&                                & 2015-07-06    & 206$\pm$2   &179$\pm$4    &175$\pm$1        &181$\pm$3 &$\cdots$   &$\cdots$ \\ \hline
\textbf{LMC}    &
LHA 120-S59                    & 2015-12-07    & 293$\pm$4        &$\cdots$        &$\cdots$        &$\cdots$        &$\cdots$        &$\cdots$        \\
&                & 2016-12-05        & 298$\pm$3        &301$\pm$6        &$\cdots$        &$\cdots$        &$\cdots$        &$\cdots$        \\
&                & 2016-12-06        & 295$\pm$3        &292$\pm$4        &$\cdots$        &$\cdots$        &$\cdots$        &$\cdots$         \\
                \hline 
\end{tabular}
\end{table*}

%----------------------------------------------------------------------
%----------------------------------------------------------------------
\subsection{Spectral  Classification}
\label{subsec:Spectral Classification}

The determination of the spectral type and luminosity class for objects with the B[e] phenomenon is rather complicated due to the absence, in general, of photospheric lines and the contamination by circumstellar emission. Thus, we need to deal with indirect methods, which have different levels of uncertainty.

\begin{table*}
        \centering
        \caption{Spectral type, luminosity class, intrinsic color index, and effective temperature for some stars of our sample, obtained using the different methods described in the text.}
                \label{table:SpectralType}
\begin{tabular}{llcccccc}
                \hline \hline
                \multicolumn{7}{c}{\textbf{Method 1 } }                   \\
                Star& \multicolumn{3}{l}{Sp.type: L.C.}       & $(B-V)_0$       & $T_{\rm eff}$ [K] &  \\  \hline
Hen 3-938   &\multicolumn{3}{l}{B0-B1: I}         & -0.21$\pm$0.02  &23400$\pm$2600             & \\
IRAS 07080+0605 & \multicolumn{3}{l}{A0-A1: II}       & -0.01$\pm$0.02  &9700$\pm$400               & \\
IRAS 07455-3143 & \multicolumn{3}{l}{B0-B1: II/III/V} &-0.28$\pm$0.02   &26000$\pm$4000             & \\                
V* FX Vel       & \multicolumn{3}{l}{B8-B9: III/V}    &-0.09$\pm$0.02   &11500$\pm$900              & \\
IRAS 17449+2320 &\multicolumn{3}{l}{A1-A2: II/III}    &0.03$\pm$0.02    &9200$\pm$300               & \\
{[}MA93{]} 1116  &\multicolumn{3}{l}{B1-B2: II/III/V}  & -0.25$\pm$0.01  &21600$\pm$3000             & \\
LHA 115-N82 &\multicolumn{3}{l}{B8-B9: II/V}      & -0.09$\pm$0.02  &11200$\pm$700              & \\
                & \multicolumn{3}{l}{A0-A2: III}      & 0.06$\pm$0.09   &9100$\pm$1000              & \\
ARBD 54     &\multicolumn{3}{l}{A0-A1: I}         & 0.01$\pm$0.02   &9500$\pm$200               & \\ \hline \hline
                %------------------
                \multicolumn{7}{c}{\textbf{Method 2}}                   \\
                Star               & \multicolumn{3}{l}{Date}   &  Mg\,{\sc ii}~4482\,\AA / He\,{\sc i}~4471\,\AA  & Sp.type  & $T_{\rm eff}$ [K] \\ \hline
IRAS 07377-2523 &\multicolumn{3}{l}{2008-12-21}  &1.08$\pm$0.05       &B8-B9   & 12000$\pm$1000\\
IRAS 07455-3143 &\multicolumn{3}{l}{2016-04-13}  &0.97$\pm$0.03       &$\sim$B8         & 12500$\pm$500\\
V* FX Vel             &\multicolumn{3}{l}{2008-12-22}  &5.57$\pm$1.04       &$\leq$A2         & $\leq$9000 \\
                                &\multicolumn{3}{l}{2015-12-06}  &6.53$\pm$0.31       &$<$A2                        & $<$9000 \\
                                &\multicolumn{3}{l}{2016-03-21}  &4.34$\pm$0.26       &A0-A2               & 9500$\pm$500 \\
                                &\multicolumn{3}{l}{2016-04-13}  &4.85$\pm$1.28       &$\sim$A2         & $\sim$9000 \\  
IRAS 17449+2320 &\multicolumn{3}{l}{2016-04-13}  &4.29$\pm$0.73       &A0-A2               & 9500$\pm$500 \\
 \hline \hline
                %------------------------------
\multicolumn{7}{c}{\textbf{Method 3}}                   \\
Star               & \multicolumn{3}{l}{Date}   & He\,{\sc i}~ 4713\,\AA / Si\,{\sc ii}~6347\,\AA & He\,{\sc i}~5875\,\AA / Si\,{\sc ii}~6347\,\AA & $T_{\rm eff}$ [K] \\ \hline
IRAS 07080+0605          &\multicolumn{3}{l}{2015-12-07}       &$\cdots$                     &0.62$\pm$0.30          & 10500$\pm$1000    \\
IRAS 07377-2523   &\multicolumn{3}{l}{2008-12-21}    &0.23$\pm$0.09       &1.15$\pm$0.20           & 12000$\pm$1000  \\  
IRAS 17449+2320    &\multicolumn{3}{l}{2016-04-13} &0.11$\pm$0.02       &0.75$\pm$0.05           &10700$\pm$1000\\
%------------------------------------
\hline
\end{tabular}
\end{table*}

One of these methods was described by \citet{Borges-Fernandes-2009}, where through the observed color indices, it is possible to derive the intrinsic ones, such as $(U-B)_0$ and $(B-V)_0$, and the total extinction of each object (hereafter Method 1).

Based on empirical spectroscopic criteria, using equivalent width ratios of photospheric lines, we can also estimate the spectral classification for B- and A-type stars in the Galaxy and in the Magellanic Clouds. We chose the relation that associates the spectral type to the Mg\,{\sc ii}~4482\,\AA / He\,{\sc i}~4471\,\AA \ equivalent width ratio (hereafter Method 2), as done by \citet{Lennon-1997}, \citet{Evans-2003} and \citet[][their fig. 3]{Kraus-2008}. In order to estimate the effective temperature, we also used the He\,{\sc i}~4713\,\AA / Si\,{\sc ii}~6347\,\AA \ and He\,{\sc i}~5875\,\AA / Si\,{\sc ii}~6347\,\AA \ equivalent width ratios, as in \citet*[][their fig. 3]{Khokhlov-et-al-2017} (hereafter Method 3). Both Methods 2 and 3 are only used if these lines are of photospheric origin, i.e., they are in absorption without contamination from the wind or the circumstellar emission. The results from these different methods can be seen in Table~\ref{table:SpectralType} and a detailed analysis for each star is seen in Sect.~\ref{sec:Discussion  of the  nature of our objects}.

%----------------------------------------------------------------------------------
%----------------------------------------------------------------------------------
\subsection{Interstellar, circumstellar and total extinction}
\label{subsec:interstellar-extinction}

Stars with the B[e] phenomenon have a complex circumstellar structure, making it difficult to disentangle the interstellar and circumstellar contributions from the total extinction.

Therefore, in order to determine the interstellar extinction or color excess, $E(B-V)_{\text{IS}}$, of our Galactic objects, we used the diffuse interstellar band (DIB) at 5780~\AA \ and the empirical relation described by \citet{Herbig-1993}. For the objects that do not present this DIB and objects located in the SMC and LMC, we used values from \href{http://irsa.ipac.caltech.edu/applications/DUST/}{IRSA/Galactic Dust Reddening and
Extinction}\footnote{\textsf{http://irsa.ipac.caltech.edu/applications/DUST/}} \citep{Schlafly-2011}. In addition, for objects with declination of $\delta \gtrsim-30^{\circ}$, we also used \href{http://argonaut.skymaps.info/}{3D dust mapping}\footnote{\textsf{http://argonaut.skymaps.info/}} \citep{Green-et-al-2018}. 

In order to derive the visual interstellar extinction, $A_V$, we assumed $A_V/E(B-V)_{IS}=$ 3.1 for Galaxy \citep*[e.g.,][]{Cardelli-et-al-1989}, 2.74 for SMC  and 2.76 for LMC stars \citep[e.g.,][]{Gordon-et-al-2003}, see Table~\ref{table:interstellar-extinction}. For the total extinction of each object, we used the relation: $E(B-V)_{\text{T}} = (B-V)-(B-V) _0$. These color indices are obtained from Tables~\ref{table:objectsphotometry} and~\ref{table:SpectralType}, respectively. For the circumstellar extinction, we used the relation: $E(B-V)_{\text{CS}} = E(B-V)_{\text{T}} - E(B-V)_{\text{IS}}$. Our results can be seen in Table~\ref{table:interstellar-extinction} and in Sect.~\ref{sec:Discussion  of the  nature of our objects}.

\subsection{Modeling optical forbidden emission lines}
\label{subsec:fl}

Optical forbidden lines have been used to describe the kinematics of the circumstellar medium of stars 
with the B[e] phenomenon \citep[e.g.,][]{2005A&A...441..289K, Aret_2016}. These lines are 
optically thin, and their profiles mirror the kinematics within the line-forming regions.
But in contrast to the forbidden emission lines that are usually seen in low-density nebulae, the lines
of [O\,{\sc i}] $\lambda\lambda$5577,6300,6363 and [Ca\,{\sc ii}] $\lambda\lambda$7291,7323 are often 
found associated with the high-density (quasi-)Keplerian circumstellar or circumbinary 
rings or disks of the B[e] stars \citep{2010A&A...517A..30K, Kraus-2016, 2017AJ....154..186K, 
2012MNRAS.423..284A, 2018A&A...612A.113T, 2018MNRAS.480..320M}. To extract the information about the 
dynamics of the atomic and ionized gas around our objects and to search for indication of 
circumstellar disks/rings, we model the profiles of the forbidden emission lines, focusing on the 
[Ca\,{\sc ii}] and [O\,{\sc i}] lines.  

\begin{figure*}
\centering
\includegraphics[scale=0.8]{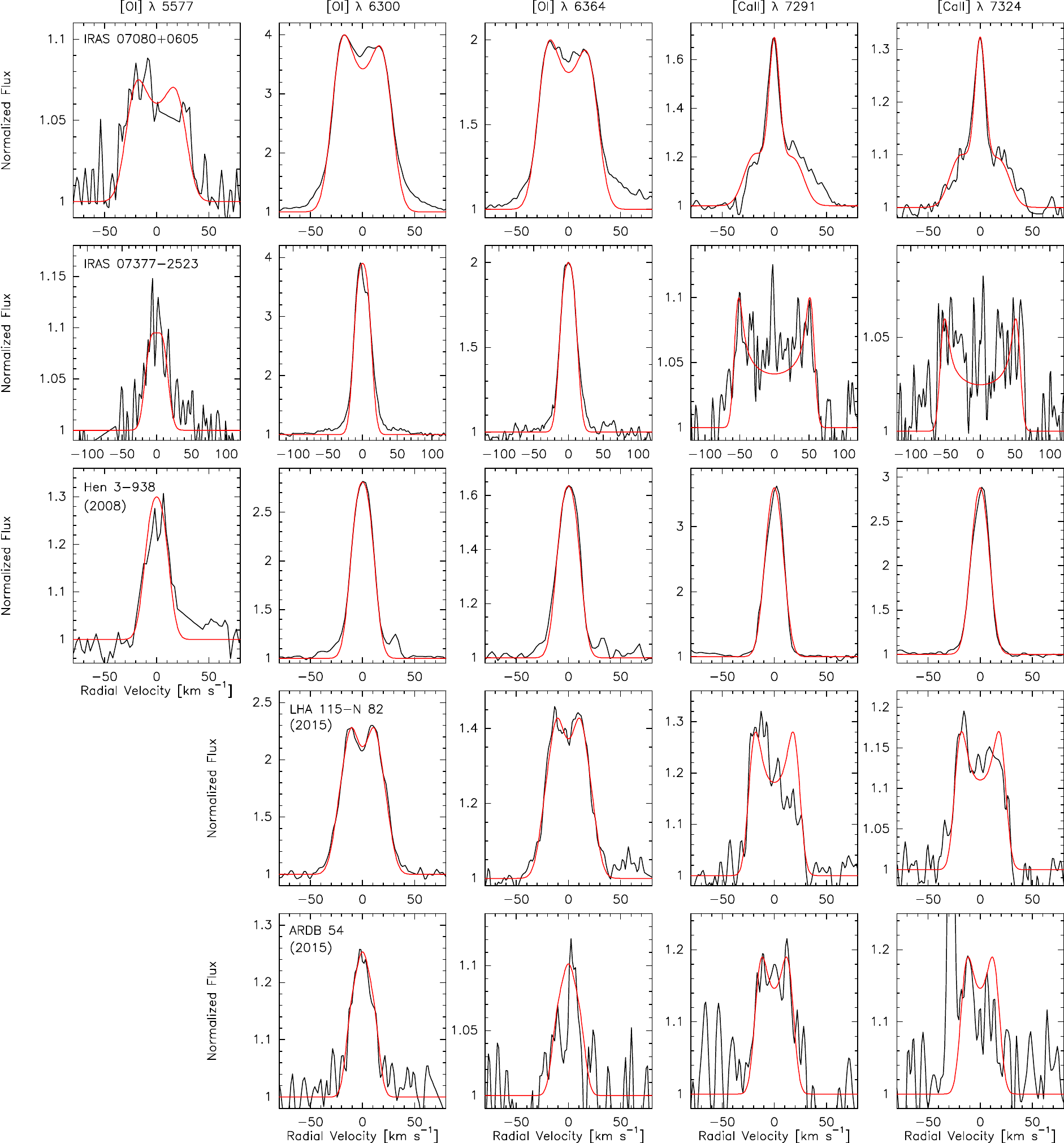}
\caption{Fit (red) to the observed (black) profiles for stars that have both sets of forbidden lines,  
[O\,{\sc i}] and [Ca\,{\sc ii}].}
\label{fig:fits-OI-CaII}
\end{figure*}

\begin{figure}
\centering
\includegraphics[scale=0.92]{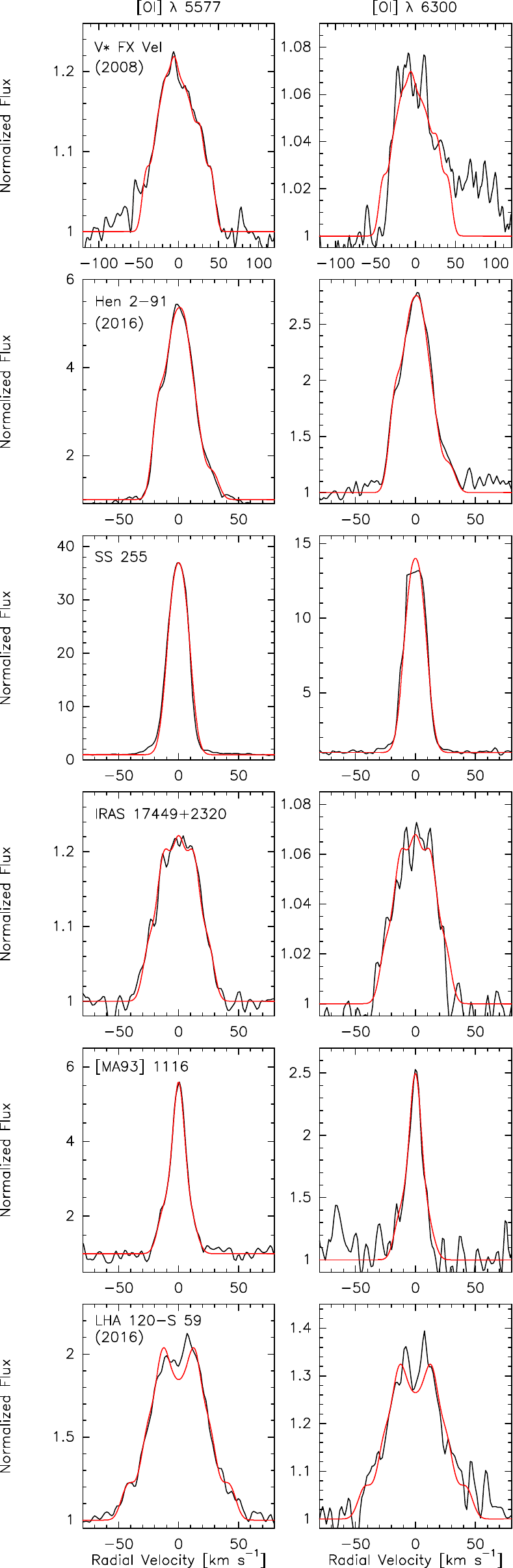}
\caption{Fit (red) to the observed (black) profiles for stars with only [O\,{\sc i}] forbidden lines.}
\label{fig:fits-OI}
\end{figure}

\begin{figure}
\centering
\includegraphics[scale=0.92]{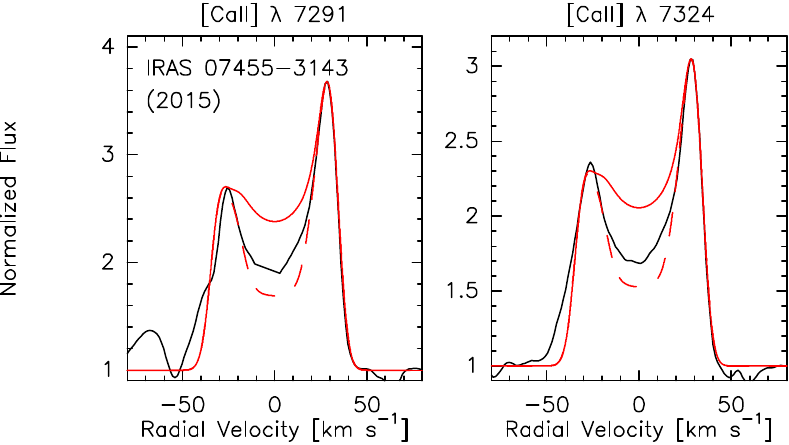}
\caption{Fits (red) to the observed (black) [Ca\,{\sc ii}] profiles for IRAS\,07455-3134, using a ring model with one (solid) and two (dashed) gaps.}
\label{fig:fits-CaII}
\end{figure}

Five of our objects display both sets of disk-tracing lines (Fig.\,\ref{fig:fits-OI-CaII}) whereas in 
six objects only the [O\,{\sc i}] lines were detected (Fig.\,\ref{fig:fits-OI}), and one object only displays the [Ca\,{\sc ii}] lines (Fig.\,\ref{fig:fits-CaII}). The absence of the 
[Ca\,{\sc ii}] lines in half of our sample could indicate that the density in their environments is 
lower than in the other objects. This conclusion is in line also with the absence of [O\,{\sc i}] 
$\lambda$5577 in these stars, which requires also higher densities than the [O\,{\sc i}] 
$\lambda\lambda$6300,6363 lines, but not as high as the [Ca\,{\sc ii}] lines, to generate 
measurable amounts of emission. No trend is seen regarding the presence or absence of individual sets of lines with respect to the lower metallicity of the Magellanic Cloud stars. This implies that the density structure within the circumstellar environment of these stars is not a direct consequence of the stellar mass-loss rate via a smooth wind which is known to be metallicity dependent.

The shapes of the profiles of the forbidden lines are either single-, double-, or multiple-peaked. 
For the single-peaked lines, a pure Gaussian component cannot fit the shape. The profiles require a 
non-Gaussian component, which might indicate that the gas revolves the central object. 
For simplicity we utilize  a pure kinematic model to reproduce the profile shapes and hence
the kinematics of the circumstellar gas. We assume that the emission originates from a thin ring
of material revolving the central star. To compute the profile function, we need to specify two 
velocity components: the component of the rotational velocity projected to the line of sight 
$v_{\rm rot, los}$, and a Gaussian component $v_{\rm gauss}$. The latter combines the broadening 
contributions from thermal motion, which is on the order of 1--2\,km\,s$^{-1}$, and from possible 
turbulent motion of the gas, which can be on the order of a few km\,s$^{-1}$. The resulting line profile is 
convolved to the spectral resolution of 6.5\,km\,s$^{-1}$ of FEROS.
In cases where a single ring is insufficient to reproduce the observed profile shape, 
we add one or more rings.
In some cases where the profiles are very asymmetric, we allow for gaps in the
rings. We note, however, that especially those multi-component models might not provide unique
solutions, and other scenarios might result in similar profiles. 

Our results are included in Figs.\,\ref{fig:fits-OI-CaII} - \ref{fig:fits-CaII} and the parameters 
needed for the model fits are listed in Tables\,\ref{tab:velocities-1} and \ref{tab:velocities-2}. For 
an easier comparison with the models, we centered the observed line profiles around zero by correcting 
for the radial velocities listed in Tables\,\ref{table:velocities} and \ref{table:velocities2}. A detailed discussion for each star is provided in Sect.~\ref{sec:Discussion  of the  nature of our objects}.

\subsection{Period analysis} \label{sec:45}

Six stars from our sample (IRAS\,07080+0605, V*FX~Vel, [MA 93] 116, LHA 115-N 82, LHA~120-S 59, and IRAS 07377-2523) were investigated in photometric surveys (Sect.~\ref{sec:Photometry}). Two of them were excluded from the present period analysis: IRAS 07377-2523 had very poor data coverage, while the LHA 115-N 82 data show that it is dominated by pronounced long-term variability that exceeds the total time span (Sect.~\ref{subsubsec:N82}).

\begin{landscape}
\begin{table}
        \caption{Interstellar, circumstellar, and total color excess, $E(B-V)$, and visual interstellar extinction, $A_V$, for the objects of our sample.}
        \label{table:interstellar-extinction}
        \centering
        \begin{tabular}{llccccccccc}
                \hline
&Star                &EW(DIBs)        &$E(B-V)_{\text{IS}}^{\text{DIBs}}$ &$E(B-V)_{\text{IS}}^{\text{IRSA}}$ &$E(B-V)_{\text{IS}}^{\text{3D}}$ &$E(B-V)_{\text{IS}}$        &$E(B-V)_{\text{CS}}$        &$E(B-V)_{\text{T}}$    &$A_V$         &$E(B-V)_{\text{lit}}$ \\
%&(m\AA)                &(mag)                        &(mag)                        &(mag)         &(mag)        &(mag)                &(mag)        &(mag) & (mag)\\
&(1)&(2)                &(3)                        &(4)                        &(5)         &(6) &(7)                &(8)        &(9) & (10)\\
                \hline        \hline        
\textbf{Galaxy}            &
Hen 3-938         &0.85        &1.64$\pm$0.02    &2.36$\pm$0.08        &$\cdots$  &1.64$\pm$0.02  &0.10$\pm$0.04    &1.74$\pm$0.02    &5.08$\pm$0.06    &0.45$^\text{a}$\\
%-----
&SS 255            &$\cdots$    &$\cdots$        &0.45$\pm$0.01        &$\cdots$   &0.45$\pm$0.01   &$\cdots$        &$\cdots$        &1.40$\pm$0.03   &\\
%----
&Hen 2-91            &1.54        &2.92$\pm$0.02    &6.56$\pm$1.34 &$\cdots$  &2.92$\pm$0.02    &$\cdots$        &$\cdots$        &9.05$\pm$0.06
    &2.34$^\text{b}$, 1.87$^\text{c}$\\
&IRAS 07080+0605                &$\cdots$                &$\cdots$                &0.14$\pm$0.01         &0.05$\pm$0.02        & 0.05$\pm$0.02        &0.11$\pm$0.04        &0.16$\pm$0.02        &0.16$\pm$0.06        & $\sim$0.10$^\text{d}$\\
%-----
&IRAS 07377-2523         &0.50               &0.98$\pm$0.02  &0.85$\pm$0.03        &0.50$\pm$0.03         &0.50$\pm$0.03    &$\cdots$        &$\cdots$        &1.55$\pm$0.09    &$\sim$0.63$^\text{e}$\\
%------------
&IRAS 07455-3143         &0.70               &1.37$\pm$0.02  &0.93$\pm$0.01         &$\cdots$                &1.15$\pm$0.22    &0.05$\pm$0.23        &1.20$\pm$0.02       &3.56$\pm$0.68    &$\sim$1.13$^\text{a}$, 1.17$^\text{f}$\\
%-------------
&V* FX Vel           &0.02                &0.05$\pm$0.03  &1.16$\pm$0.03        &$\cdots$  &0.05$\pm$0.03   &0.22$\pm$0.05    &0.27$\pm$0.08    &0.15$\pm$0.09 &\\
%------
&IRAS 17449+2320$^{*}$   &0.07        &0.14$\pm$0.03    &0.07$\pm$0.01        &0.05$\pm$0.02 &0.05$\pm$0.02  &$\cdots$        & $\cdots$       &0.16$\pm$0.06&\\
\hline
\textbf{SMC}   &
{[}MA93{]} 1116&$\cdots$    &$\cdots$        &0.42$\pm$0.07$^{**}$        &$\cdots$  &0.42$\pm$0.07   &0.10$\pm$0.08        &0.52$\pm$0.01        &1.15$\pm$0.22   &0.35$^\text{g}$\\
%-----
&LHA 115-N82    &$\cdots$    &$\cdots$        &0.04$\pm$0.01         &$\cdots$ &0.04$\pm$0.01   &0.17$\pm$0.03    &0.21$\pm$0.02    &0.11$\pm$0.03    &0.03$^\text{h}$, 0.12$^{\text{i}}$\\
\hline
\textbf{LMC}   &
ARDB 54        &$\cdots$    &$\cdots$        &0.11$\pm$0.01 &$\cdots$  &0.11$\pm$0.01   &0.13$\pm$0.03    &0.24$\pm$0.02    &0.30$\pm$0.03    &0.15$^\text{j}$\\
%------
&LHA 120-S59    &$\cdots$    &$\cdots$        &0.40$\pm$0.01                &$\cdots$  &0.40$\pm$0.01  &$\cdots$        &$\cdots$        &1.10$\pm$0.03 & 0.15$^\text{j}$, 0.05$^\text{k}$\\
\hline 
\end{tabular}
\begin{tablenotes}
        \item \textbf{Notes 1.} Column information: (1) name of the object; (2) equivalent width of the DIB at 5780~\AA~ in m\AA; (3) interstellar color excess derived from the DIB; (4) interstellar color excess taken from \href{http://irsa.ipac.caltech.edu/applications/DUST/}{IRSA}; (5) interstellar color excess taken from \href{http://argonaut.skymaps.info/}{3D dust mapping}; (6) interstellar color excess adopted in this work; (7) circumstellar color excess;  (8) total color excess; (9) visual interstellar extinction, $A_V$; (10) interstellar color excess from the literature. 
        \item \textbf{Notes 2.} ($^*$) the total color excess is negative using the colors from Table\,\ref{table:objectsphotometry}, thus, we decided to consider only its interstellar extinction; ($^{**}$) minimum value of interstellar color excess from IRSA.
        \item \textbf{References.} (a) \citet{Vieira_2011};
(b) \citet{Pereira_2003};
(c) \citet{Cidale_2001};
(d) \citet{Miroshnichenko_2007};
(e) \citet*{Chen-et-al-2016};
(f) \citet{Orsatti_1992};
(g) \citet{Wisniewski_2007};
(h) \citet{Kamath_2014};
(i) \citet{Heydari-Malayeri-1990};
(j) \citet{Levato_2014};
(k) \citet{Gummersbach_1995}.
\end{tablenotes}
\end{table}

\begin{table}
\centering
\caption{Velocities needed for fitting the line profiles for the stars with both [O\,{\sc i}] and 
[Ca\,{\sc ii}] forbidden lines. Multiple rows per line list the multiple fitting components. Parameters for the [O\,{\sc i}] $\lambda\lambda$6300,6364 lines are identical. All velocities have units km\,s\,$^{-1}$.}
\label{tab:velocities-1}
\begin{tabular}{lccccccccccccccc}
\hline
\hline
Line  & & \multicolumn{2}{c}{IRAS\,07080+0605} & & \multicolumn{2}{c}{IRAS\,07377-2523} & & \multicolumn{2}{c}{Hen\,3-938} & & \multicolumn{2}{c}{LHA\,115-N\,82} & & \multicolumn{2}{c}{ARDB\,54}\\
      & & $v_{\rm rot, los}$ &  $v_{\rm gauss}$  & &  $v_{\rm rot, los}$ &  $v_{\rm gauss}$  & &  $v_{\rm rot, los}$ &  $v_{\rm gauss}$  & &  $v_{\rm rot, los}$ &  $v_{\rm gauss}$ & &  $v_{\rm rot, los}$ &  $v_{\rm gauss}$ \\
\hline 
\protect{[O\,{\sc i}] $\lambda$5577}  & & 25$\pm$0.5$^{a}$ & 10$\pm$1 & & 13$\pm$0.5 & 7.5$\pm$0.5 & & 9$\pm$0.5 & 6.0$\pm$0.5 & & --- & --- & & --- & ---\\
\protect{[O\,{\sc i}] $\lambda$6300}  & & 25$\pm$0.5$^{a}$ & 10$\pm$1 & & 10$\pm$0.5 & 6.0$\pm$0.5 & & 9$\pm$0.5 & 6.0$\pm$0.5 & & 22$\pm$0.5 & 7.5$\pm$0.5 & & 13$\pm$0.5 & 2.5$\pm$0.5\\
                                      & &            &          & &            &            & &                          
&      & & 13$\pm$0.5 & 2.5$\pm$0.5 & &  3$\pm$0.5 & 2.5$\pm$0.5\\ 
\protect{[Ca\,{\sc ii}] $\lambda$7291}  & & 25$\pm$0.5$^{a}$ & 10$\pm$1 & & 55$\pm$1 & 2.5$\pm$0.5 & & 7$\pm$0.5 & 6.0$\pm$0.5 & & 22$\pm$0.5 & 2.5$\pm$0.5 & & 16$\pm$0.5 & 2.5$\pm$0.5\\
                                        & &  --- & 2.5$\pm$0.5 & &      &            & &                          
&      & &         &          & &          &         \\                                     
\protect{[Ca\,{\sc ii}] $\lambda$7324}  & & 25$\pm$0.5$^{a}$ & 10$\pm$1 & & 55$\pm$1 & 2.5$\pm$0.5 & & 7$\pm$0.5 & 6.0$\pm$0.5 & & 22$\pm$0.5 & 2.5$\pm$0.5 & & 16$\pm$0.5 & 2.5$\pm$0.5\\
                                        & &  --- & 2.5$\pm$0.5 & &      &            & &                          
&      & &         &          & &          &         \\
   \hline 
\end{tabular}
\\
$^{a}$ Ring with a gap. For details see text.
\end{table}
\end{landscape}

\begin{table}
\centering
\caption{Velocities needed for fitting the line profiles for the stars with only the [O\,{\sc i}] 
forbidden lines. Multiple rows per object list the multiple fitting components. Parameters for the 
[O\,{\sc i}] $\lambda\lambda$6300,6364 lines are identical. All velocities have units km\,s\,$^{-1}$.}
\label{tab:velocities-2}
\begin{tabular}{lccc}
\hline
\hline
Object                 & & $v_{\rm rot, los}$ & $v_{\rm gauss}$ \\
\hline
V*\,FX\,Vel            & & 43$\pm$0.5 & 1$\pm$0.5 \\
                       & & 29$\pm$0.5 & 1$\pm$0.5 \\
                       & & 19$\pm$0.5 & 1$\pm$0.5$^{a}$ \\
                       & &  8$\pm$0.5 & 1$\pm$0.5$^{a}$ \\ 
\hline                       
SS\,255           & & 7.5$\pm$0.5 & 4.5$\pm$0.5 \\
\hline
IRAS\,17449+2320       & & 27$\pm$0.5 & 1$\pm$0.5 \\
                       & & 15$\pm$0.5 & 1$\pm$0.5 \\
                       & &  --- & 1$\pm$0.5 \\ 
\hline
Hen\,2-91         & & 31$\pm$0.5 & 1$\pm$0.5$^{a}$ \\
                       & & 19$\pm$0.5 & 1$\pm$0.5$^{a}$ \\
                       & &  7$\pm$0.5 & 1$\pm$0.5 \\
\hline
\protect{[MA93]\,1116} & & 15$\pm$0.5 & 1$\pm$0.5$^{a}$ \\
                       & &  --- & 2.5$\pm$0.5 \\
\hline
LHA\,120-S\,59     & & 45$\pm$0.5 & 2.5$\pm$0.5 \\
                       & & 28$\pm$0.5 & 2.5$\pm$0.5 \\
                       & & 16$\pm$0.5 & 2.5$\pm$0.5 \\                       
\hline 
\end{tabular}
\smallskip

$^{a}$ Ring with a gap. For details see text.
\end{table}

In order to find periodic variabilities, we used the Lomb-Scargle algorithm \citep{Lomb76,Scar82} designed to perform period analysis for unequally spaced data, with the implementation affiliated to the \textsc{astropy} package\footnote{\textsf{http://www.astropy.org/}}\citep{VaCo12,VaIz15}. With this tool, we prepared the periodograms on the time series of the filter that is better covered by data. The frequency powers were normalised to unity by the residuals of least-square fits of the data around their mean value (generalised Lomb-Scargle periodogram). 

\begin{figure}\centering
\includegraphics[scale=.73,clip,trim=3mm 3mm 3mm 0mm]{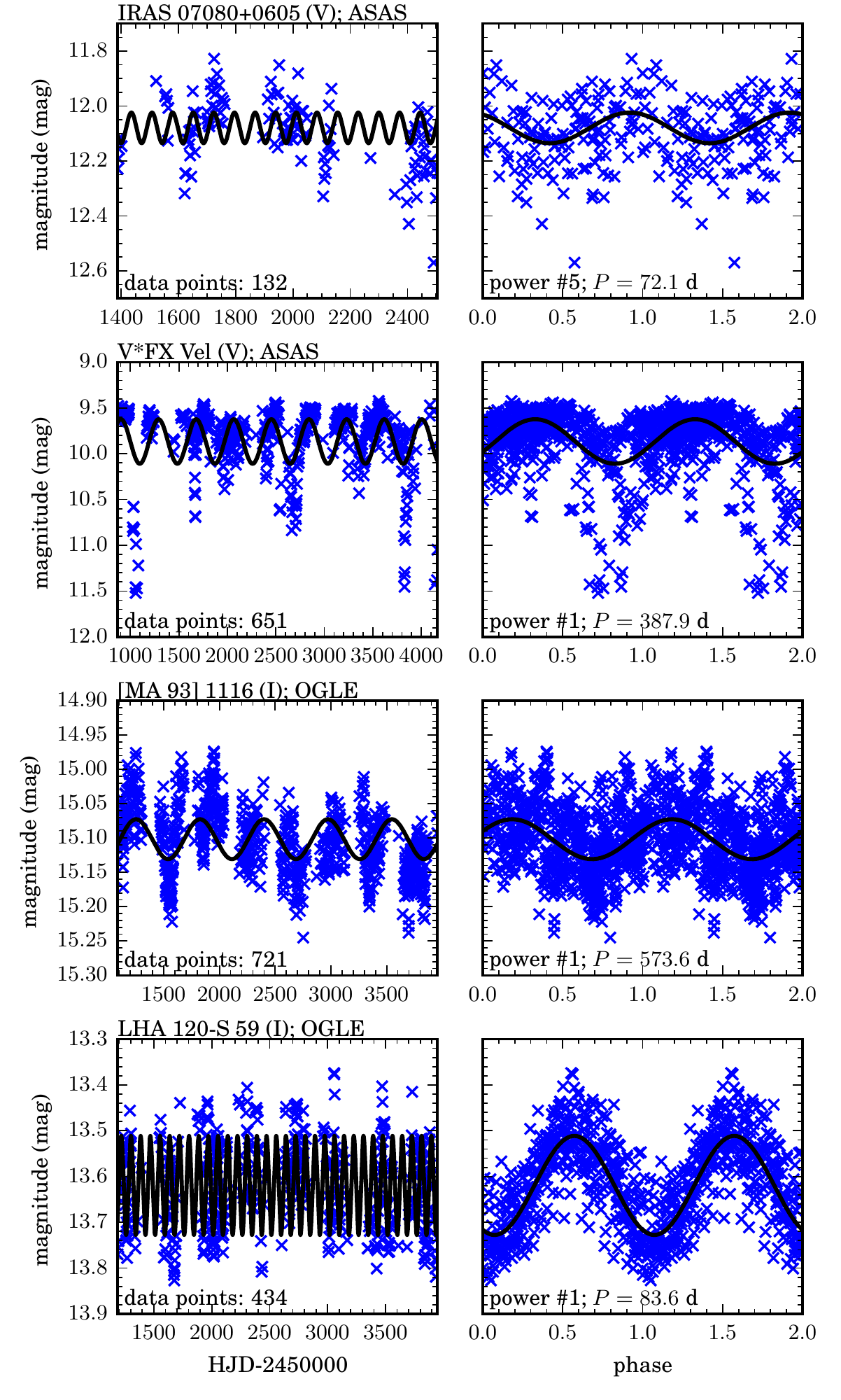}
\caption{The results of the period analysis of three stars, one per row. The star, filter and photometric survey are indicated on the top of the left panels, which plot the light curve. The most probable period (labelled on the bottom of the right panels) was used to fit a harmonic curve to the data points (markers) on the left. The phased data points are plotted on the right.}\label{f:main}
\end{figure}

The most reliable results (due to the quality of original data) of this study are concentrated in Fig.~\ref{f:main}.
As there are no previous variability studies of these stars, we decided to compare the results of our period analysis with the period reported for each star in VizieR, which probably is some period automatically inferred from Vizier data. A discussion for each star can be found in Sect.~\ref{sec:Discussion  of the  nature of our objects}.

%----------------------------------------------------------------
%----------------------------------------------------------------
\section{The (possible) nature of our objects}
\label{sec:Discussion  of the  nature of our objects}

Combining the information obtained from the spectral features and their variabilities, the stellar parameters and extinction, the forbidden line dynamics, and the period analysis we are now in the position to discuss the possible nature of our sample stars. To supplement the most plausible evolutionary state of the individual objects, we plot in Fig.~\ref{fig:traks-solar-lmc-art} the HR diagrams including evolutionary tracks for solar, SMC, and LMC metallicities (top row), and pre-main sequence evolutionary tracks (bottom row) for solar and SMC metallicities. We present the objects following the outline of Table\ref{table:objects}, embedding and combining our results with what is known from the literature. In addition, the summary of the physical parameters is provided in Table~\ref{table:Physical-parameters}, and the SEDs of the objects, displaying clear infrared excess emission, are compiled in Fig.~\ref{fig:SEDs-model}.

\subsection*{First group}
\subsection{Galactic stars}
\subsubsection{Hen 3-938}

Hen 3-938 (Hen 938, PDS 67, IRAS 13491-6318) was cataloged by \citet{Allen-Swings-1976} as a peculiar Be  star, because it exhibited NIR excess and forbidden emission lines in the optical spectrum. These authors also reported the presence of TiO bands in absorption. \citet{Gregorio-Hetem_1992}, through the analysis of medium-resolution spectra (0.4~\AA/pixel), classified it as a probable  Herbig Ae/Be star, also reporting the presence of P-Cygni profiles in the Balmer lines. Later, \citet{Miroshnichenko_1999} analyzed photometric and spectroscopic data (R$\sim$1000) and determined a B0 spectral type for this object, also reporting the presence of [O\,{\sc i}] and Fe\,{\sc ii} emission lines, and He\,{\sc i} lines with P-Cygni profiles, but not TiO bands. These authors also suggested that Hen 3-938 has more similarities with B[e] supergiants than with Herbig Ae/Be stars, but it might be a star evolving towards the planetary nebula stage, due to its similarities with HD 51585, a post-AGB star \citep{Arkhipova-1992}.

From Method 1, we classified Hen 3-938 as a \mbox{B0-1\,I} star, probably being a B[e] supergiant, in agreement with \citet{Miroshnichenko_1999}. This classification is reinforced by the high extinction that we derived from our DIB, and the high luminosity, determined from the distance of $\sim 6.2$\,kpc, obtained from Gaia DR2, even considering the high uncertainty (see Table~\ref{table:Physical-parameters}). In addition, from the HR diagram (left panel, Fig.~\ref{fig:traks-solar-lmc-art}), we derived $M_{\rm ZAMS} \sim$ 20 M$_\odot$.

This scenario of a hot and luminous star is also favoured by our FEROS spectra, where we could identify the presence of Balmer, Fe\,{\sc ii}, and especially He\,{\sc i} lines showing P-Cygni profiles (Sect.~\ref{sec:35}).

The narrow single-peaked profiles of the forbidden lines might contain a slight rotation component. For the [O\,{\sc i}] lines we find $v_{\rm rot, los} = 9\pm 0.5$\,km\,s$^{-1}$, whereas for the [Ca\,{\sc ii}] lines it would be with $v_{\rm rot, los} = 7\pm 0.5$\,km\,s$^{-1}$ slightly lower but still comparable. A Gaussian component of $v_{\rm gauss} = 6\pm 0.5$\,km\,s$^{-1}$ is needed in all lines (see Table~\ref{tab:velocities-1} and Fig.~\ref{fig:fits-OI-CaII}). The lines from 2016 are very similar, but less intense (see Fig.\,\ref{fig:Hen-3-938-Bep-art}). If the star has a Keplerian disk or ring, the system might be seen close to pole-on.

Our spectra taken in 2005 and 2016 show a variation in the line intensities, which has also been reported in the literature. Such variability is not common in B[e] supergiants, but we could not find any signature of a companion. On the other hand, we cannot completely discard a pre-main sequence scenario for this object, as proposed by \citet{Gregorio-Hetem_1992} and \citet{Vieira_2003}, who also suggested that Hen 3-938 could be associated to a star-forming region in Centaurus.

%----------------------------------------------------------------------------------------

\subsubsection{SS 255}

Not much is known about SS 255 (IRAS 14100-6655, 2MASS J14135896-6709206). It was discovered as an H$\alpha$ emission-line star by \citet{Stephenson-and-Sanduleak-1977}, who listed it as number 255 in their catalog. Recently, \citet{Miszalski-Mikolajewska-2014a} suggested that SS 255 strongly resembles a B[e] star, however, its NIR colors do not show evidence of hot dust. Thus, due to the similarities with SS73 24, these authors suggested the classification of SS 255 as a Be star.

Due to the scarcity of photometric measurements, we could not apply the Method 1 and derive its spectral type and luminosity class. However, based on the presence of He\,{\sc i} lines in emission, its spectral type can be B2 or earlier \citep[e.g.,][]{Zickgraf-1986,Miroshnichenko-2007A}. Thus, for our analysis we assume a B2 spectral type and a mean effective temperature of 19500$\pm$2500\,K.

Due to the lack of knowledge about its luminosity class, we assume a mean value for the bolometric correction ($-1.68$), considering the values provided by \citet{Humphreys-McElroy-1984}.

SS 255 is the most distant Galactic object of our sample, $\sim$10.3\,kpc (although this value is not so reliable, due to the high uncertainty of its parallax), but it has a relatively low extinction. 

In addition, the [O\,{\sc i}] lines are narrow, symmetric and extremely intense, being modeled with a single rotating ring ($v_{\rm rot, los} = 7.5\pm 0.5$\,km\,s$^{-1}$, $v_{\rm gauss} = 4.5\pm 0.5$\,km\,s$^{-1}$), see Table~\ref{tab:velocities-2} and Fig.~\ref{fig:fits-OI}. If these lines are originated from a gaseous disk, this disk must be seen close to pole-on.

%------------------------------------------

\begin{landscape}
\begin{figure}
        \centering
        \begin{tabular}{@{}ccc@{}}
                \includegraphics[width=78mm]{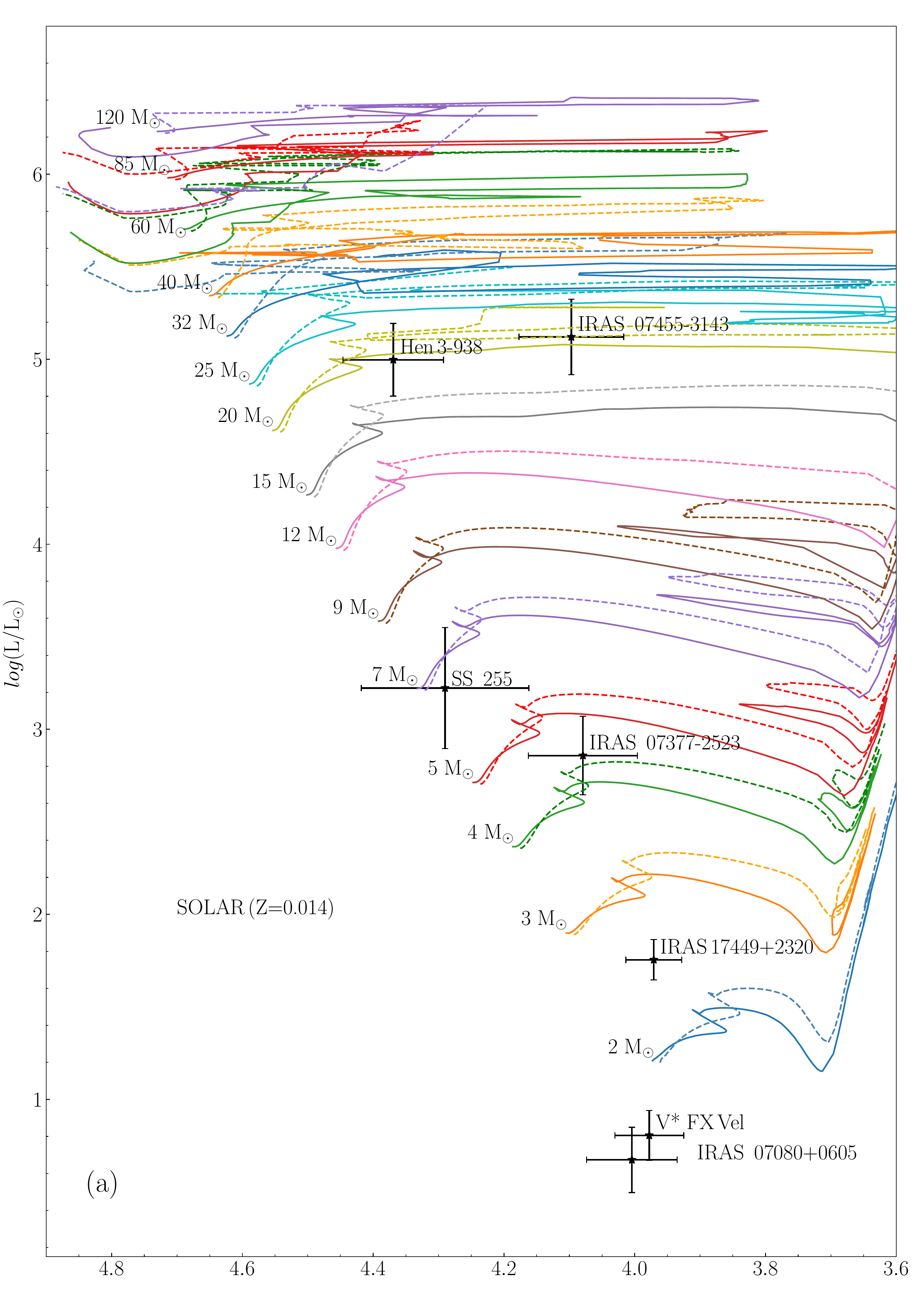} &
                \includegraphics[width=78mm]{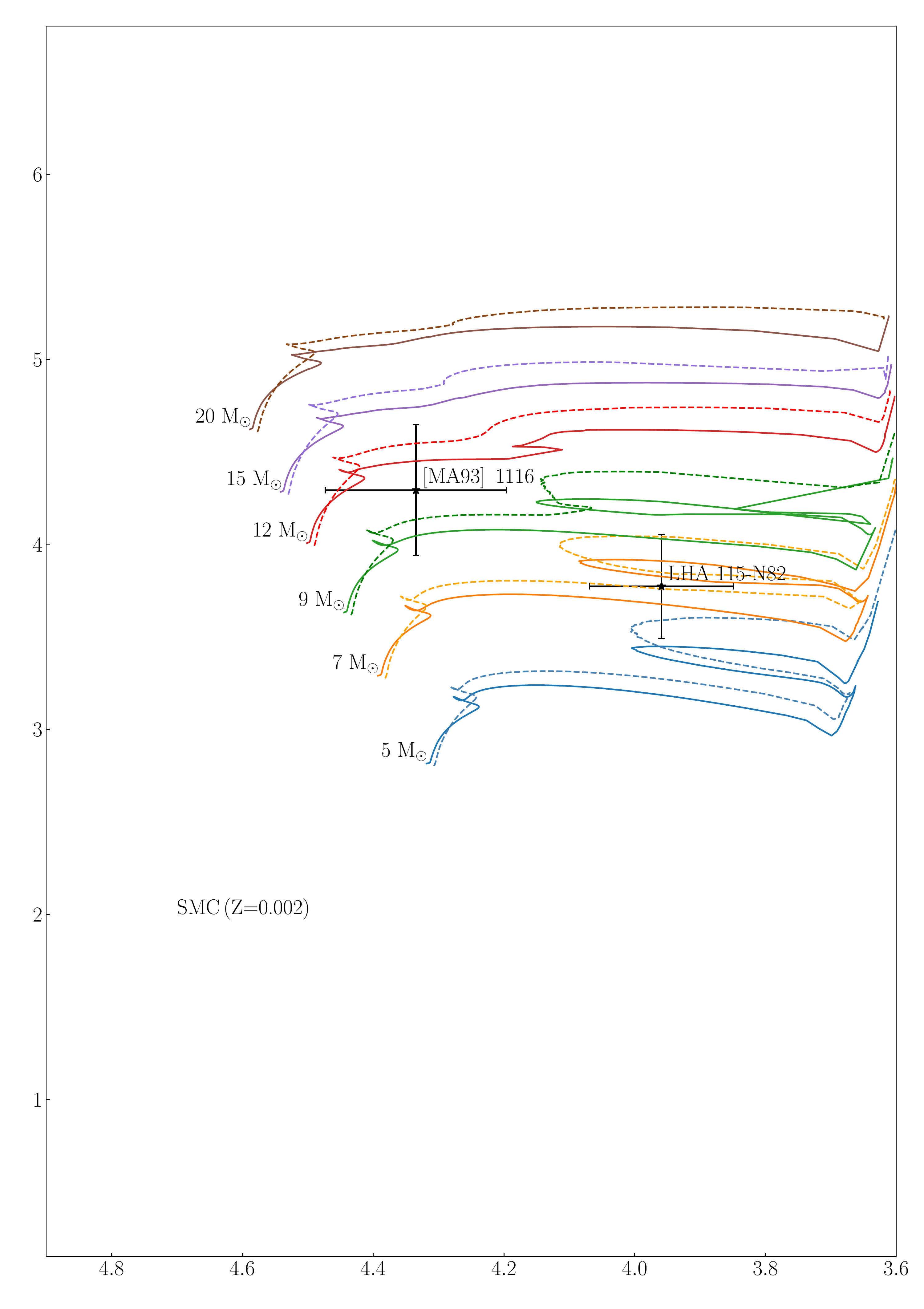} & 
                \includegraphics[width=78mm]{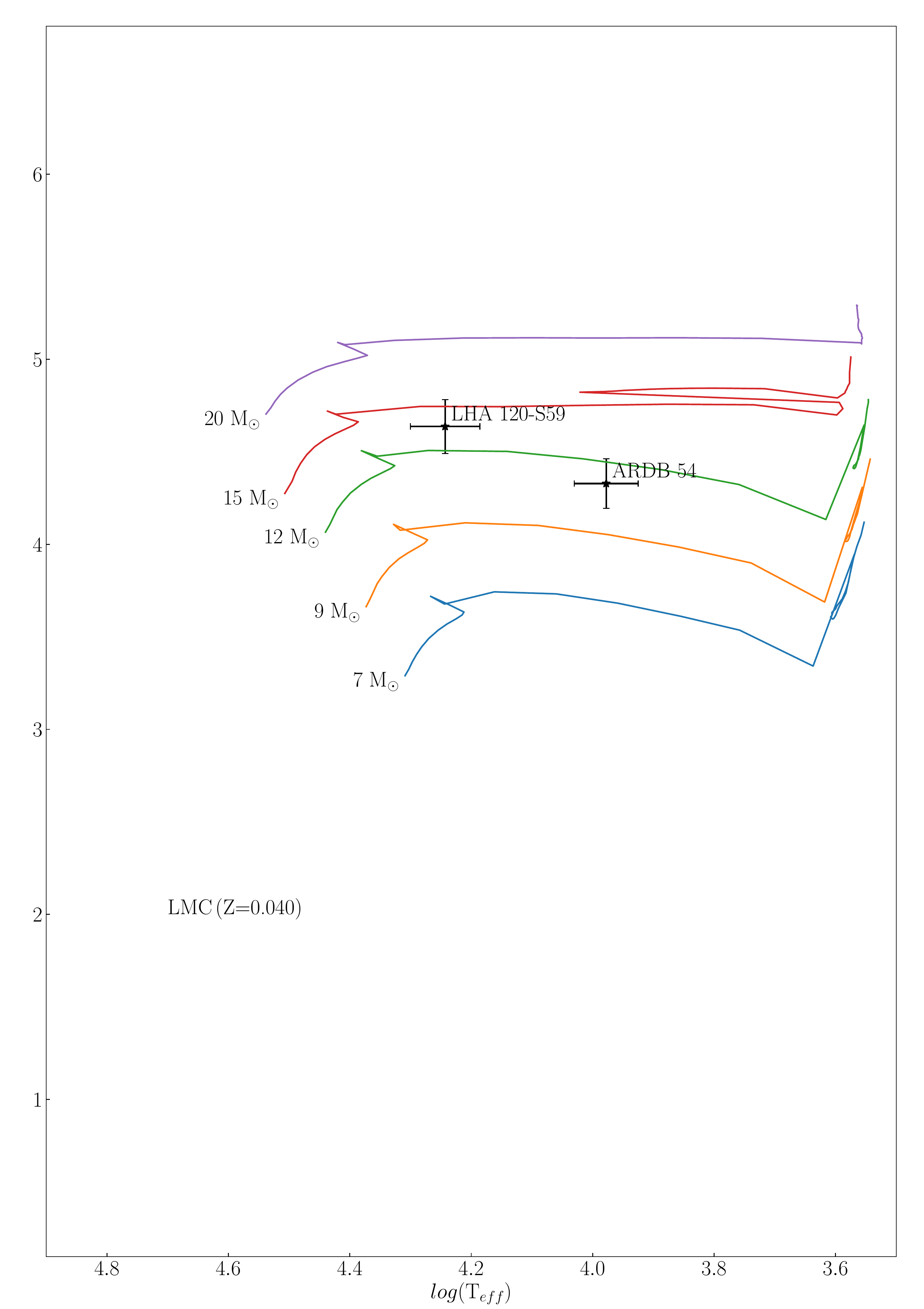} \vspace{-5.2mm} \\ 
                \includegraphics[width=78mm]{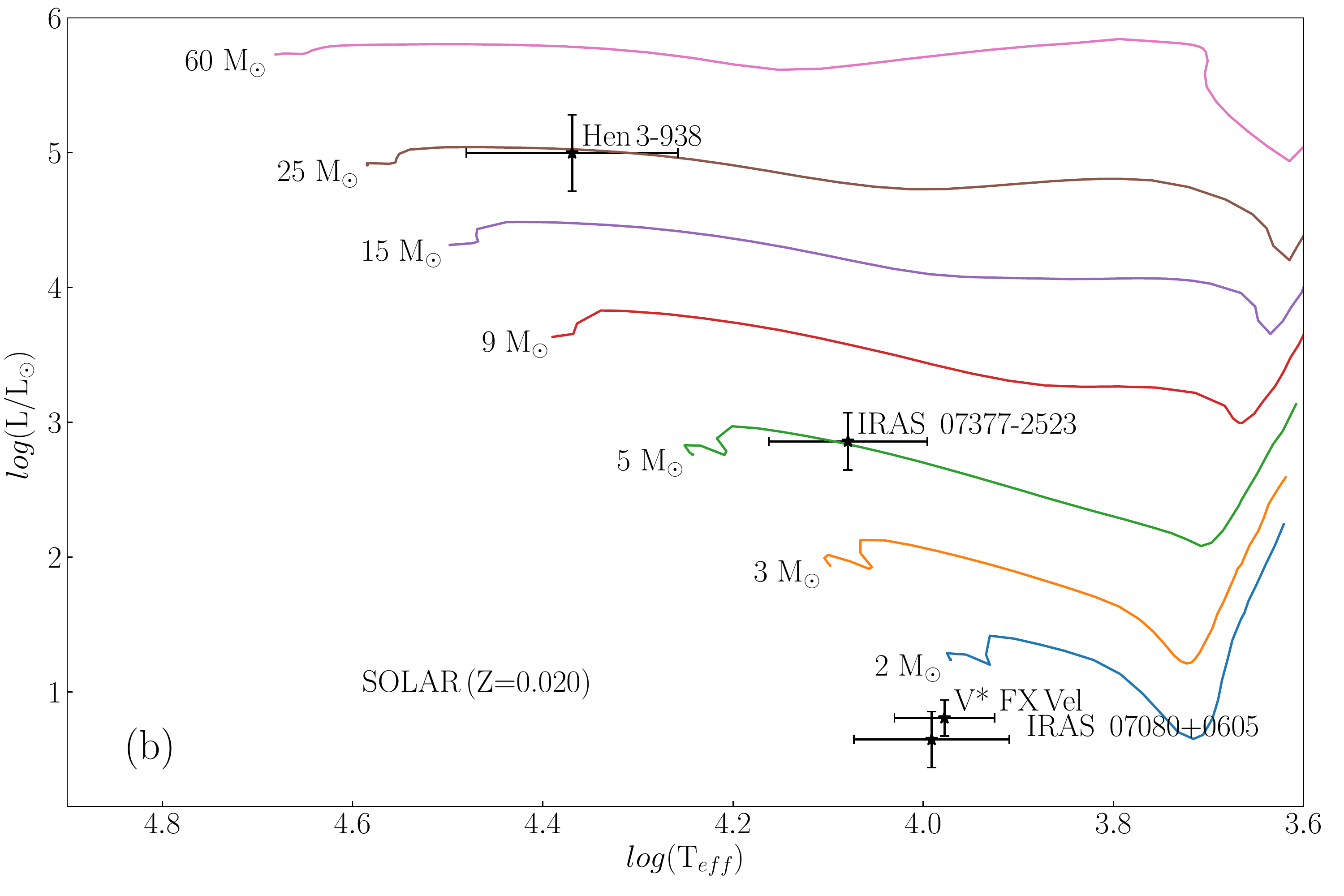}&
                \includegraphics[width=78mm]{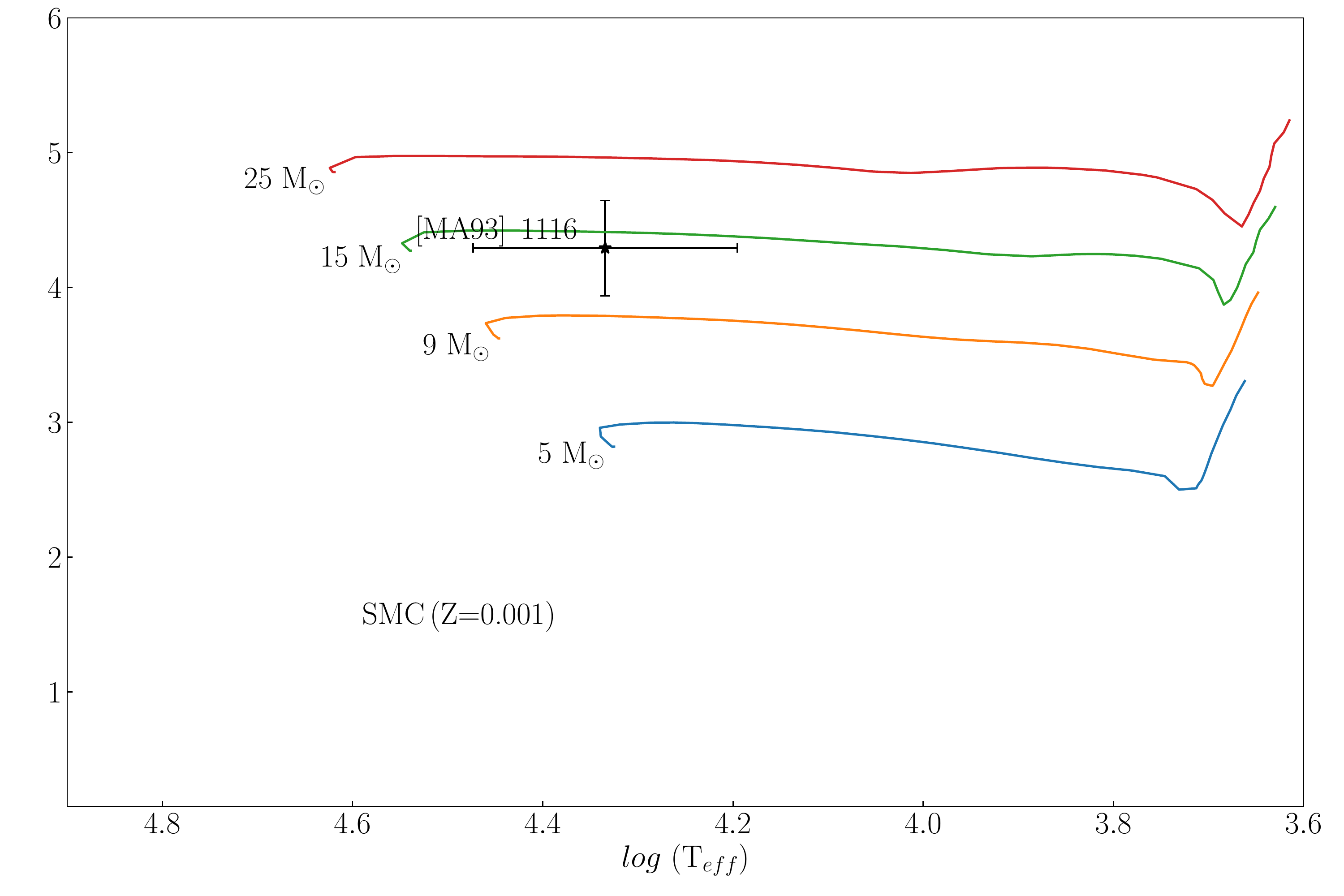}
        \end{tabular}
        \caption{Position of the stars of our sample in the HR diagram, considering (a) evolutionary tracks with (dashed lines) and without (solid lines) rotation for solar (left panel) and SMC metallicities (middle panel) from \citet{Georgy-2013}, and without rotation for LMC metallicity (right panel) from \citet{Schaerer-et-al-1993}; and (b) pre-main sequence tracks for similar metallicities to the solar and SMC ones from \citet{Bernasconi-Maeder-1996} in the left and middle panels, respectively.}
        \label{fig:traks-solar-lmc-art}
\end{figure}
\end{landscape}

Even considering the high uncertainties for SS 255, we estimated for the first time its parameters. Based on its position in the HR diagram, this star may have a ZAMS mass between 5 and 7 M$_\odot$, being at the end of the main sequence or close to it (left panel, Fig.~\ref{fig:traks-solar-lmc-art}). However, based on the intense emission lines, especially  nebular lines, like [O\,{\sc ii}], [S\,{\sc ii}] and [N\,{\sc ii}], in our spectra (Sect.~\ref{sec:36}), a scenario as a post-AGB star not hot enough to excite [O\,{\sc iii}] lines, seems very favourable.

\subsubsection{Hen 2-91}

The nature of Hen 2-91 (SS73 39, IRAS 13068-6255, MN7, THA 17-18) is very uncertain. Some articles have classified it either as a planetary nebula \citep*{Webster-1966,Henize-1967, Allen-1973a,Frew-2013}, an M star with emission \citep{MacConnell-1983, Bidelman-1998}, an emission-line star \citep{The-1962, Weaver-1974}, a peculiar Be star \citep{Allen-Swings-1976,Allen-1982}, a B[e] star \citep{Lamers-1998}, or as a FS CMa candidate \citep{Miroshnichenko-2007A}. Hen 2-91 is also in the catalogues of OB stars \citep{Reed-2003}, and evolved massive stars \citep*{Gvaramadze-2010}. 

The very intense and double-peaked Balmer lines may indicate an extended nebula or even a circumstellar disk. The FEROS spectra taken in five different nights in 2016 (in a period of 4 months) do not present sensible variations (Sect.~\ref{sec:38}). 

No significant variations are seen in the [O\,{\sc i}] line profiles of Hen\,2-91 observed between April 
and August 2016. The wiggly profiles suggest at least three ring components
($v_{\rm rot, los} = 31\pm 0.5$; 19$\pm$0.5; and 7$\pm$0.5\,km\,s$^{-1}$) of which two 
are asymmetric. We implemented a big symmetric gap around the blue peak, into the high-velocity ring, 
excluding velocities smaller than $-5.4$\,km\,s$^{-1}$ and a gap, symmetric around the red peak, into
the medium-velocity ring, excluding velocities larger than 17.2\,km\,s$^{-1}$. No noticeable Gaussian 
component is needed for the fit (Table~\ref{tab:velocities-2} and Fig.~\ref{fig:fits-OI}).

Hen 2-91 is located at $\sim$ 5\,kpc, but this measurement is very uncertain. Its interstellar extinction is high based on the DIB present in our spectra. However, the value provided by IRSA is much higher and seems to be very imprecise. Thus, we decided to assume the value obtained from our spectra. 

Unfortunately, there are only few photometric measurements available in the literature and no diagnostic absorption line is visible in our FEROS spectra. Thus, we could not apply any of the three methods to obtain the physical parameters of Hen 2-91. 

Based on the BCD method \citep{Barbier-Chalonge1941, Chalonge-Divan_1952}, \citet{Cidale_2001} derived $T_{\rm eff} =$ 32500$\pm$2600\,K and B0 type for Hen 2-91, which are not in agreement with the spectral features that we have identified, especially the presence of He\,{\sc i} lines in absorption, and the absence of He\,{\sc ii}, Si\,{\sc iv} and other high-ionization lines. 

Thus, based on the high uncertainty in its distance and extinction, and the lack of any reliable stellar parameter, it is not possible a deeper discussion about the nature of this star.

\begin{table*}
	\centering
	\caption{Physical parameters of the stars of our sample. }
	\label{table:Physical-parameters}
	\begin{tabular}{llcccccc}
		\hline   \hline                                                                                 
& Star         		&Distance$^{*}$     			&BC$^{**}$ 			&$M_{\rm bol}$ &$T_{\rm eff}$  &$\log (L/L_\odot$) &$R/R_\odot$ \\
	&		&(pc)				&(mag)				&(mag)		  &(K)		     &			 &  \\	
\hline \hline
\multicolumn{8}{c}{\textbf{First group}} \\ \hline
\textbf{Galaxy}      &
Hen 3-938	&6228$^{+1409}_{-1010}$ 	&-2.20  	&-7.75$\pm$0.58  &23400$\pm$2600    &5.00$\pm$0.20	&19$\pm$3\\
%----
&SS 255		&10321$^{+2524}_{-1818}$	&-1.68		&-3.31$\pm$0.45  &19500$\pm$2500		  &3.22$\pm$0.33	&$\sim$4\\
%---		
\hline
\textbf{SMC}      &
LHA 115-N82		  &	$18.95\pm0.07${$^{***}$} 			 &-0.13  	&-4.69$\pm$0.30  &9100$\pm$1000		&3.77$\pm$0.28	&31$\pm$7\\	
\hline
\textbf{LMC}      &     
ARDB54		&$18.22\pm0.05${$^{***}$} 			&-0.27  	&-6.08$\pm$0.13  &9500$\pm$200		&4.33$\pm$0.13	&54$\pm$6\\
&LHA 120-S59		&				&-1.55  	&-6.85$\pm$0.07  &17500$\pm$500$^\text{L}$	&4.63$\pm$0.15	&23$\pm$3\\
\hline   \hline
\multicolumn{8}{c}{\textbf{Second group}} \\ \hline
\textbf{Galaxy}      & 
IRAS 07080+0605	 &535$^{+15}_{-14}$  	  	&-0.29  	&3.06$\pm$0.12	&10100$\pm$700	   &0.67$\pm$0.18  & $\sim$1\\
&IRAS 07377-2523	 &4100$^{+ 521}_{-418}$  	&-0.59  	&-2.40$\pm$0.33	&12000$\pm$1000	   &2.86$\pm$0.21  &6$\pm$1\\
&IRAS 07455-3143	 &12262$^{+3154}_{-2327}$	&-0.59  	&-8.07$\pm$1.12	&12500$\pm$1000    &5.12$\pm$0.20  &78$\pm$12 \\
&V*FX Vel	 &353$^{+6}_{-6}$		&-0.17  	&2.73$\pm$0.13   &9500$\pm$500     &0.81$\pm$0.13  &$\sim$1\\
%----
&IRAS 17449+2320	 &740$^{+22}_{-21}$	  	&-0.14		&0.35$\pm$0.07   &9350$\pm$400	   &1.75$\pm$0.11	&$\sim$3\\
%----
\hline
\textbf{SMC}     &
{[}MA93{]} 1116	 &$18.95\pm0.07${$^{***}$} 		&-1.80  	&-5.99$\pm$0.27  &21600$\pm$3000	&4.29$\pm$0.26	&10$\pm$2\\
\hline
\end{tabular}
\begin{tablenotes}
        \item \textbf{Notes.} ($^{*}$) Distances from Gaia DR2 for Galactic stars provided by \citet{Bailer-Jones}. We caution that the values for objects further away than 4 kpc have high uncertainties. These values might considerably change with Gaia DR3; ($^{**}$) Bolometric correction ($BC$) from \cite{Humphreys-McElroy-1984}; 
        ($^{***}$) distance modulus for SMC taken from \cite{Graczyk-et-al-2014}, and for LMC taken from \cite{Udalski-et-al-1998}. $^\text{(L)}$ from \cite{Levato_2014}. 
\end{tablenotes}
\end{table*}

\begin{figure}
        \centering
        \includegraphics[width=\linewidth, clip]{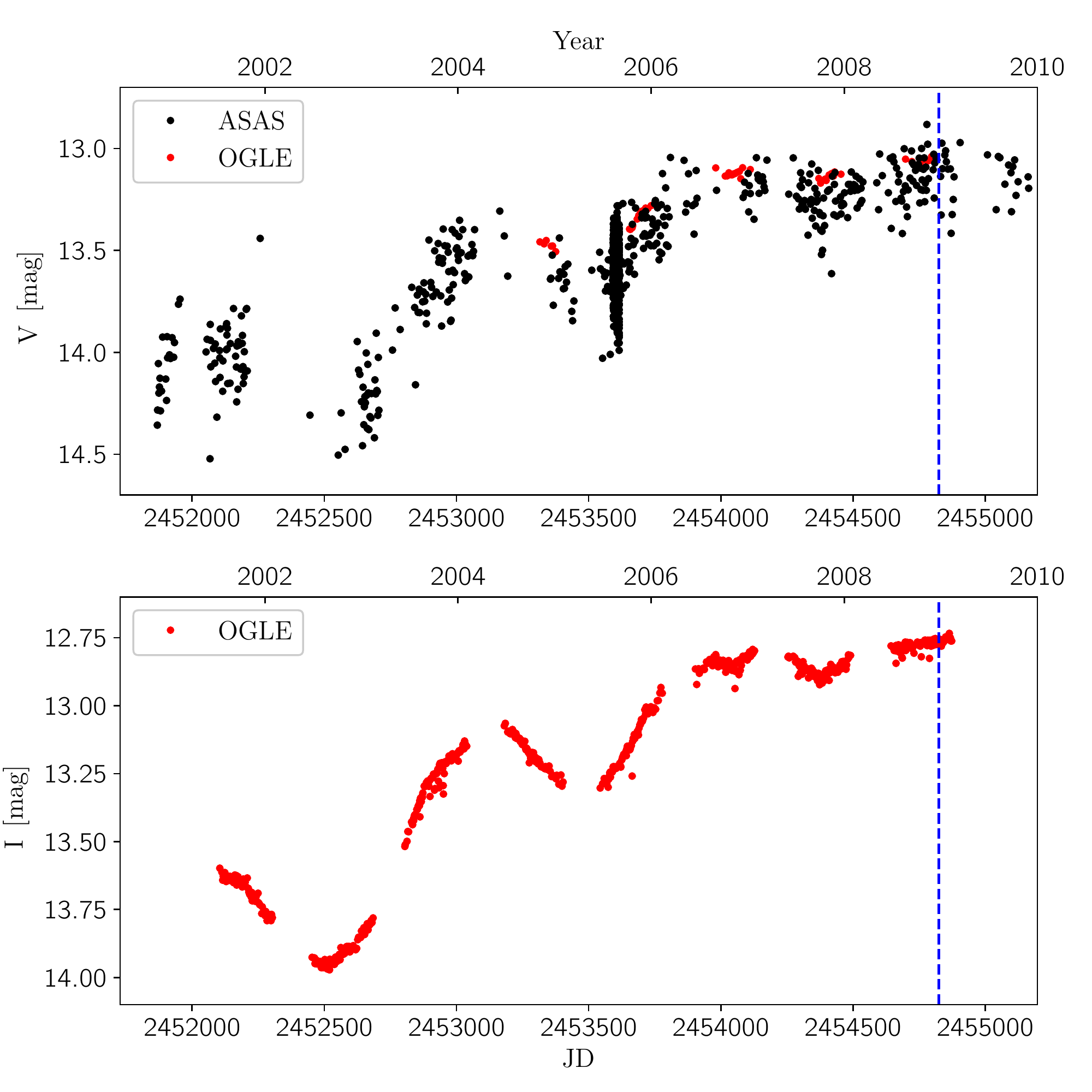}
        \caption{Light curves of LHA 115-N82 obtained from ASAS and OGLE III surveys, taken from 2001 until 2010, in the V- and I-bands. The vertical dashed lines indicate the FEROS spectra taken in 2008.}
        \label{fig:LHA_155_N82_ASAS-OGLEII_band_V_band_I}
\end{figure}

%----------------------------------------------------------------------------------------
\subsection{SMC star}

\subsubsection{LHA 115-N82}
\label{subsubsec:N82}

LHA 115-N 82 (LIN 495, N82, 2dFS 2837) was originally identified as a nebula with H$\alpha$ in emission by \citet{Henize-1956}, who gave the designation of N82 to this object. In the Catalog of Stars with Emission Lines and Planetary Nebulae of \citet{Lindsay-1961}, LHA 115-N 82 was listed as LIN 495. Its evolutionary stage was firstly suggested by \citet{Heydari-Malayeri-1990}, who based on photometric and spectroscopic data, classified it as a B[e] supergiant of spectral type B7-8. Later \citet{Evans_2004}, in a spectroscopic survey (low-resolution, R$\backsimeq$1500) in the SMC (2dF survey of SMC), classified this star with a possible composed spectra: AF/B[e]. Recently \citet{Kamath_2014}, based on the analysis of low-resolution spectra, classified this object as a post-AGB/RGB candidate.

Our FEROS spectra taken in 2008 and 2015 show strong variability in the line profiles, radial velocities and {\it V/R} (for the profiles with double peaks), as described in Sect.~\ref{sec:310}. 

Based on the Method 1, using photometric measurements taken in 1989-1999 (see Table~\ref{table:objectsphotometry}), we could derive two possible sets of parameters for LHA 115-N82, a late-B or an early-A type star. Based on very weak or even absent He\,{\sc i} lines and strong Mg\,{\sc ii} lines in absorption seen in our FEROS spectra taken in 2008 and 2015, the classification as an early-A star seems to be more favourable. Thus, assuming the distance of SMC and the low extinction obtained from IRSA, we derived the parameters for LHA 115-N82 (see Table~\ref{table:Physical-parameters}). 

From the HR diagram, considering the evolutionary tracks for SMC stars, we note the post-main sequence nature of this star, with $M_{\rm ZAMS}$ of 7$-$9\,M$_\odot$ (middle panel, Fig.~\ref{fig:traks-solar-lmc-art}).

On the other hand, the light curves of LHA 115-N82 in the V- and I-band (Fig.~\ref{fig:LHA_155_N82_ASAS-OGLEII_band_V_band_I}) show a long-term increase of brightness. In the V-band, there is an increase from $\sim$14 mag in 2002 to $\sim$13.2 mag in 2010. In the I-band, the star goes from $\sim$13.5 mag around 2001 to $\sim$12.75 mag in 2009. In addition, due to the low dispersion of the data in the I-band, we can see two minima around 2003 and 2005, followed by two brightness increases. This behaviour of the light curves is similar to that one seen in Luminous Blue Variables (LBV) during their eruptions, especially as recently reported for R40, also a SMC star \citep{Campagnolo-et-al-2018}. 

The effect of these eruptions is seen in the spectrum of LBVs, in a transition from a B-type during the quiescence (without eruptions) to an A- or even late-F type, depending on how strong the eruption is and the amount of matter that is ejected, forming a pseudo-photosphere. 

For the Method 1, we used photometric data taken $\sim$10 years before the ASAS and OGLE III data, indicating an even lower brightness for LHA 115-N82 (14.25-14.75 mag in the V-band), and a probably higher effective temperature during the quiescent stage, as a B-type star. However, our spectrum taken in 2008, during its brightest phase, does not show a sensible variation, being of an early A-type star, indicating a not so strong eruption in 2005. Due to the absence of photometric data later than 2010, we cannot confidently say at which stage (quiescence or eruption) the 2015 spectra were observed. However, based on their characteristics, it seems that this star is still under the effect of an eruption. In addition, the presence of Paschen lines in absorption may also indicate a cool and dense photosphere, typical of LBVs after eruption \citep{Mehner-et-al-2013}.

The behaviour of the Balmer and Fe\,{\sc ii} lines is also interesting, showing a more intense blue emission and inverse P-Cygni profiles in 2008, and a more intense red emission and P-Cygni profiles in 2015 (see Fig.~\ref{fig:Bep-LHA 115-N82-a}). The absorption component of these lines show radial velocity variation, being blueshifted in 2015 and redshifted in 2008. This is also seen in the He\,{\sc i} absorption profiles (Fig.~\ref{fig:Bep-LHA 115-N82-a}), indicating the presence of rotating absorbing material around the star. 

The two sets of observations of LHA\,115-N\,82 reveal similar profiles of the forbidden lines with no significant 
variability, indicating a stable emitting region. Due to the better quality, we show the fits to the 2015 data. The [Ca\,{\sc ii}] lines 
suffer from low SNR and possible remnants of telluric pollution. The profiles might be modeled 
with a single ring ($v_{\rm rot, los} = 22\pm 0.5$\,km\,s$^{-1}$, $v_{\rm gauss} = 2.5\pm 
0.5$\,km\,s$^{-1}$). The same ring, but with a higher Gaussian component of $v_{\rm gauss} = 7.5\pm 
0.5$\,km\,s$^{-1}$ is seen in the profiles of the [O\,{\sc i}] lines. However, these lines require a 
second, lower velocity ring component ($v_{\rm rot, los} = 13\pm 0.5$\,km\,s$^{-1}$, $v_{\rm gauss} = 
2.5\pm 0.5$\,km\,s$^{-1}$) for a reasonable fit, see Table~\ref{tab:velocities-1} and Fig.~\ref{fig:fits-OI-CaII}. The decreasing velocity with decreasing density is typical for Keplerian disks or rings.

This scenario is really a puzzle, because LHA 115-N82 is not massive enough to be a LBV. In addition, LBVs do not show [O\,{\sc i}] lines, but LHA 115-N82 does. Thus, we are maybe observing a post-main sequence B[e] star, showing instabilities that cause eruptions, like an ``LBV impostor". More simultaneous spectroscopic and photometric data are necessary for a better comprehension of this star.

%------------------------------------
\subsection{LMC stars}
\subsubsection{ARDB 54}

ARDB 54 (SOI 720) was first observed by \citet{Ardeberg_1972}, who based on UBV fluxes, suggested that it might be a multiple star or an emission-line object. Later, \citet*{Stock_1976}, from objective prism spectrum, classified ARDB 54 as a B9 Ib star. Recently, \citet{Levato_2014} analysed medium-resolution spectra and photometric data and classified it as a B[e] supergiant with an effective temperature of 10000\,K and $\log$($L$/L$_\odot$) of 4.57.

Our FEROS spectra taken in 2014 and 2015 do not show variability (Sect.~\ref{sec:311}). However, in comparison with the spectra taken by \citet{Levato_2014} in 2011, there is a noticable spectral variation.

The absence of Paschen lines, weak forbidden lines and weak IR excess probably indicate a small amount of ionized circumstellar material. 

The spectra of ARDB\,54 are very noisy and we limit our model attempts to the profiles seen in 2015. 
The [Ca\,{\sc ii}] $\lambda$7291 line appears double-peaked, which can be approximated with a single
ring with $v_{\rm rot, los} = 16\pm 0.5$\,km\,s$^{-1}$. The same model is used for the [Ca\,{\sc ii}] 
$\lambda$7324 line, but this line is very noisy and contaminated on its blue edge possibly with a cosmic 
ray so that this fit can only be regarded as suggestive. The triangular shape of the [O\,{\sc i}] lines 
imply in multi-components. We achieved a reasonable fit using two rings with $v_{\rm rot, los} = 13\pm 
0.5$\,km\,s$^{-1}$ and $v_{\rm rot, los} = 3\pm 0.5$\,km\,s$^{-1}$. All lines have the same Gaussian 
contribution of $v_{\rm gauss} = 2.5\pm 0.5$\,km\,s$^{-1}$. In total, this object has three rings traced 
by the forbidden lines (Table~\ref{tab:velocities-1} and Fig.~\ref{fig:fits-OI-CaII}). The decrease in density with rotation velocity hints towards a Keplerian disk scenario.

In our analysis, using the Method 1, we concluded that ARDB 54 is actually an A0-1I ($T_{\rm eff} \sim$ 9500\,K) star, which is in agreement with the spectral features of our FEROS spectra, especially the weakness or the absence of He\,{\sc i} lines. Assuming the LMC distance and the low extinction derived from IRSA, we derived the physical parameters of ARDB 54. From the HR diagram, using the evolutionary tracks for LMC stars from \citet{Schaerer-et-al-1993}, we confirm that this star is an A[e] supergiant with $M_{\rm ZAMS}$ = 10$-$12 M$_\odot$ (Fig.~\ref{fig:traks-solar-lmc-art}). Thus, ARDB 54 is the third A[e] supergiant already identified, the first one in the LMC. The other two A[e] supergiants are the SMC star LHA 115-S23 \citep{Kraus-2008} and the Galactic object HD 62623 \citep{Meilland-et-al-2010}.

%------------------------------------

\subsubsection{LHA 120-S59}

LHA 120-S59 (S59, AL 415, OGLE LMC-LPV-83573) was first identified by \citet{Henize-1956} as an H$\alpha$ emission star. \citet{Gummersbach_1995}, based on spectroscopic and photometric data, suggested that LHA 120-S 59 is a B[e] supergiant with B5II spectral type and effective temperature of 14000\,K. Recently, \citet{Levato_2014},  based on the analysis of medium-resolution spectra and OGLE photometric data, also suggested a B[e] supergiant classification, but with B2-3 spectral type, effective temperature of 19000\,K and $\log(L/$L$_\odot$) of 4.64. These authors also reported radial velocity variations associated to variable ($B-V$) and UV excess, suggesting the presence of a companion. From our FEROS spectra taken in 2015 and 2016, we also noticed line profile and radial velocity variations (Sect.~\ref{sec:312}). 

If this scenario is correct, the orbital period might be the period of 83.6 d inferred from period analysis (Fig.~\ref{f:main}), which is very close to the VizieR value (83.4 d).

Concerning the physical parameters of this star, we decided to assume a mean value for the effective temperature of 17500$\pm$500\,K, as provided in the literature \citep{Levato_2014}, due to the absence of diagnostic lines in absorption (Mg\,{\sc ii} and Si\,{\sc ii}) and the impossibility of convergence for a set of parameters using the Method 1. Assuming the distance of LMC and the extinction for this object, as derived by IRSA, we obtained its bolometric correction, bolometric magnitude, luminosity and radius, as seen in Table~\ref{table:Physical-parameters}. From the HR diagram, considering the evolutionary tracks for LMC stars from \citet{Schaerer-et-al-1993}, we classify LHA 120-S59 as a B[e] supergiant with $M_{\rm ZAMS} =$ 12$-$15 M$_\odot$ (Fig.~\ref{fig:traks-solar-lmc-art}, right panel).

The high temperature, IR excess and broad Balmer lines (the broadest of our sample) suggest that this star probably has a large amount of ionized circumstellar gas. 

The [O\,{\sc i}] lines can be modeled with a combination of at least three rotating rings ($v_{\rm rot, los} = 45\pm 1$; $28\pm0.5$; and $16\pm0.5$\,km\,s$^{-1}$), each with a Gaussian component of $v_{\rm gauss} = 2.5\pm 0.5$\,km\,s$^{-1}$ (see Table~\ref{tab:velocities-2}), supporting the picture of a Keplerian disk around LHA 120-S59. We show the fit to the
data taken in 2016 in Fig.~\ref{fig:fits-OI}, but the same model reproduces the line profiles in the spectra taken in 2015.

In addition, molecular emission was also seen in its environment, being in line with the classification of this star as a B[e] supergiant \citep{2013A&A...558A..17O}.

\subsection*{Second group}

\subsection{Galactic stars}

\subsubsection{IRAS 07080+0605}

IRAS 07080+0605 (TYC 175-3772-1, HBHA 717-01) was detected by \citet{Kohoutek-1999} in a survey for stars with H$\alpha$ in emission. Later \citet{Miroshnichenko_2007}analysed high-resolution spectra (R$\sim$70000) and identified the presence of the B[e] phenomenon, classifying it as a FS CMa A-star with low luminosity. These authors also identified a strong IR excess, suggesting a binary nature with mass transfer. However, no direct evidence of a companion was found, especially due to the absence of radial velocity variations.

It is the second closest star (535 pc) of our sample. This is in agreement with the low interstellar extinction obtained from the 3D dust mapping (Table~\ref{table:interstellar-extinction}), which we consider more reliable than the value provided by IRSA.

From Method 1, an A0-1II type was derived, which is, in principle, corroborated by the absence of the He\,{\sc i} $\lambda$4471 line in our spectrum. However, the presence of other He\,{\sc i} lines may weaken this classification. Unfortunately, we could not use other common diagnostic lines for A-type stars, such as Ca\,{\sc ii} H \& K lines and H$\epsilon$, because they are contaminated by wind emission.  

Assuming this possible classification and  a mean effective temperature of 10100$\pm$700\,K, we derived the bolometric magnitude (M$_{\text{bol}}$), luminosity and radius of IRAS 07080+0605 (Table~\ref{table:Physical-parameters}) in agreement with the results of \citet{Miroshnichenko_2007}.

The spectral variability, as described in Sect.~\ref{sec:31}, is not sufficient to confirm a binary nature for this object, as proposed by \citet{Miroshnichenko_2007}. The presence of inverse P-Cygni profiles, seen in some Ca\,{\sc ii} and Fe\,{\sc ii} lines, may indicate that an accretion process is ongoing. The triple peaked components and asymmetries seen in some line profiles, and also cited by \citet{2018PASP..130k4201A}, imply in a complex circumstellar environment, possible composed of a disk and a nebular component.

The [O\,{\sc i}] lines in IRAS\,07080+0605 are clearly double-peaked although the 5577\,\AA \ line is
rather noisy. All three lines can be modeled with a single ring with a rotational velocity, 
projected to the line-of-sight, of $v_{\rm rot, los} = 25\pm 0.5$\,km\,s$^{-1}$ and an additional 
Gaussian component of $v_{\rm gauss} = 10\pm 0.5$\,km\,s$^{-1}$ (Table~\ref{tab:velocities-1}). This extra broadening might be either 
ascribed to some turbulence related to the accretion flow of the gas or, alternatively, might be interpreted 
as an indication for a certain width of the emitting ring and hence a slight variation of the rotation 
velocity, in contrast to the infinitesimal thin ring with constant rotation velocity used in the model.
The slight depression of the red peak (Fig.~\ref{fig:fits-OI-CaII}) can be achieved if we allow for a gap in the ring around the 
maximum radial velocity ($>$24.8\,km\,s$^{-1}$). The [Ca\,{\sc ii}] lines appear composite. While the 
same ring model can be used to approximate the broad component, an additional pure Gaussian component with $v_{\rm gauss} = 2.5\pm 0.5$\,km\,s$^{-1}$ is needed to account for the narrow central peak (Table~\ref{tab:velocities-1}).

Our period analysis, excluding periods nearly equal to our total time span, indicates as the most powerful period the one of 248.2 d, very close to the 248.7 d of VizieR. None the less, the phased data with this period are empty for $\sim50\%$ of this potential variability cycle. Given that the five highest powers have similar values (Fig.~\ref{f:app}), we suggest that the most probable period is that of 72 d, noting however the scarcity of data.

These results, in association with the strong IR excess, as seen in the Spitzer \citep{2004ApJS..154...18H} spectrum of IRAS 07080+0605 with the presence of intense PAHs bands (Fig~\ref{fig:SEDs-model}), and the presence of a cold molecular cloud along the line-of-sight detected in the $K$-band spectrum of IRAS 07080+0605 \citep{2018PASP..130k4201A}, may reinforce a young nature for this object. This scenario seems to be favoured by the position of IRAS 07080+0605 in the HR diagram (left panel, Fig.~\ref{fig:traks-solar-lmc-art}).

%----------------------------------------

\subsubsection{IRAS 07377-2523}

IRAS 07377-2523 (SS 147) was detected by \citet{Stephenson-and-Sanduleak-1977} in their survey searching for H$\alpha$-emitting stars. \citet*{Parthasarathy-2000}, based on low resolution spectroscopy, classified it as a B8 III-IVe star. It was also selected as a massive young stellar object (YSO) candidate by \citet{Mottram-2007}. In the same year, based on the analysis of high-resolution spectra, IRAS 07377-2523 was classified as a FS CMa star by \citet{Miroshnichenko_2007}, who suggested a B8/A0 spectral type.

Through the analysis of the FEROS spectra, we could also derive a B8-B9 spectral type with an effective temperature of 12000$\pm$1000 K, based on different equivalent width ratios (Methods 2 and 3).  

The presence of shell-type profiles seen in the Balmer and Fe\,{\sc ii} lines may indicate a circumstellar environment seen edge-on (Sect.~\ref{sec:32}). 

From the modeling of the forbidden lines, IRAS 07377-2523 seems to be surrounded by at least three rotating rings, one for each density tracer (Table~\ref{tab:velocities-1}). The [Ca\,{\sc ii}] lines, though very noisy (Fig.~\ref{fig:fits-OI-CaII}), display the highest rotation velocity, ($v_{\rm rot, los} = 55\pm 1$\,km\,s$^{-1}$, $v_{\rm gauss} = 2.5\pm 0.5$\,km\,s$^{-1}$) followed by the [O\,{\sc i}] $\lambda$5577 line ($v_{\rm rot, los} = 13\pm 0.5$\,km\,s$^{-1}$, $v_{\rm gauss} = 7.5\pm 0.5$\,km\,s$^{-1}$) and the [O\,{\sc i}] $\lambda\lambda$6300,6363 lines ($v_{\rm rot, los} = 10\pm 0.5$\,km\,s$^{-1}$, $v_{\rm gauss} = 6\pm 0.5$\,km\,s$^{-1}$). This trend of decreasing velocity with decreasing density is what is typically seen in Keplerian disks.

Its distance is around 4.1 kpc and adding this to an interstellar color excess of $E(B-V)_{IS} = 0.5$, obtained from the 3D dust mapping, we derived some physical parameters of IRAS 07377-2523 (Table~\ref{table:Physical-parameters}). Thus, placing it in the HR diagram reveals a post-main sequence scenario, as suggested by \citet{Parthasarathy-2000}, for a star with roughly 5$\pm$1 M$_\odot$. However, a pre-main-sequence nature cannot be discarded (see left panel of Fig.~\ref{fig:traks-solar-lmc-art}). 

%-------------------------------------------

\subsubsection{IRAS 07455-3143}
\label{subsubsec:IRAS07455}

IRAS 07455-3143 (CD-31 5070, ALS 782, Hen 3-78) was classified by \citet{Orsatti_1992}, based on UBV photometry, as an early B-type star. Later, \citet{Miroshnichenko_2007} classified it as FS CMa star of spectral type B7/B8 based on an analysis of their high-resolution spectra. Due to the presence of Li\,{\sc i} and Ca\,{\sc i} lines, they also suggested the presence of a late-type companion (K-type). 

Based on our FEROS spectra, this is the only object of our sample, for which we could not confirm the presence of the B[e] phenomenon, due to the absence of [O\,{\sc i}] lines (Sect.~\ref{sec:33}).

From Method 1, we classified IRAS 07455-3143 as a B0-1 II/III/V star with $T_{\rm eff} \sim$ 25500\,K, in agreement with \citet{Orsatti_1992}, who suggested an early B-type star. However, this classification is hampered, due to the absence of He\,{\sc ii} and Si\,{\sc iv} lines in our spectra, and also He\,{\sc i} lines in emission. On the other hand, from Method 2, we derived a B8-type with $T_{\rm eff} \sim$ 12500\,K, in agreement with the spectral features seen in our spectra and with \citet{Miroshnichenko_2007}.

IRAS 07455-3143 exhibits an intense spectral variability, as seen in our spectra taken in four different nights in 2008, 2015, and 2016 (Sect.~\ref{sec:33}), which may indicate a binary scenario for this star.

In our spectra only the Ca\,{\sc i} line at 6717.7~\AA \ was identified. The presence of the Li\,{\sc i} line at 6707.7~\AA\ is doubtful. Thus, the presence of the Ca\,{\sc i} line, associated to radial velocity variations (Table~\ref{table:velocities2}), may indicate a complex scenario. There, absorption lines from He\,{\sc i}, Mg\,{\sc ii}, and Si\,{\sc ii} may come from the primary B star, the Ca\,{\sc i} line may come from a cool companion and the stable forbidden emission lines may come from a circumbinary disk or rings. 

IRAS\,07455-3134 displays only the [Ca\,{\sc ii}] 
lines but lacks [O\,{\sc i}]. The profiles of the [Ca\,{\sc ii}] lines are double-peaked with the red 
peak more intense than the blue one (Fig.\,\ref{fig:fits-CaII}). If we interpret the double-peaks as due 
to rotation, then this star might be surrounded by a very compact, high-density ring of gas. To model 
the profile shape, we apply a ring with $v_{\rm rot, los} = 32\pm 0.5$\,km\,s$^{-1}$ and a negligible 
Gaussian component ($v_{\rm gauss} = 1\pm 0.5$\,km\,s$^{-1}$). To suppress the blue peak, we implement 
an asymmetric gap to exclude velocities from  -31.8\,km\,s$^{-1}$ over the maximum of 32\,km\,s$^{-1}$
and reaching values of -24.5\,km\,s$^{-1}$. However, the central part of the profile does not fit and the implementation of a 
second gap, symmetric around zero velocity (from -18.35\,km\,s$^{-1}$ to 18.35\,km\,s$^{-1}$) reduces
too much of the intensity (Fig.\,\ref{fig:fits-CaII}). Therefore, we conclude that the ring does
not necessarily have gaps, but displays density inhomogeneities. Such inhomogeneities might cause
variabilities of the [Ca\,{\sc ii}] profiles, which change in line with the [Fe\,{\sc ii}] lines shown
in Fig.\,\ref{fig:Bep-IRAS07455-3143}.

In addition, the Fe\,{\sc ii} lines in our spectra show shell-type profiles that may suggest an edge-on orientation of this circumbinary environment. 

IRAS 07455-3143 is the most distant object of our sample with a high uncertainty of its value of 12.3 kpc due to so far very imprecise parallax measurement (0.008$\pm$0.026 mas). Assuming this uncertainty, a B8-type, and the mean interstellar extinction, obtained from the DIB and from IRSA, we estimated the physical parameters of this star (Table~\ref{table:Physical-parameters}). According to the HR diagram (Fig.~\ref{fig:traks-solar-lmc-art}, left panel), it can be a post-main sequence (supergiant) star with $M_{\rm ZAMS} \sim$ 20 M$_\odot$. 

%-------------------------------------------

\subsubsection{V* FX Vel}
\label{subsubsec:FXVel}

V* FX Vel (IRAS 08307-3748, WRAY 15-231)  was classified by \citet*{Strohmeier-1968}, \citet{Kukarkin-1972} and \citet{Malkov-2006} as an eclipsing binary. However, \citet{Eggen-1978} questioned the eclipsing nature and classified it as a B9 III-IV star. Later, V* FX Vel was classified as a FS CMa star by \citet{Miroshnichenko_2007} who, based on the analysis of high-resolution spectra, suggested a binary scenario composed of an A and a K stars. Recently \citet{Tisserand-2013}, based on the analysis of low-resolution (R$\sim$3000$-$7000) spectra suggested an A3III type. \citet*{Avvakumova-2013} classified V*~FX~Vel as an eclipsing variable again.

We classified this star as a B8-9 III/V ($T_{\rm eff} \sim$ 11500\,K) from Method 1, and as an A0-2 ($T_{\rm eff} \sim$ 9500\,K or even lower) from Method 2. However, the B8-9 classification can be discarded, as the Mg\,{\sc ii} line at 4482 \AA\, is much stronger than the He\,{\sc i} line at 4471 \AA\ . This is typical for A-type stars, further favouring the A0-2 classification in agreement with \citet{Miroshnichenko_2007}.

Through the analysis of our FEROS spectra of 2008, 2015 and 2016, we noticed a strong variability in the line profiles and radial velocities (Sect.~\ref{sec:34} and Table~\ref{table:velocities2}). These variations, associated with the identification of Li\,{\sc i} and Ca\,{\sc i} lines in just one of the four spectra (Fig.~\ref{fig:Bep-FXVel}), reinforce the suggestion of an eclipsing binary for this object, as suggested in the literature. 

If the variations are indeed due to binarity, the photometric period inferred from the period analysis could be the binary period. We found that the highest-power period is 387.9 d. The VizieR-inferred period (286 d) is also within the periodogram of V*FX Vel, but with much smaller power (by one order of magnitude), i.e.\ higher probability of not being the main variability period (Fig.~\ref{f:app}).

Actually, the complex spectra of V* FX Vel suggest that the absorption lines from He\,{\sc i}, Mg\,{\sc ii}, and Si\,{\sc ii} may come from a primary A star, the Ca\,{\sc i} and Li\,{\sc i} lines may come from a cool companion and the forbidden emission lines and shell-type Fe\,{\sc ii} lines from a circumbinary disk or rings. 

Only [O\,{\sc i}] lines, and none [Ca\,{\sc ii}] lines, are displayed in the spectra of V*\,FX\,Vel. The 
spectrum from 2008 has the highest quality (see Fig.\,\ref{fig:Bep-FXVel}), and we limit our modeling 
to the lines from that year. The profiles are clearly asymmetric, implying multiple components. Our best 
fit model (Table\,\ref{tab:velocities-2} and Fig.\,\ref{fig:fits-OI}) consists of four rings with velocities $v_{\rm rot, los} = 
43\pm 0.5$; $29\pm0.5$; $19\pm0.5$; and $8\pm0.5$\,km\,s$^{-1}$. For all rings, the Gaussian component 
is negligible ($v_{\rm gauss} = 1\pm 0.5$\,km\,s$^{-1}$). To account for the asymmetry, we need to 
implement a gap into the two rings with the lowest velocities. For the ring with 19\,km\,s$^{-1}$, this gap is 
symmetric around the red peak, excluding velocities $>$17.2\,km\,s$^{-1}$, whereas the ring with 
8\,km\,s$^{-1}$ requires a large ($>$ one quarter) gap to suppress partly the red peak and the central
region. This gap corresponds to a lack in velocities starting from 7.88\,km\,s$^{-1}$ over the maximum 
red peak and reaching down to -1.4\,km\,s$^{-1}$. In a scenario with a close companion, these rings might originate from previous interaction phases.

From its SED and Spitzer spectrum (Fig.~\ref{fig:SEDs-model}), we confirm an intense IR excess and the presence of the silicate band at 10 $\mu$m. 

V* FX Vel is the closest star of our sample ($\sim$ 353\,pc), in agreement with the low interstellar extinction derived from the DIB present in our spectra. From its parameters and position in the HR diagram (Fig.~\ref{fig:traks-solar-lmc-art}, left panel), we note that is a low-mass star, with $M_\text{ZAMS} <$ 2 M$_\odot$. However, we cannot discard a pre-main sequence nature for V* FX Vel.

%----------------------------------------------------------------------------------------

\subsubsection{IRAS 17449+2320}

IRAS 17449+2320 (BD+23 3183) was detected for the first time by \citet{Stephenson-1986A} as a new H$\alpha$ emission line star. \citet{Downes-Keyes-1988} obtained a low-resolution spectrum and classified it as a Be star. Later \citet{Miroshnichenko_2007}, based on the analysis of their high-resolution spectra, identified the presence of the B[e] phenomenon, and classified it as an A0V star. The B[e] phenomenon in IRAS 17449+2320 was furthermore confirmed by the medium resolution (R$=$13000$-$18000) spectra of \citet*{Aret_2016} as well as by our FEROS spectra.

From Methods 1, 2, and 3, we also derived a spectral type A0-A2, with a mean effective temperature of \mbox{$9350\pm400$}~K, in agreement with \citet{Miroshnichenko_2007}.

We noticed, from the comparison with the literature, a high spectral variability on a time scale of few days \citep{Sestito-2017}. However, we could not confirm this, because we have spectra taken in just one night.

The broad absorption, associated to a central double-peaked emission, seen in the Balmer lines and the shell-type profiles of Fe\,{\sc ii} lines (Sect.~\ref{sec:37}) may indicate the presence of a circumstellar disk seen edge-on. On the other hand, the [O\,{\sc i}] emission lines are broad and almost flat-topped. This kind of profile might be 
reproduced with a combination of three components, of which only two contain a rotation velocity 
($v_{\rm rot, los} = 27\pm 0.5$ and $15\pm0.5$\,km\,s$^{-1}$). No significant Gaussian component is 
needed ($v_{\rm gauss} = 1\pm 0.5$\,km\,s$^{-1}$), see Table~\ref{tab:velocities-2} and Fig.~\ref{fig:fits-OI}. In the absence of a second density indicator, it remains open whether these rings represent rotating or equatorially outflowing gas.

IRAS 17449+2320 also displays an IR excess with a weak silicate band in emission at 10\,$\mu$m, as seen in its Spitzer spectrum (Fig.~\ref{fig:SEDs-model}). 

IRAS 17449+2320 is one of the closest stars of our sample ($\sim$ 740\,pc), in agreement with the very low extinction obtained from the 3D dust mapping (Table~\ref{table:interstellar-extinction}). From the parameters obtained by us (Table~\ref{table:Physical-parameters}) and the position in the HR diagram, IRAS 17449+2320 has $M_{\rm ZAMS} = 2-3$ M$_\odot$, being still in the main-sequence or close to its end (left panel, Fig.~\ref{fig:traks-solar-lmc-art}).

%------------------------------------------
\subsection{SMC star}

\subsubsection{[MA93] 1116}

[MA93] 1116 (Cl* NGC 346 KWBBE 200, 2MASS J00590587-7211270) belongs to the SMC and was classified as a compact H\,{\sc ii} source by \citet{Meyssonnier-Azzopardi-1993}. Based on a photometric survey, [MA93] 1116  was later classified as a classical Be star located in the open cluster NGC 346\footnote{NGC 346 is an open cluster with intense star formation \citep{Nota_2006, Sabbi_2007}.} by \citet*{Keller-1999}. 

The presence of the B[e] phenomenon was mentioned for the first time by \citet{Wisniewski_2007}, based on the analyze of high-resolution spectra. The same authors, using photometric data and Kurucz atmospheric models, derived its physical parameters: $T_{\rm eff}\sim$19000~K, $\log$($L$/L$_\odot$)$\sim$4.4, and $R_{\rm star}\sim$14 R$_\odot$. From these characteristics, \citet{Wisniewski_2007} suggested that [MA93] 1116 would be a B[e] supergiant. Later, \citet{Whelan-2013} analysed the Spitzer spectrum and identified silicate emission and strong PAH bands at 6.2 $\mu$m, 7.7 $\mu$m, 8.6 $\mu$m and 11.3$\mu$m. Due to these characteristics, [MA93] 1116 was classified as a class A PAH spectrum, which is typical of non-isolated Herbig Ae/Be stars. \citet{Kamath_2014} analyzed low-resolution spectra and identified the presence of forbidden lines, classifying [MA93] 1116 as a planetary nebula candidate. Recently \citet{Paul-et-al-2017}, based on the spectral energy distribution and using theoretical spectral templates, derived a B0.5 spectral type, $T_{\rm eff}\sim$29000~K, $\log(L/$L$_\odot)\sim$4.41 and age of 2.5 Myr. Due to these physical parameters, these authors suggested a classification as HAeBe star for [MA93] 1116.

Our spectra has a richness of emission lines and no absorption lines were identified (Sect.~\ref{sec:39}). From the literature, it is possible to see a sensible variability in the line profiles. 

The [O\,{\sc i}] lines in [MA93]\,1116 display a narrow, symmetric central component which can be fit 
with a pure Gaussian ($v_{\rm gauss} = 2.5\pm 0.5$\,km\,s$^{-1}$). This Gaussian is superimposed on a 
broader asymmetric component, for which we find a good fit using a ring ($v_{\rm rot, los} = 15\pm 
0.5$\,km\,s$^{-1}$, $v_{\rm gauss} = 1\pm 0.5$\,km\,s$^{-1}$) with a symmetric gap around the red peak, 
excluding velocities higher than 14.1\,km\,s$^{-1}$ (see Table~\ref{tab:velocities-2} and Fig.~\ref{fig:fits-OI}).

From its light curve, the period of 573.6 d for [MA 93] 1116 is reported in the present paper for the first time (Fig.~\ref{f:main}).

Based on Method 1, we classified it as a B1-2 star with a $T_{\rm eff}$ = 21600$\pm$3000~K. This classification is in agreement with the spectral features identified in our FEROS spectra, especially the presence of He\,{\sc i}, O\,{\sc ii} and [O\,{\sc ii}] lines in emission and the absence of He\,{\sc ii} lines. Thus, considering the SMC distance and the minimum interstellar extinction obtained from IRSA, we derived $\log(L_*/$L$_\odot) = 4.29\pm0.35$ and a radius of $10\pm3$\,$R_\odot$ (see Table~\ref{table:Physical-parameters}). From the HR diagram, considering the evolutionary tracks for SMC \citep{Georgy-2013} the classification of [MA93] 1116 as a B[e] supergiant with $M_{\rm ZAMS} =$ 9$-$12 \,M$_\odot$ (middle panel, Fig.~\ref{fig:traks-solar-lmc-art}), seems to be more favourable, being in a good agreement with the results of \citet{Wisniewski_2007}.

On the other hand, the classification as a Herbig Ae/B[e] star cannot be discarded, due to the presence of silicates and PAHs bands in the Spitzer spectrum (Fig.~\ref{fig:SEDs-model}), as cited by \citet{Whelan-2013}. Thus, from the HR diagram, considering the pre-main sequence tracks with SMC metallicity from \citet{Bernasconi-Maeder-1996}, [MA93] 1116 has $M_{\rm ZAMS}$ around 15 \,M$_\odot$ (middle panel, Fig.~\ref{fig:traks-solar-lmc-art}).

%-------------------------------------------
%-------------------------------------------
\section{Conclusions}
\label{sec:Conclusions}

We analysed photometric and high-resolution spectroscopic data for a sample of 12 unclassified B[e] stars and candidates: 8 from the Galaxy, 2 from LMC and 2 from SMC. For six of them (Hen\,3-938, Hen 2-91, SS\,255, LHA 115-N82, ARDB\,54, and LHA 120-S59) the analysis of high-resolution spectra was done for the first time. For the other six (IRAS 07080+0605, IRAS 07377-2523, IRAS 07455-3143, IRAS 17449+2320, V* FX Vel, and [MA93] 1116), our analysis of new high-resolution data provided more information about their nature, variability and/or binarity.

We confirmed the presence of the B[e] phenomenon for all objects, except for IRAS 07455-3143. For eight stars, we obtained spectra taken in more than one night, being possible to identify, for most of them, variabilities in the line profiles, radial velocities and {\it V/R}. Even for stars observed for just one night, it was possible to identify variabilities from the comparison with the literature.

Based on different methods and considering the distance provided by Gaia DR2, we derived the effective temperature (spectral type), bolometric magnitude, luminosity, radius (luminosity class), and interstellar extinction for most of our stars. For LHA 120-S 59, we assumed the effective temperature from the literature and derived the other parameters. For Hen 2-91, due to the absence of reliable parameters, we could obtain no further information on its nature. 

Based on the SED, we identified that all stars of our sample have IR excess. From the Spitzer spectra of some of them, we found indication of dust, hinting at a dense and complex circumstellar environment. Our analysis of [Ca\,{\sc ii}] and [O\,{\sc i}] line profiles reveals that all stars have indication for one or more gaseous rings in (quasi-) Keplerian rotation around the central star or binary system. It is important to notice that no trend was seen as a function of the metallicity, indicating that the circumstellar density structure of these stars is not  metallicity dependent.

From the period analysis of light curves of four objects, we found that for two of them, namely V* FX Vel and LHA 120-S59, the estimated photometric periods are in good agreement with the literature. For IRAS 07080+0605, we found a different period compared to the literature, although dubious due to scarcity of data. For [MA93] 1116, its period was obtained for the first time. The presence of these periodic variabilities may indicate binarity, however, except for V* FX Vel, our spectroscopic analysis did not find evidence of companions.

By comparison of the position of our stars in the HR diagram to evolutionary tracks for solar, SMC and LMC metallicities, we found that: (i) IRAS 07080+0605 and V*~FX~Vel are A[e] stars with uncertain classification, probably either main sequence or pre-main sequence objects; (ii) IRAS 07377-2523 is a B[e] star, either in a post- or a pre-main sequence phase; (iii) IRAS 17449+2320 is a B[e] star probably at the main sequence or close to its end; (iv) SS\,255 can be another B[e] star in a similar stage as IRAS 17449+2320, but based on its spectral features, a post-AGB nature seems more favourable; (v) LHA 120-S59 is a B[e] supergiant; (vi) Hen\,3-938, and [MA93] 1116 can be B[e] supergiants or HAeB[e]; and (vii) IRAS 07455-3143 is a B supergiant. However, our most remarkable results are the identification of ARDB\,54 as the third A[e] supergiant, the first one in the LMC, and of LHA 115-N82, as an intermediate mass and post-main sequence B[e] star, but with a light curve showing eruptions similar to LBVs, i.e. a ``LBV impostor".

More observations using different techniques, including interferometry and polarimetry, associated to simultaneous high-resolution spectroscopy and photometry are necessary to confirm the nature of these peculiar objects, their variability and binary fraction. This will certainly allow for a better comprehension of the B[e] and also A[e] phenomena in environments with different metallicities.

%-------------------------------------------------------------------------------------------------------------
\section*{Acknowledgements}

We thank the anonymous referee for his/her very constructive comments that helped us to improve this paper.

CAHC acknowledges financial support from Coordena\c c\~ao de Aperfei\c coamento de Pessoal de N\'ivel Superior (Brazil-CAPES) through PhD. grant. MK acknowledges financial support from GA\,\v{C}R (grant number 17-02337S). The Astronomical Institute Ond\v{r}ejov is supported by the project RVO:67985815. DP acknowledges financial support from Conselho Nacional de Desenvolvimento Cient\'ifico e Tecnol\'ogico (CNPq - Brazil) through grant 300235/2017-8. 

This study was financed in part by the Coordena\c c\~ao de Aperfei\c coamento de Pessoal de N\'ivel Superior - Brasil (CAPES) - Finance Code 001. Parts of the observations also obtained with the MPG 2.2-m telescope were supported by the Ministry of Education, Youth and Sports of the Czech Republic, project LG14013 (Tycho Brahe: Supporting Ground-based Astronomical Observations). We would like to thank the observers (S. Ehlerova and A. Kawka) for obtaining the data.

Part of this project has received funding from the European Union's Framework Programme for Research and Innovation Horizon 2020 (2014-2020) under the Marie Sk\l{}odowska-Curie Grant Agreement No. 823734.

This research has made use of the VizieR catalogue access tool, CDS, Strasbourg, France. This research has also made use of Astropy, a community-developed core Python package for Astronomy \citep{aspy13}.

This research has made use of the NASA/ IPAC Infrared Science Archive, which is operated by the Jet Propulsion Laboratory, California Institute of Technology, under contract with the National Aeronautics and Space Administration.

\addcontentsline{toc}{section}{Acknowledgements}

%%%%%%%%%%%%%%%%%%%%%%%%%%%%%%%%%%%%%%%%%%%%%%%%%%



%\bibliographystyle{mnras}
\bibliography{template} % if your bibtex file is called example.bib


%%%%%%%%%%%%%%%%%%%%%%%%%%%%%%%%%%%%%%%%%%%%%%%%%%

%%%%%%%%%%%%%%%%% APPENDICES %%%%%%%%%%%%%%%%%%%%%

\appendix

\section{Spectral Description}
\label{Apendix:Spectral Descriptions}
We analyzed the FEROS spectra of our sample, in order to identify the spectral and variability features, as well as to check the presence of the B[e] phenomenon.

\subsection*{First group}

\subsection{Galactic stars}

%---------------------------------------------------

\subsubsection{Hen 3-938}
\label{sec:35}

\begin{figure}
        \centering
        \includegraphics[width=\linewidth, clip]{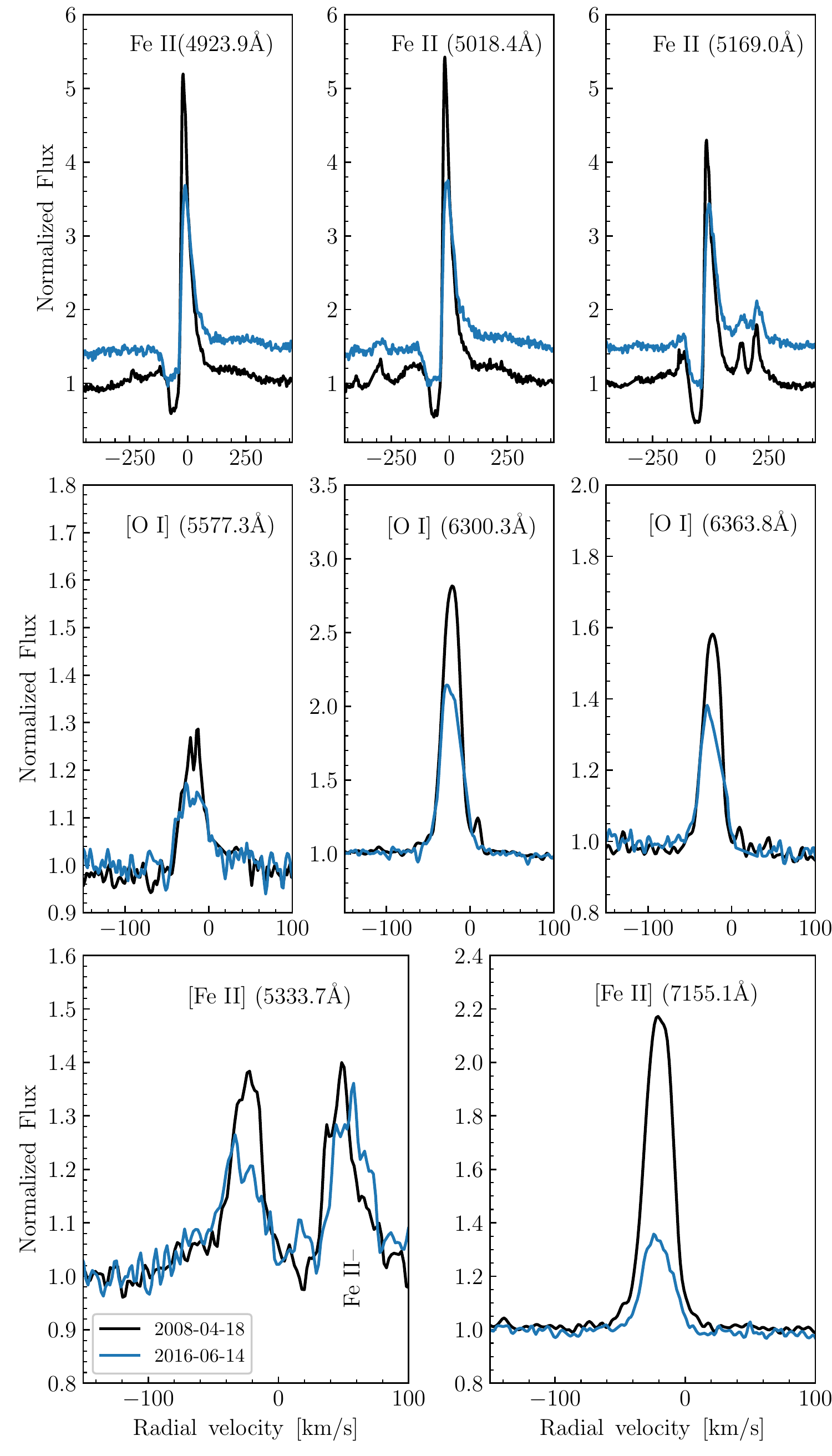}
        \caption{Line profiles variability seen in the FEROS spectra of Hen 3-938: in the top row, Fe\,{\sc ii} lines of multiplet 42 showing P-Cygni profiles; in the middle row, [O\,{\sc i}] lines showing single-peaked profiles; and in the bottom row, [Fe\,{\sc ii}] also presenting single-peaked profiles. }
        \label{fig:Hen-3-938-Bep-art}
\end{figure}

We observed Hen 3-938 with FEROS on 2005-04-18 and 2016-06-14 and in both dates the Balmer lines showed P-Cygni profiles (Fig.~\ref{fig:Balmer-lines}), with the exception of H$\alpha$, which is doulbe-peaked. In both dates, the blue wing of the absorption component of H$\beta$ extends to around $-200$~km~s$^{-1}$.
On the other hand, the blue wing of H$\alpha$ extends to $-1300$~km~s$^{-1}$. Concerning the red wing of both lines, it extends to $\sim500$~km~s$^{-1}$ for H$\beta$ and to $\sim1500$~km~s$^{-1}$ for H$\alpha$. The absorption component of H$\beta$ is centered at $-40$~km~s$^{-1}$ and the central absorption of H$\alpha$ at $-50$~km~s$^{-1}$. We also noted that these lines are more intense in 2005 than in 2016. Concerning the Paschen lines, they have P-Cygni profiles in 2005 and 2016.

Comparing with the literature, we found that H$\alpha$ was observed with a single-peaked profile by \citet{Miroshnichenko_1999} and \citet{Sartori_2010}. This was probably caused by their low-resolution spectra. In addition, there are also equivalent width ($EW$) variations in H$\alpha$, as measured by \citet[][$EW\sim$77~\AA]{Gregorio-Hetem_1992}, \citet[][$EW=$150~\AA]{Miroshnichenko_1999}, \citet[][$EW=$154~\AA]{Sartori_2010}, \citet[][$EW=90$~\AA]{Vieira_2011} and by us ($EW\sim$144~\AA~ and 95~\AA\ in 2005 and 2016, respectively).

Most of the lines that we have identified for Hen 3-938 are from Fe\,{\sc ii} and they are also more intense in 2005 than in 2016 (see Fig.~\ref{fig:Hen-3-938-Bep-art}). The permitted lines clearly show P-Cygni profiles, while the forbidden ones are narrow and show single-peaked profiles. Hen 3-938 also displays the complete [O\,{\sc i}] triplet ($\lambda\lambda$ 5577.3, 6300.3, 6363.8) showing single-peaked emission, confirming the presence of the B[e] phenomenon. Regarding the O\,{\sc i} permitted lines, the IR triplet ($\lambda \lambda$ 7772.0, 7774.2, 7775.4) presents a complex profile in 2005, with a combination of two absorption and two emission components, but in 2016, it seems like a P-Cygni profile with 3 absorption components. On the other hand, the line at 8446.8\AA \ is very intense with a single peak emission in both spectra.

We have also identified Ca\,{\sc ii} and [Ca\,{\sc ii}] lines, which are variable. The IR triplet ($\lambda\lambda$~ 8498.0, 8542.1, 8662.1) observed in 2005 has the bluest line in emission with a single peak, but the reddest emission seems to be double-peaked (the line at 8542.1 \AA \ is affected by the gap in the FEROS coverage), however, in 2016 these lines are double-peaked. The Ca\,{\sc ii} H (3933.7 \AA) and K (3968.5 \AA) lines are in absorption in 2005, but they are not visible in 2016. The [Ca\,{\sc ii}] lines ($\lambda\lambda$ 7291.5, 7323.9) observed in 2005 show single-peaked profiles and those ones observed in 2016 show double-peaked profiles. 

We have also identified He\,{\sc i} lines with P-Cygni profiles, as reported by \citet{Miroshnichenko_1999}, permitted emission lines from  Mg\,{\sc i}, Mg\,{\sc ii}, Si\,{\sc ii},  Ti\,{\sc ii}, Cr\,{\sc ii}, S\,{\sc ii} and  N\,{\sc ii} and forbidden emission lines from  N\,{\sc ii}, V\,{\sc ii} and Cr\,{\sc ii}.

%----------------------------------------
\subsubsection{SS 255}
\label{sec:36}
In our FEROS  spectra, we have identified the presence of intense Balmer emission-lines. Five of them, H$\epsilon$, H$\delta$, H$\gamma$, H$\beta$ and H$\alpha$, can be seen  in Fig.~\ref{fig:Balmer-lines}. The wings of  H$\beta$ and H$\alpha$ extend from -150~km~s$^{-1}$ to +300~km~s$^{-1}$ and from -3000~km~s$^{-1}$ to +2500~km~s$^{-1}$, respectively. We can see the formation of a blue peak or a shoulder in H$\epsilon$ that becomes progressively more intense for the lower members of the Balmer series. Thus, H$\gamma$, H$\beta$ and H$\alpha$ are clearly double-peaked. Regarding the Paschen lines, they have single peaks.
        
Fe\,{\sc ii} is the ion with the majority of lines, including forbidden and permitted ones, which show narrow and single-peaked profiles. We have also identified the presence of intense [O\,{\sc i}] emission lines, showing single-peaked profiles, and confirming the B[e] phenomenon. The presence of intense double-peaked [O\,{\sc ii}] lines ($\lambda\lambda$ 7319, 7330) is remarkable. Concerning the O\,{\sc i} permitted lines, the IR triplet has single peak emissions and the line at 8446.8\AA \ is very intense and also single-peaked. No Ca\,{\sc ii} lines were identified.
        
We have also identified lines from He\,{\sc i}, Mn\,{\sc i}, Mn\,{\sc ii}, Cr\,{\sc ii}, Si\,{\sc ii}, S\,{\sc ii},  N\,{\sc ii}, [N\,{\sc ii}], [S\,{\sc ii}], [Cr\,{\sc ii}] and [Ti\,{\sc ii}], all of them in emission. 

%-------------------------------------------------
\subsubsection{Hen 2-91}
\label{sec:38}
\cite*{Pereira_2003} based on the analysis of medium-resolution (R$\sim$ 1.9~\AA) spectra, reported that Hen 2-91 is a reddened Be star, showing H$\beta$, H$\alpha$, weak iron and [O\,{\sc i}] lines in emission. 

We analyzed FEROS spectra taken by us on 2016-04-12 and from the ESO archive taken on 2016-08-14, 15, 16 and 17. Due to their very low {\it S/N}, H$\beta$ and H$\alpha$ are the only Balmer lines identified. They are very intense and show double-peaked profiles, without any sensible temporal variation (Fig.~\ref{fig:Balmer-lines}). These very intense lines suggest a very extended nebula, in agreement with \citet{Gvaramadze-2010}, who using the Multiband Imaging Photometer from
Spitzer, classified Hen 2-91 as an ellipsoidal nebula with a size of $10\arcsec \times 20\arcsec$. The wings of H$\beta$ and H$\alpha$ lines do not show a variation and extend from $-300$~km~s$^{-1}$ to +200~km~s$^{-1}$ and from -2000~km~s$^{-1}$ to +2000~km~s$^{-1}$, respectively. Concerning the Paschen lines, they are also in emission, possibly with double peaks, and without any variability.

Once again, Fe\,{\sc ii} lines, including forbidden and permitted ones, are the most numerous in our spectra. The permitted lines are weak and appear to have double-peaked, shell or even P-Cygni profiles (due to the very low {\it S/N} of our spectra a better description is not possible). The forbidden lines, including O\,{\sc i} lines, show narrow and single-peaked profiles. The O\,{\sc i} permitted lines are in emission with single or double peaks. The Ca\,{\sc ii} H and K lines are not present, and the IR triplet, if present, is blended with the Paschen lines. Similarly to the hydrogen lines, these lines (Fe\,{\sc ii}, [Fe\,{\sc ii}], O\,{\sc i} and [O\,{\sc i}]) do not present variations.

We also identified absorption lines from  He\,{\sc i}, Mn\,{\sc i}, Mn\,{\sc ii},  Na\,{\sc i} and Si\,{\sc ii}, and emission lines from  N\,{\sc ii} and  [N\,{\sc ii}], without a temporal variability.

%----------------------------------------------------
%---------------------------------------------------
\subsection{SMC stars}

\subsubsection{LHA 115-N82}
\label{sec:310}

LHA 115-N82 was observed with FEROS in 2008 and 2015. The latter consists of public spectra from the ESO archive. The Balmer lines from H$\epsilon$ to H$\gamma$ show a broad absorption, probably of photospheric origin, with wings extending from $\sim-600$ km~s$^{-1}$ to $\sim$+800 km~s$^{-1}$. H$\epsilon$ also shows a narrow absortion probably due to Ca II H line. H$\delta$, H$\gamma$, and H$\beta$ show broad absorptions associated to shell-type profiles with the blue emission more intense in 2008 and the red emission more intense in 2015. H$\alpha$ is double-peaked, presenting the same {\it V/R} variation, and with wings extending from -1800 km~s$^{-1}$ to +2000 km~s$^{-1}$ (see Fig.~\ref{fig:Balmer-lines}). It is important to cite that H$\alpha$ in 2015 is more intense than in 2008. Concerning the Paschen lines, they are in absorption in both years.

\begin{figure}
\includegraphics[width=\linewidth, clip,trim=0mm 3mm -1mm 0mm]{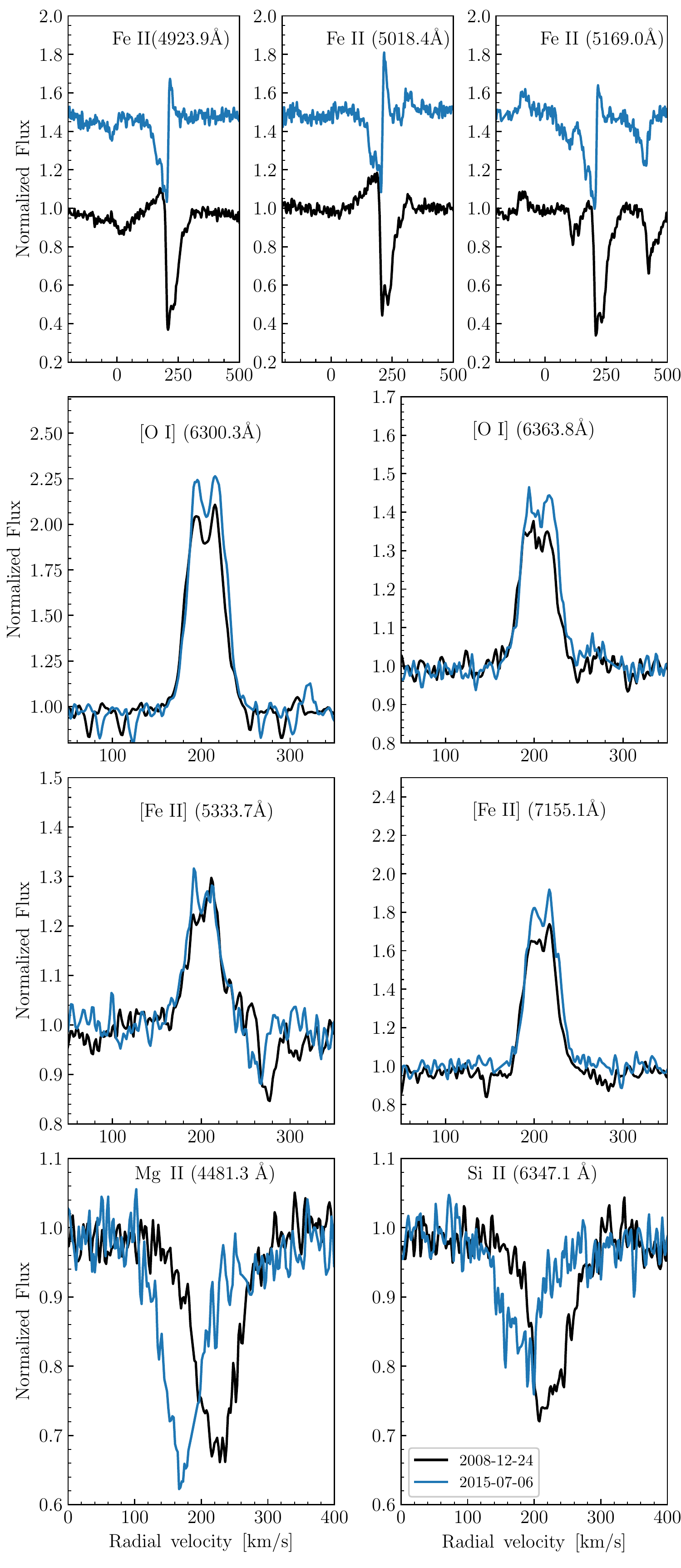}
\caption{Line profile variability seen in the FEROS spectra of LHA 115-N82: in the top row, Fe\,{\sc ii} lines of the multiplet 42 showing normal and inverse P-Cygni profiles; in the two middle rows, forbidden lines of O\,{\sc i} and Fe\,{\sc ii} showing double-peaked profiles; and in the bottom row, absorption lines of Mg\,{\sc ii} and Si\,{\sc ii} showing high variability.}
\label{fig:Bep-LHA 115-N82-a}
\end{figure}

In the literature, LHA 115-N82 was also observed by \citet{Heydari-Malayeri-1990} in 1989, who reported that H$\gamma$, H$\beta$ and H$\alpha$ had P-Cygni profiles. In 1998 and 2001, LHA 115-N82 was also observed by \citet{Evans_2004}, who obtained low resolution spectra with R$\backsimeq$1500 and reported that these two spectra were indistinguishable.

As typically seen in stars with the B[e] phenomenon, most of the lines that we have identified in our spectra are from Fe\,{\sc ii}, including forbidden and permitted lines. The permitted Fe\,{\sc ii} lines show similar behaviour compared to the Balmer lines, showing in 2008 a more intense blue emission and inverse P-Cygni profiles, and in 2015 a more intense red one and P-Cygni profiles (see Fig.~\ref{fig:Bep-LHA 115-N82-a}). On the other hand, the forbidden lines show similar double-peaked profiles, being more intense in 2015. LHA 115-N82 also shows O\,{\sc i} permitted lines in absorption (except for the line at 8446.8\AA \ that seems to be in emission) and [O\,{\sc i}] lines with double-peaked profiles ($\lambda\lambda$ 5577.3 is very noisy), being also more intense in 2015 (see Fig.~\ref{fig:Bep-LHA 115-N82-a}).

We also identified Ca\,{\sc ii} permitted lines in absorption and [Ca\,{\sc ii}] lines in emission with single-peaked profiles. In addition, we have identified absorption lines from He\,{\sc i},  Mg\,{\sc i}, Mg\,{\sc ii} and Si\,{\sc ii}, where the latter two clearly show high variability in their radial velocities (Fig.~\ref{fig:Bep-LHA 115-N82-a}). We have also identified  permitted  and forbidden emission lines from  Cr\,{\sc ii}, Ti\,{\sc ii}, S\,{\sc ii},  [N\,{\sc ii}] and [S\,{\sc ii}].

\citet{Heydari-Malayeri-1990} derived a mean radial velocity of 204.8$\pm$5.6 km~s$^{-1}$, using the [Fe\,{\sc ii}] lines. This value is in good agreement with our values, 206$\pm$2 km~s$^{-1}$ obtained in 2008 and 2015, combining both [Fe\,{\sc ii}] and [O\,{\sc i}] lines. Thus, we can note that from 1989 until 2015, there is no sensible variation in the radial velocities derived from the forbidden lines.
On the other hand, this is not true when considering the absorption lines from He\,{\sc i}, Mg\,{\sc ii} and Si\,{\sc ii}, where a $\Delta v_{\rm rad}\sim$44 km~s$^{-1}$ is seen from 2008 until 2015 (see Fig.~\ref{fig:Bep-LHA 115-N82-a}).

%--------------------------------------------------------------------------
\subsection{LMC stars}
\subsubsection{ARDB 54}
\label{sec:311}

This object was observed with FEROS in two different dates, in 2014 and in 2015. However, both spectra have low {\it S/N}.
The Balmer lines are shown in Fig.~\ref{fig:Balmer-lines}, and H$\epsilon$, H$\delta$ and H$\gamma$ show broad absorptions, probably of photospheric origin, with wings extending from $-400$ km~s$^{-1}$ to +900 km~s$^{-1}$ in H$\epsilon$, -200 km~s$^{-1}$ to +900 km~s$^{-1}$ in H$\delta$ and H$\gamma$ (both lines presenting assymetries in the center of the absorption, caused by another component, probably in emission).
H$\beta$ has a central P-Cygni profile superimposed on a broad photospheric absorption with wings extending from $-400$~km~s$^{-1}$  to +800~km~s$^{-1}$. H$\alpha$ also clearly shows a P-Cygni profile, where the blue wing of the absorption component extends to +50~km~s$^{-1}$ in 2014 and $-10$~km~s$^{-1}$ in 2015, and the red wing of the emission component has similar velocity of +550~km~s$^{-1}$ in both dates (2014 and 2015). The Paschen lines are probably in absorption or not present in our spectra.

Comparing the  H$\gamma$ and H$\beta$ in our spectra with the spectrum observed by \citet{Levato_2014} in 2011, we note that H$\gamma$ is deeper and the emission component of H$\beta$ is more intense in our spectra. Concerning the radial velocity of ARDB 54, \citet{Levato_2014} derived $216\pm7$~km~s$^{-1}$, but from the forbidden lines present in our FEROS spectra, we derived $\sim238$ km~s$^{-1}$ (Table~\ref{table:velocities}).

We could also identify weak Fe\,{\sc ii} lines probably showing P-Cygni profiles. The Fe\,{\sc ii} lines of the multiplet 42 are possibly blended with He\,{\sc i} lines, as suggested by \citet{Levato_2014}, although the line at 5169\,\AA \ also presents a similar profile. In addition, differently than cited by \citet{Levato_2014}, we could identify few weak [Fe\,{\sc ii}] lines in emission and not just one. This star has only the [O\,{\sc i}] line at 6300\,\AA \ in 2014. In 2015, both lines ($\lambda\lambda$ 6300, 6364) can be identified. The IR triplet of O\,{\sc i} is in absorption.

ARBD 54 exhibits the Ca\,{\sc ii} IR triplet and [Ca\,{\sc ii}] lines in emission probably with double-peaks, but this is uncertain due to the low {\it S/N} of our spectra. 

We have also identified absorption lines from He\,{\sc i}, Mg\,{\sc ii} and Si\,{\sc ii} and permitted and forbidden emission lines from Ti\,{\sc ii}, N\,{\sc ii}, and S\,{\sc ii}.

%------------------------------------------------------------
\subsubsection{LHA 120-S59}
\label{sec:312}

\begin{figure}
        \centering
        \includegraphics[width=\linewidth, clip,trim=-1mm 3mm -1mm 0mm]{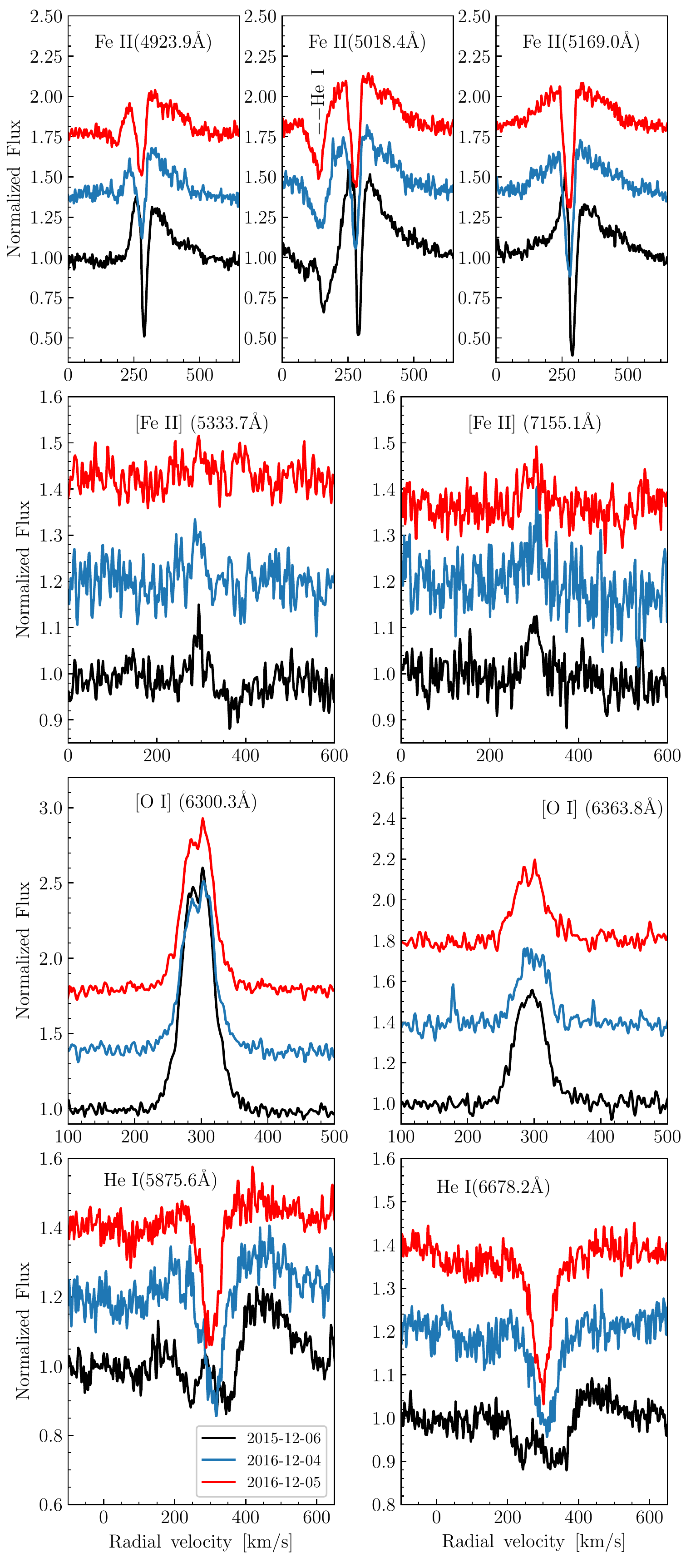}
        \caption{Line profile variability seen in the FEROS spectra of LHA 120-S59: in the top row, the permitted lines of Fe\,{\sc ii} (multiplet 42) showing shell-type profiles; in the middle rows, the [Fe\,{\sc ii}] lines showing single-peaked profiles and [O\,{\sc i}] lines also showing double-peaked profiles; and in the bottom row, the He\,{\sc i} lines in absorption showing high variability.}
        \label{fig:S59-Bep-art}
\end{figure}

Based on our analysis, we noticed that Balmer lines, from H$\epsilon$ to H$\beta$, show shell-type profiles, with {\it V/R} $<$ 1, in our FEROS spectra taken in three different dates in 2015 and 2016. H$\alpha$ has double peaks in both years, with also $V/R < 1$, but being more intense in 2015. The wings of H$\beta$ and H$\alpha$ lines extend from $-900$ km~s$^{-1}$ to +1200 km~s$^{-1}$, and from $-4000$ km~s$^{-1}$ to +5200 km~s$^{-1}$ (the broadest in our sample), respectively (see Fig.~\ref{fig:Balmer-lines}). The central absorption of both lines has a velocity around 270 km~s$^{-1}$, while \cite{Gummersbach_1995} in 1991 measured a velocity of -50 km~s$^{-1}$ for H$\beta$. On the other hand, the double peaks of H$\gamma$, H$\beta$, and H$\alpha$ have a separation around $\sim$137 km~s$^{-1}$, $\sim$147 km~s$^{-1}$ and $\sim$135 km~s$^{-1}$, respectively. However, the spectrum observed by \citet{Levato_2014} in 2011 had a separation of 170 km~s$^{-1}$ for H$\gamma$ and H$\beta$, with {\it V/R}$<$1. 

Concerning the Paschen lines, they are in emission with double peaks and {\it V/R}$>$1. The same happens for the O\,{\sc i} line at 8446.8\AA. On the other hand, the IR triplet is in absorption.

Most of the lines that we identified in our spectra are from Fe\,{\sc ii}, including forbidden and permitted ones. The permitted lines show shell type profiles, with an intense central absorption, which show variations in their radial velocities (Fig.~\ref{fig:S59-Bep-art}): 289 km~s$^{-1}$ (2015-12-06), 278 km~s$^{-1}$ (2016-12-04), and 275 km~s$^{-1}$ (2016-12-05). On the other hand, [Fe\,{\sc ii}] lines are narrow, less intense and with single-peaked profiles, being stronger in the spectrum in 2015. 

LHA 120-S59 also exhibits [O\,{\sc i}] lines ($\lambda\lambda$ 5577.3, 6300.3, 6363.8) showing double-peaked profiles with {\it V/R}$\sim$1, which are also more intense in the spectrum taken in 2015 (see Fig.~\ref{fig:S59-Bep-art}). Regarding the Ca\,{\sc ii} H and K lines, they are in absorption, but the IR triplet and [Ca\,{\sc ii}] lines are not present in all spectra.

We also identified that He\,{\sc i} lines present line profile and radial velocity variations in a timescale of hours/days (Table~\ref{table:velocities2}), showing P-Cygni profiles with two absorption components in 2015 and shell-type or absorption profiles in the spectra taken in two consecutive nights in 2016 (Fig.~\ref{fig:S59-Bep-art}).
In addition,  we have also identified emission lines from  Mg\,{\sc ii},  N\,{\sc ii}, S\,{\sc ii}, Ti\,{\sc ii}, [S\,{\sc ii}] and [N\,{\sc ii}] that do not present important variabilities in our spectra. However, we found a strong variability in radial velocities compared to the literature, even considering the forbidden lines, like [Fe\,{\sc ii}]: $249\pm10$ km~s$^{-1}$ \citep{Levato_2014} and $295\pm5$ km~s$^{-1}$ \citep{Gummersbach_1995}, being the latter in agreement with our values.

\subsection*{Second group}

\subsection{Galactic stars}

\subsubsection{IRAS 07080+0605}

\begin{figure}
        \includegraphics[width=\linewidth, clip]{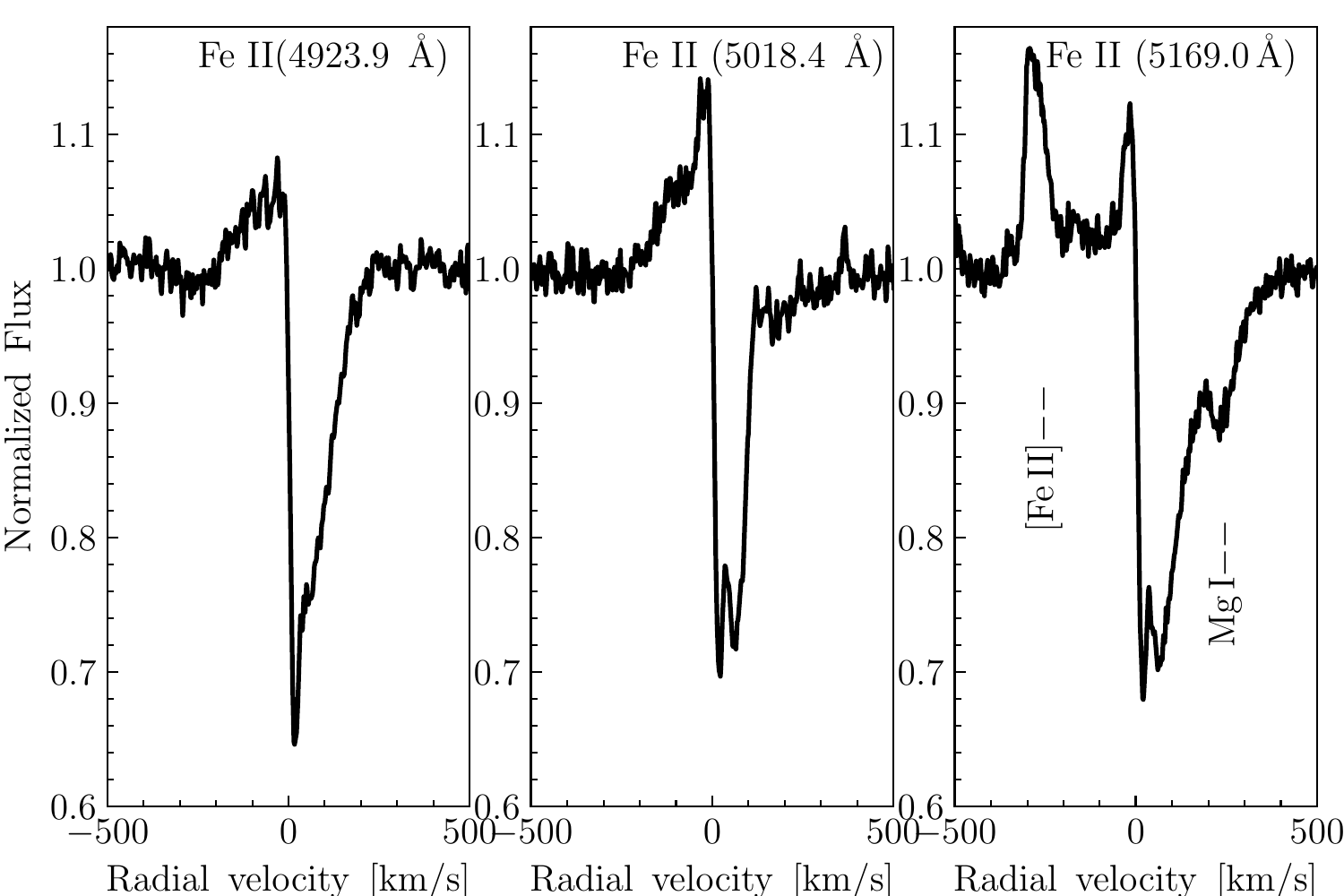}
        \caption{Fe\,{\sc ii} permitted lines of the multiplet 42 showing inverse P-Cygni profiles in the FEROS spectrum of IRAS 07080+0605.}
        \label{fig:IRAS07080+0605-FeII-art2}
\end{figure}
\label{sec:31}

In our 2015 spectra, H$\epsilon$, H$\delta$ and H$\gamma$ show a broad absorption, probably of photospheric origin (H$\epsilon$ is blended with the \ion{Ca}{ii} H line in emission). H$\beta$ and H$\alpha$ also present broad absorption components, but superimposed with double and triple peaked emissions, respectively, which are formed in the circumstellar medium (see Fig.~\ref{fig:Balmer-lines}). The wings of all Balmer lines reach $\sim$1500$-$2000 km~s$^{-1}$. Compared to H$\beta$ and H$\alpha$ profiles observed by \citet{Miroshnichenko_2007} in 2004, our profiles are less intense. For example, the equivalent width of H$\alpha$ in our spectra is 48~\AA \, while it is 68~\AA \,in their spectra. Concerning the Paschen lines, we could only identify P14 and P12 in absorption. P13 is probably blended with Ca\,{\sc ii}.

As usually seen in stars with the B[e] phenomenon, most of the lines identified in our spectra are from Fe\,{\sc ii}, including forbidden and permitted ones. The permitted lines are in absorption, except the lines from multiplet 42, which have inverse P-Cygni profiles (Fig.~\ref{fig:IRAS07080+0605-FeII-art2}). The forbidden lines show narrow and asymmetrical single-peaked profiles. IRAS 07080+0605 also has O\,{\sc i} permitted lines in absorption and forbidden lines showing double-peaked emission. 

The Ca\,{\sc ii} K line shows an inverse P-Cygni profile with an intense triple peak emission. The Ca\,{\sc ii} H line also shows a triple peaked emission, but blended with H$\epsilon$ (see Fig.~\ref{fig:Balmer-lines}). The IR Ca\,{\sc ii} triplet also seems to have inverse P-Cygni profiles, similar to Fe\,{\sc ii} lines of the multiplet 42, but with very weak emission component. On the other hand, [Ca\,{\sc ii}] lines are in emission with single-peaked profiles. 

Concerning other lines, we have identified absorption lines from He\,{\sc i}, Mg\,{\sc i}, Mg\,{\sc ii} and Si\,{\sc ii} and forbidden emission lines from N\,{\sc ii}, Ni\,{\sc ii} and S\,{\sc ii}. He\,{\sc i} and Si\,{\sc ii} lines have asymmetric absorption profiles with more extended red wings, whereas Mg\,{\sc ii} lines are symmetric. We could also identify variation in the He\,{\sc i} $\lambda$4471 line, which is stronger than the Mg\,{\sc ii} $\lambda$4482 line in the spectrum described by \citet{Miroshnichenko_2007}, and absent in our spectra.

%-----------------------------------------------------------
\subsubsection{IRAS 07377-2523}
\label{sec:32}

Fig.~\ref{fig:Balmer-lines} shows some Balmer line profiles (from H$\epsilon$ to H$\alpha$) present in our FEROS spectra. The line profiles of H$\epsilon$, H$\delta$, H$\gamma$ and H$\beta$  show broad absorption components (probably of photospheric origin) and central emission components like shell-type profiles, similar as seen for Be stars \citep*{Rivinius-et-al-2013}, and probably indicating an edge-on orientation of the circumstellar material. The wings of these lines extend up to $\sim$
1000 km~s$^{-1}$.  On the other hand, H$\alpha$ shows a triple-peaked emission profile and its wings extend from $-1900$ km~s$^{-1}$ to +2000 km~s$^{-1}$. Concerning the Paschen lines, they seem to be double-peaked.  

Compared to the literature, H$\beta$ is slightly more intense in the spectrum of \citet{Miroshnichenko_2007}, while the intensity of H$\alpha$ is higher in our spectrum. In addition, the equivalent width of H$\alpha$ is 90~\AA \,in our spectrum and 65~\AA\, in theirs.

Similarly to IRAS 07080+0605, most of the lines that we have identified are from Fe\,{\sc ii}, permitted and forbidden ones. The former show shell-type profiles with a very intense central absorption. On the other hand, [Fe\,{\sc ii}] lines are narrow and show single-peaked profiles. This object also shows forbidden and permitted  O\,{\sc i}  lines. The IR triplet is in absorption, but inserted in an emission component, like a shell profile. The O\,{\sc i} line at 8447 \AA \ shows a double-peaked profile and the forbidden O\,{\sc i} lines show single-peaked profiles.

The Ca\,{\sc ii} H and K lines exhibit three absorption components, similarly as seen for Na\,{\sc i} D lines ($\lambda \lambda$ 5890.0, 5895.9), being probably a combination of interstellar and circumstellar contributions. The Ca\,{\sc ii} IR triplet is in emission with double-peaked profiles, being more intense than the Paschen lines. On the other hand, [Ca\,{\sc ii}] lines are in emission with single-peaked profiles. 

We have also identified absorption lines from He\,{\sc i}, Mg\,{\sc i}, Mg\,{\sc ii} and Si\,{\sc ii}, and  emission lines from  Cr\,{\sc ii}, Ti\,{\sc ii}, [N\,{\sc ii}], and [S\,{\sc ii}].

%-------------------------------------------------

\subsubsection{IRAS 07455-3143}
%\label{Sect:iras07455}
\label{sec:33}

\begin{figure}
        \includegraphics[width=\linewidth, clip]{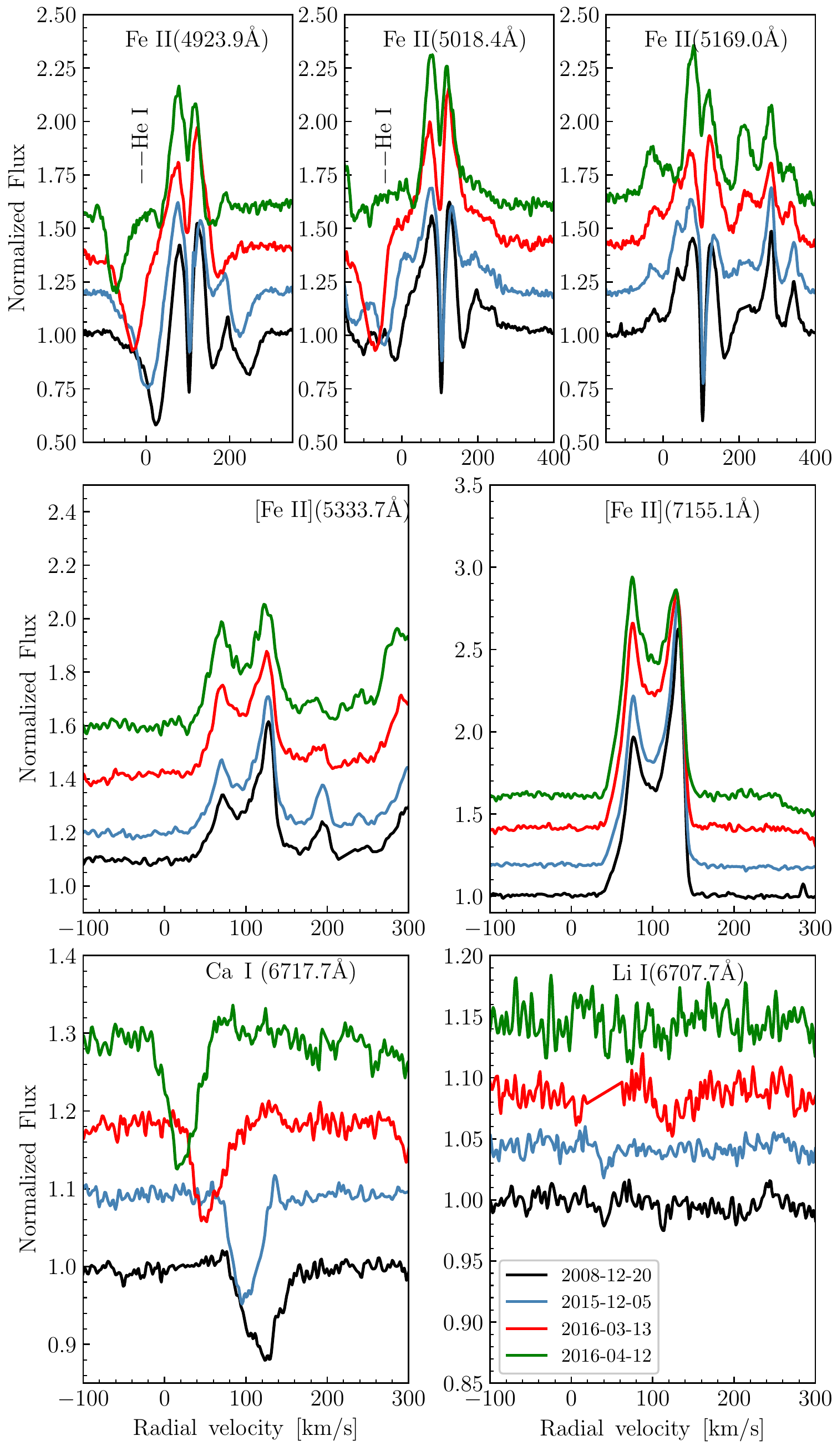}
        \caption{Line profile variability seen in the FEROS spectra of IRAS 07455-3143: in the top row, Fe\,{\sc ii} lines of the multiplet 42 showing shell-type or double-peaked profiles (also He\,{\sc i} lines in absorption showing radial velocity variations); in the middle row, the forbidden lines of Fe\,{\sc ii} showing double-peaked profiles; and in the bottom row, the Ca\,{\sc i} line in absorption and the absent Li\,{\sc i} line.}
        \label{fig:Bep-IRAS07455-3143}
\end{figure}

From spectra taken in four nights in 2008, 2015 and 2016, the presence of intense line profile (mainly in the emission lines) and radial velocity (especially in the absorption lines) variabilities were noted. 

The Balmer lines observed on these dates present both variations. As seen in Fig.~\ref{fig:Balmer-lines}, H$\epsilon$ and  H$\delta$ are in absorption and show radial velocity variations.  The profile of H$\gamma$ observed on 2008-12-20 and 2015-12-05 appears to show inverse P-Cygni profile and the ones observed on  2016-03-13 and 2016-04-12 are in P-Cygni.  On the other hand, H$\beta$ and H$\alpha$ are in emission, showing variations from double-peaked to multiple-peaked profiles. The wings of
these lines do not show any variation and they extend from $-100$ km~s$^{-1}$ to +350 km~s$^{-1}$ in H$\beta$, and from $-1000$ km~s$^{-1}$ to +1000 km~s$^{-1}$ in H$\alpha$.
From the literature, we noted that H$\alpha$ and H$\beta$ were double-peaked in 2004 \citep{Miroshnichenko_2007}, where H$\alpha$ was less intense than seen in our spectra.
 
We identified Fe\,{\sc ii} lines, which similarly to the Balmer lines, present high variability. In Fig.~\ref{fig:Bep-IRAS07455-3143}, we can see the Fe\,{\sc ii} lines of the multiplet 42 that show shell-type profiles, except for those ones observed on 2016-04-12 that have double peaks. It is interesting to note that the shell lines have {\it V/R} $<$ 1, differently than the double peaks with {\it V/R} $>$ 1. The [Fe\,{\sc ii}] lines show double-peaked profiles, which are more intense than the permitted ones, and also present {\it V/R} variation. In addition, IRAS 07455-3143 presents permitted O\,{\sc i} lines in absorption, but curiously the [O\,{\sc i}] lines are absent, making the presence of the B[e] phenomenon doubtful. 
 
We have identified the Ca\,{\sc ii} IR triplet in emission with multiple-peaked profiles. However, the Ca\,{\sc ii} H and K lines present four absorption components of interstellar and stellar origin. On the other hand, [Ca\,{\sc ii}] lines show double-peaked profiles. Forbidden and permitted lines from Ti\,{\sc ii}, Cr\,{\sc ii}, N\,{\sc ii}, and [N\,{\sc ii}] were also identified.

In addition, we identified absorption lines from He\,{\sc i}, Si\,{\sc i}, Si\,{\sc ii}, Mg\,{\sc i}, Mg\,{\sc ii} lines that present high variability in their radial velocities seen in the four different dates. These velocities are different from ion to ion, and there is a decreasing trend from 2008 until 2016 (see Table~\ref{table:velocities2}). The mean radial velocity obtained from He\,{\sc i} lines is lower than the velocities obtained from Mg\,{\sc ii} and Si\,{\sc ii} lines. On the other hand, the radial velocity obtained from the forbidden lines is relatively stable (1$\sigma$). 

The presence of the \ion{Ca}{i} line at 6717.7 \AA\, also reported by \citet{Miroshnichenko_2007}, was confirmed in our spectra. It has a lower radial velocity than the velocities obtained from the other aborption lines, and also shows variability (see Fig.~\ref{fig:Bep-IRAS07455-3143}). However, the identification of the Li\,{\sc i} line at 6707.7 \AA\, is doubtful.

%------------------------------------------------------------------------
\subsubsection{V* FX Vel} \label{sec:34}

\begin{figure}
        \includegraphics[width=\linewidth, clip,trim=0mm 3mm 0mm 0mm]{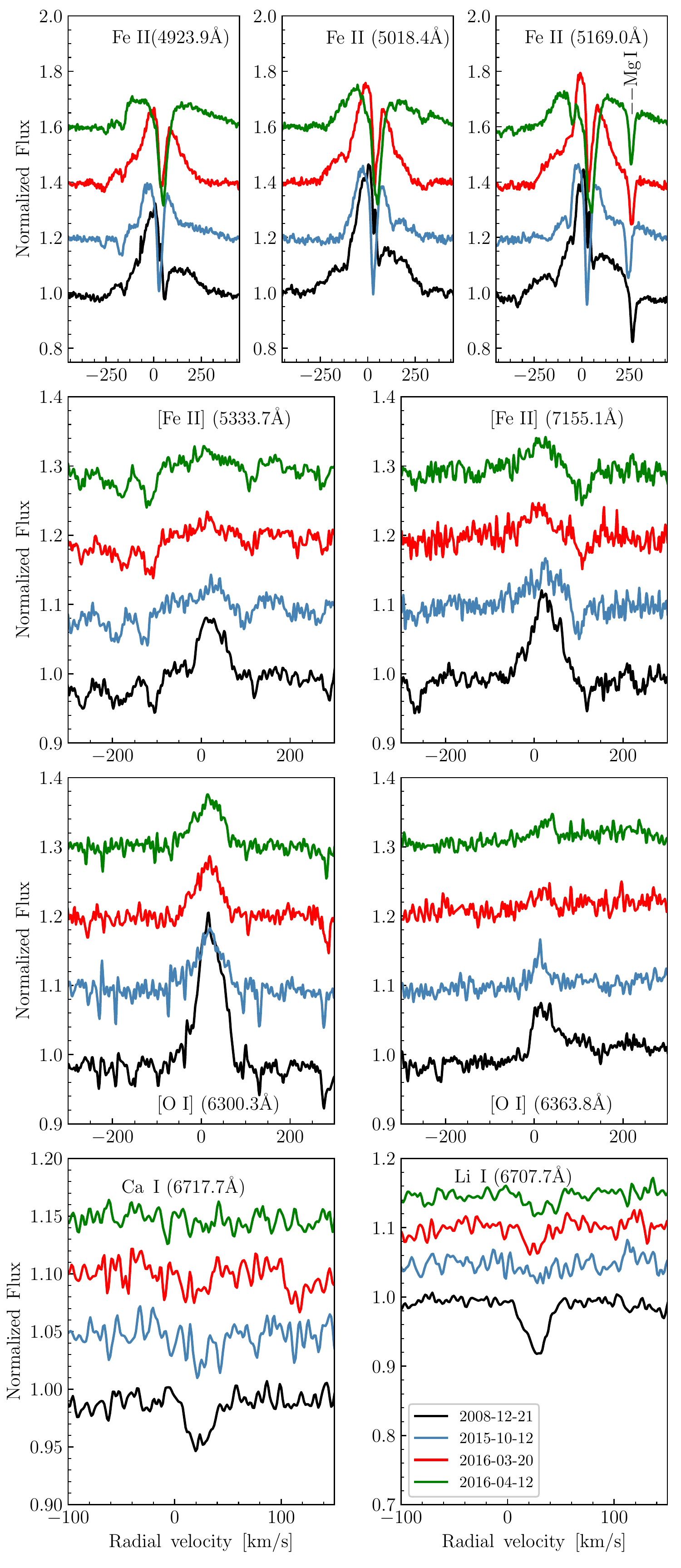}
        \caption{Line profile variability seen in the FEROS spectra of V* FX Vel: in the top row, Fe\,{\sc ii} lines showing shell-type or double-peaked profiles; in the two middle rows, the [Fe\,{\sc ii}] and [O\,{\sc i}] lines showing single-peaked profiles; and in the bottom row, the Ca\,{\sc i} and Li\,{\sc i} absorption lines.}
        \label{fig:Bep-FXVel}
\end{figure}

We observed  V* FX Vel with FEROS on 2008-12-21, 2015-10-12, 2016-03-20 and 2016-04-12, and similarly to IRAS 07455-3143, its spectrum also shows clear variability. The line profiles of H$\epsilon$, H$\delta$ and H$\gamma$ show broad absorption components, superimposed by a narrow component that appears either in emission or in absorption (see Fig.~\ref{fig:Balmer-lines}). The wings of these lines do not show variation. H$\beta$ and H$\alpha$ also present broad absorption components, but
superimposed by double-peaked emissions that exhibit strong variation. Similarly to the other Balmer lines, the wings of H$\beta$ and H$\alpha$ do not show variation, extending from $-2500$ km~s$^{-1}$ to +2500 km~s$^{-1}$ for H$\beta$  and H$\alpha$. It is important to note that our H$\beta$ and H$\alpha$ profiles are different compared to those ones observed by \citet{Miroshnichenko_2007} in 2004. Concerning the Paschen lines, they present single peaks in 2016, but they are absent in 2008.

The permitted Fe\,{\sc ii} lines present in the V* FX Vel spectra also show strong variability, showing shell-type profiles with {\it V/R} $>$ 1 (see Fig.~\ref{fig:Bep-FXVel}). Even the [Fe\,{\sc ii}] lines, which are weak and with single-peaked profiles, present a noticable variation, being more intense in the spectrum taken in 2008. This is also seen for [O\,{\sc i}] ($\lambda\lambda$ 6300 and 6364) lines (Fig.~\ref{fig:Bep-FXVel}). However, the mean radial velocity derived from the forbidden lines remains constant within 1$\sigma$. Concerning the permitted O\,{\sc i} lines, they also present sensible variability, but have shell-type profiles. 

Regarding the Ca\,{\sc ii} H and K lines, they have shell-type profiles with strong variability, and the IR triplet has double-peaks. The forbidden lines are absent in the spectra of V*~FX~Vel. 
%---------------------------------------------

\begin{landscape}
\begin{table}
        \caption{The main spectral features and their line profiles seen in the FEROS spectra of our sample. There, `a` means absorption, `e` single-peaked, `d` double-peaked, `m` multiple-peaked, `s` shell-type, `P Cyg` and `P Cyg$^*$` P-Cygni and inverse P-Cygni profiles, respectively, `a:`  absortion with an emission  component,`d:` double-peaked with an absorption component, `m:` multiple-peaked with an absorption component, `P Cyg:` P-Cygni profile with an absorption component, `$?$` uncertain due to low $S/N$, and `0` indicates the feature is not observed.}
        \label{table:atlas}
        \centering
        {\setlength{\tabcolsep}{0.20em}
\begin{tabular}{c@{}cccccc@{}ccccccccccccccc}
\hline
Star            & Date       & H$\epsilon$ & H$\delta$ & H$\gamma$ & H$\beta$ & H$\alpha$ & Paschen & Fe\,{\sc ii}   & {[}Fe\,{\sc ii}{]} & {[}O\,{\sc i}{]} & {[}O\,{\sc i}{]} & {[}O\,{\sc i}{]} & Ca\,{\sc ii}  & {[}Ca\,{\sc ii}{]} & {[}Ca\,{\sc ii}{]} & He\,{\sc i}  & He\,{\sc i}   & Mg\,{\sc ii}   & Si\,{\sc ii}   & Ca\,{\sc i}    & Li\,{\sc i}    \\
         &            & 3970.1      & 4101.7    & 4340.5    & 4861.3   & 6562.8    &    & (m42) & 5333.7     & 5577.3   & 6300.3    & 6363.8    &IR triplet & 7291.5      & 7323.9      & 4471.7 & 5875.6 & 4481.3 & 6347.1 & 6717.7 & 6707.7 \\
\hline
\hline
\multicolumn{22}{c}{\textbf{First group}} \\
\hline
\textbf{Galaxy} \\
Hen 3-938       & 2005-04-18 & P Cyg           & P Cyg         & P Cyg         & P Cyg        & d         & P Cyg   & P Cyg      & e           & e         & e         & e         & e      & e           & e           & P Cyg      & P Cyg      & e      & e      & 0      & 0      \\
                & 2016-06-14 & P Cyg           & P Cyg         & P Cyg         & P Cyg        & d         & P Cyg   & P Cyg      & e           & e         & e         & e         & e      & d           & d           & P Cyg      & P Cyg      & e      & e      & 0      & 0      \\
SS 255          & 2016-06-14 & e           & e         & d         & d        & d         & e   & e      & e           & e         & e         & e         & 0      & 0           & 0           & e      & e      & 0      & 0      & 0      & 0      \\
Hen 2-91        & 2016-04-12 & 0           & 0         & 0         & d        & d         & d?  & s?     & e           & 0         & e         & e         & e      & 0           & 0           & 0      & a      & 0      & e      & 0      & 0      \\
                & 2014-08-14 & 0           & 0         & 0         & d        & d         & d?  & s?     & e           & 0         & e         & e         & e      & 0           & 0           & 0      & a      & 0      & e      & 0      & 0      \\
                & 2014-08-15 & 0           & 0         & 0         & d        & d         & d?  & s?     & e           & 0         & e         & e         & e      & 0           & 0           & 0      & a      & 0      & e      & 0      & 0      \\
                & 2014-08-16 & 0           & 0         & 0         & d        & d         & d?  & s?     & e           & 0         & e         & e         & e      & 0           & 0           & 0      & a      & 0      & e      & 0      & 0      \\
                & 2014-08-17 & 0           & 0         & 0         & d        & d         & d?  & s?     & e           & 0         & e         & e         & e      & 0           & 0           & 0      & a      & 0      & e      & 0      & 0      \\
\hline
\textbf{SMC}             &            &             &           &           &          &           &     &        &             &           &           &           &        &             &             &        &        &        &        &        &        \\
LHA 115-N 82    & 2008-12-24 & a:          & a:        & a:        & d:       & d         & a   & P Cyg$^*$   & d           & d?        & d         & d         & a      & e           & e           & 0      & a      & a      & a      & 0      & 0      \\
                & 2015-07-06 & a:          & a:        & a:        & d:       & d         & a   & P Cyg      & d           & d?        & d         & d         & a      & e           & e           & 0      & a      & a      & a      & 0      & 0      \\
\hline
\textbf{LMC}    &            &             &           &           &          &           &     &        &             &           &           &           &        &             &             &        &        &        &        &        &        \\
ARDB 54         & 2014-11-24 & a:          & a:        & a:        & P Cyg:       & P Cyg         & a?  & P Cyg?     & e           & 0         & e         & e         & d?     & d?          & d?          & 0      & a      & a      & a      & 0      & 0      \\
                & 2015-12-01 & a:          & a:        & a:        & P Cyg:       & P Cyg         & a?  & P Cyg?     & e           & 0         & e         & e         & d?     & d?          & d?          & 0      & a      & a      & a      & 0      & 0      \\
LHA 120-S 59    & 2015-12-06 & s           & s         & s         & s        & d         & d   & s      & e           & e?         & d         & e         & s?     & 0           & 0           & a:     & s      & e?     & e      & 0      & 0      \\
                & 2016-12-04 & s           & s         & s         & s        & d         & d   & s      & e           & e?         & d         & d?         & s?     & 0           & 0           & a:     & s      & e?     & e      & 0      & 0      \\
                & 2016-12-05 & s           & s         & s         & s        & d         & d   & s      & e           & e?         & d         & d?         & s?     & 0           & 0           & a:     & a      & e?     & e      & 0      & 0\\
\hline
\hline
\multicolumn{22}{c}{\textbf{Second group}}\\
\hline
\textbf{Galaxy} &&&&&&&&&&&&&&&&&&&&&\\ 
IRAS 07080+0605 & 2015-12-06 & a:          & a:        & a:        & a:       & m:        & a   & P Cyg$^*$   & e           & d         & d         & d         & P Cyg$^*$?  & e           & e           & a      & a      & a      & a      & 0      & 0      \\
IRAS 07377-2523 & 2008-12-20 & a:          & a:        & a:        & d:       & m         & d   & s      & e           & e         & e         & e         & d      & e           & e           & a      & a      & a      & a      & 0      &0        \\
IRAS 07455-3143 & 2008-12-20 & a           & a         & P Cyg$^*$?     & d        & m         & e   & s      & d           & 0         & 0         & 0         & m      & d           & d           & a:     & a:     & a      & a:     & a      & 0      \\
                & 2015-12-05 & a           & a         & P Cyg$^*$?     & m        & m         & e   & s      & d           & 0         & 0         & 0         & m      & d           & d           & a:     & a:     & a      & a:     & a      & 0      \\
                & 2016-03-13 & a           & a         & P Cyg         & m        & m         & e   & s      & d           & 0         & 0         & 0         & m      & d           & d           & a      & a:     & a      & a:     & a      & a?     \\
                & 2016-04-12 & a           & a         & P Cyg         & m        & m         & e   & s      & d           & 0         & 0         & 0         & m      & d           & d           & a:     & a:     & a      & a:     & a      & 0      \\
V* FX Vel       & 2008-12-21 & a:          & a:        & a:        & d:       & d:        & e   & s      & e           & 0         & e         & e         & d      & 0           & 0           & a      & s      & a      & a      & a      & a      \\
                & 2015-10-12 & a:          & a:        & a:        & d:       & d:        & e   & s      & e           & 0         & e         & e         & d      & 0           & 0           & a      & s      & a      & a      & 0      & 0      \\
                & 2016-03-20 & a:          & a:        & a:        & d:       & d:        & e   & s      & e?           & 0         & e         & e         & d      & 0           & 0           & a      & s      & a      & a      & 0      & 0      \\
                & 2016-04-12 & a:          & a:        & a:        & d:       & d:        & e   & s      & e?           & 0         & e         & e         & d      & 0           & 0           & a      & s      & a      & a      & 0      & 0      \\
IRAS 17449+2320 &2016-04-12  & a:          & a:        & a:        & a:       & d:        & a:  & s      & e           & e         & e         & e         & s:     & 0           & 0           & a      & a      & a      & a      & 0      & 0      \\
\hline
\textbf{SMC}             &            &             &           &           &          &           &     &        &             &           &           &           &        &             &             &        &        &        &        &        &        \\
{[}MA93{]} 1116 & 2007-10-03 & P Cyg           & P Cyg         & P Cyg         & P Cyg        & P Cyg         & e   & e      & e           & e         & e         & e         & e      & e           & e           & e      & e      & 0      & 0      & 0      & 0      \\
                & 2007-10-04 & P Cyg           & P Cyg         & P Cyg         & P Cyg        & P Cyg         & e   & e      & e           & e         & e         & e         & e      & e           & e           & e      & e      & 0      & 0      & 0      & 0      \\
\hline  
        \end{tabular}
        }
\begin{tablenotes}
        \item \textbf{Notes 1.} The Fe\,{\sc ii} lines of the multiplet 42 (m42) are  centered at 4923.9\,\AA, 5018.4\,\AA~  and 5169.0\,\AA.
        The IR Ca\,{\sc ii} triplet are  centered at 8498.0\,\AA, 8542.1\,\AA~ and  8662.1\,\AA.
\end{tablenotes}
\end{table}
\end{landscape}

\begin{figure}
\includegraphics[width=\textwidth,clip,trim=2mm 0mm 2mm 0mm]{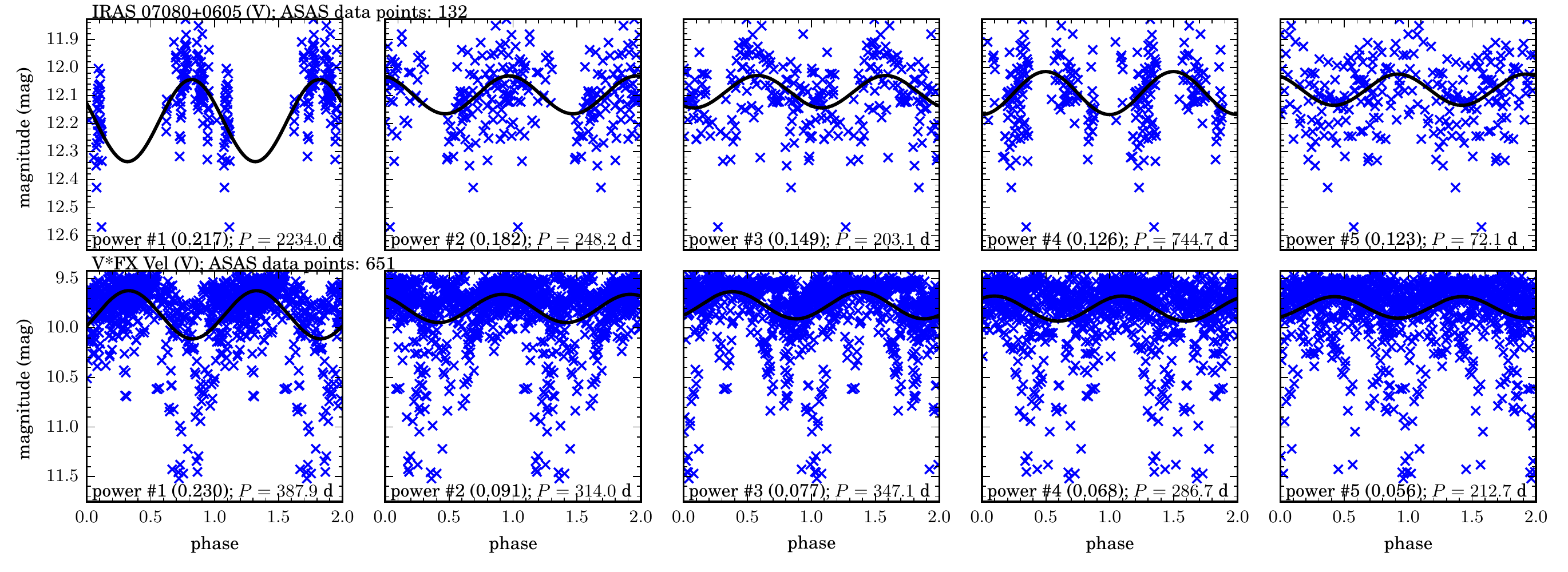}
\caption{The results of the period analysis of two stars, one per row. The star, filter and photometric survey are indicated on the top of each row. The periods with the five highest powers were used to fit a harmonic curve to the data points (markers), one period per panel (on the bottom of each the order and power of each frequency are denoted, together with the corresponding period).}\label{f:app}
\end{figure}

%----------------------------------------------------

We have also identified absorption lines from He\,{\sc i}, Si\,{\sc ii}, and Mg\,{\sc ii}, with strong variability, especially in their radial velocities. Similarly to IRAS 07455-3143, there is a decreasing of the radial velocities of He\,{\sc i}, Mg\,{\sc ii}, and Si\,{\sc ii} lines in 2015 compared to those ones taken in 2008. However a few months later, there was an increase of the radial velocities in the spectra observed in 2016 (see Table~\ref{table:velocities2}). On the other hand, emission lines from  Cr\,{\sc ii}, Ti\,{\sc ii}, S\,{\sc ii}, N\,{\sc ii}, and [N\,{\sc ii}] do not show strong variability.

We identified the presence of lines of Li\,{\sc i} at 6707.7 \AA\ and Ca\,{\sc i} at 6717.7 \AA\ in our spectrum taken in 2008. However in the other dates, these lines are very weak or absent (Fig.~\ref{fig:Bep-FXVel}). The presence of these lines and their behaviour may come from a secondary stellar companion, possibly indicating an eclipsing binary scenario, as suggested in the literature.
%--------------------------------------------

\subsubsection{IRAS 17449+2320}
\label{sec:37}

All the Balmer lines in our spectra are composed of a broad absorption, probably of photospheric origin. H$\delta$ also presents a weak emission component superimposed on the broad absorption. This emission component becomes more intense and it is clearly visible in H$\gamma$, H$\beta$ and H$\alpha$ (see Fig.~\ref{fig:Balmer-lines}). In addition, the emission components of H$\beta$ and H$\alpha$ clearly show double-peaked profiles. The double-peaks in H$\alpha$ were also noticed by
\citet{Miroshnichenko_2007} and \citet{Aret_2016}. Concerning the wings of the Balmer lines, they extend from about $-2000$ km~s$^{-1}$ to around +2000 km~s$^{-1}$ for H$\epsilon$, H$\delta$, H$\gamma$, H$\beta$, and H$\alpha$. Regarding the Paschen lines, they also present broad absorptions.

Comparing the H$\alpha$ profile in our spectra, with those showed by \cite{Miroshnichenko_2007} and \cite{Aret_2016}, we can note a high variability, being the blue peak less intense in their spectra. According to \citet{Sestito-2017}, these variations have a time scale of a few days.

\rule{0pt}{8cm}
Regarding the Fe\,{\sc ii} permitted lines, they are in absorption, but the lines of multiplet 42 show shell-type profiles with an intense central absorption. We identified a few weak [Fe\,{\sc ii}] lines in emission with single-peaked profiles. IRAS 17449+2320 also presents the O\,{\sc i} IR triplet lines showing a complex profile with three narrow absorption components and a broad one. The line at 8446.8\AA \ has double peaks. The optical [O\,{\sc i}] triplet has double-peaked emissions. 

Our spectra also show Ca\,{\sc ii} lines with shell-type profiles. However, [Ca\,{\sc ii}] and [N\,{\sc ii}] lines are not present. We have also identified absorption lines from He\,{\sc i}, Mg\,{\sc ii} and Si\,{\sc ii} and emission lines from Ti\,{\sc ii},  Cr\,{\sc ii}, N\,{\sc ii} and S\,{\sc ii}. 

%---------------------------------------------------
\subsection{SMC star}
\subsubsection{[MA93] 1116}
\label{sec:39}

The presence of the B[e] phenomenon was identified by \citet{Wisniewski_2007}, who analyzed high-resolution spectra (R$\sim$30000) and identified the presence of P-Cygni profiles in the Balmer lines and emission lines of Fe\,{\sc ii}, [Fe\,{\sc ii}] and [O\,{\sc i}]. \citet{Kamath_2014} analyzed low-resolution spectra and identified the presence of lines of [S\,{\sc ii}], [N\,{\sc ii}] and [O\,{\sc iii}].

In our FEROS spectra taken in 2007, the Balmer lines clearly show P-Cygni profiles (Fig.~\ref{fig:Balmer-lines}). Compared to the H$\alpha$ profile observed by \citet{Wisniewski_2007} on November 2004, our profile is more intense.
However, the H$\alpha$ wings in our spectra extend from $-640$~km~s$^{-1}$ to +1400~km~s$^{-1}$, while the H$\alpha$ wings in the spectra of 2004 \citep{Wisniewski_2007} extend from $-2100$~km~s$^{-1}$ to +2190~km~s$^{-1}$ with central absorption of $-260$~km~s$^{-1}$. The equivalent widths of H$\gamma$, H$\beta$ and H$\alpha$ in our spectra are $-10.5, -40.4$ and $-227$~\AA, respectively and $-5.2, -29.3$ and $-267$~\AA\ in their spectra. The Paschen lines are in emission with single peaks.

We have identified forbidden and permitted Fe\,{\sc ii} lines in emission and with single-peaked profiles. We have also identified forbidden and permitted oxygen lines in emission, such as, O\,{\sc i}, O\,{\sc ii}, [O\,{\sc i}], and [O\,{\sc ii}]. We also noticed the presence of weak Ca\,{\sc ii} and [Ca\,{\sc ii}] emission lines. In addition, we identified emission lines from other elements, such as He\,{\sc i},  S\,{\sc ii}, [Cr\,{\sc ii}],  [Ti\,{\sc ii}],  [S\,{\sc ii}] and [N\,{\sc ii}].

\begin{figure*}
\centering
\includegraphics[width=0.32\textwidth]{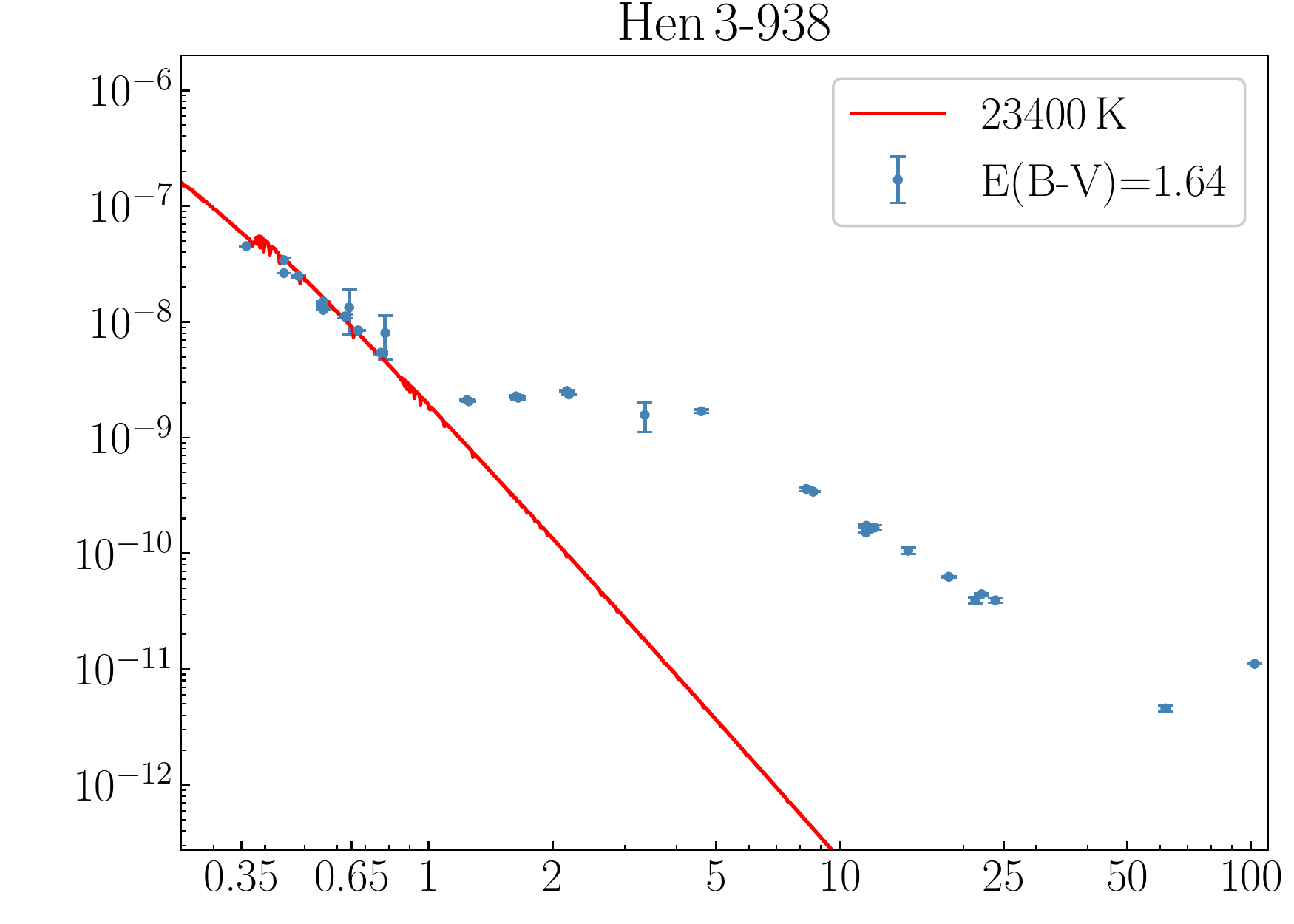}
\includegraphics[width=0.32\textwidth]{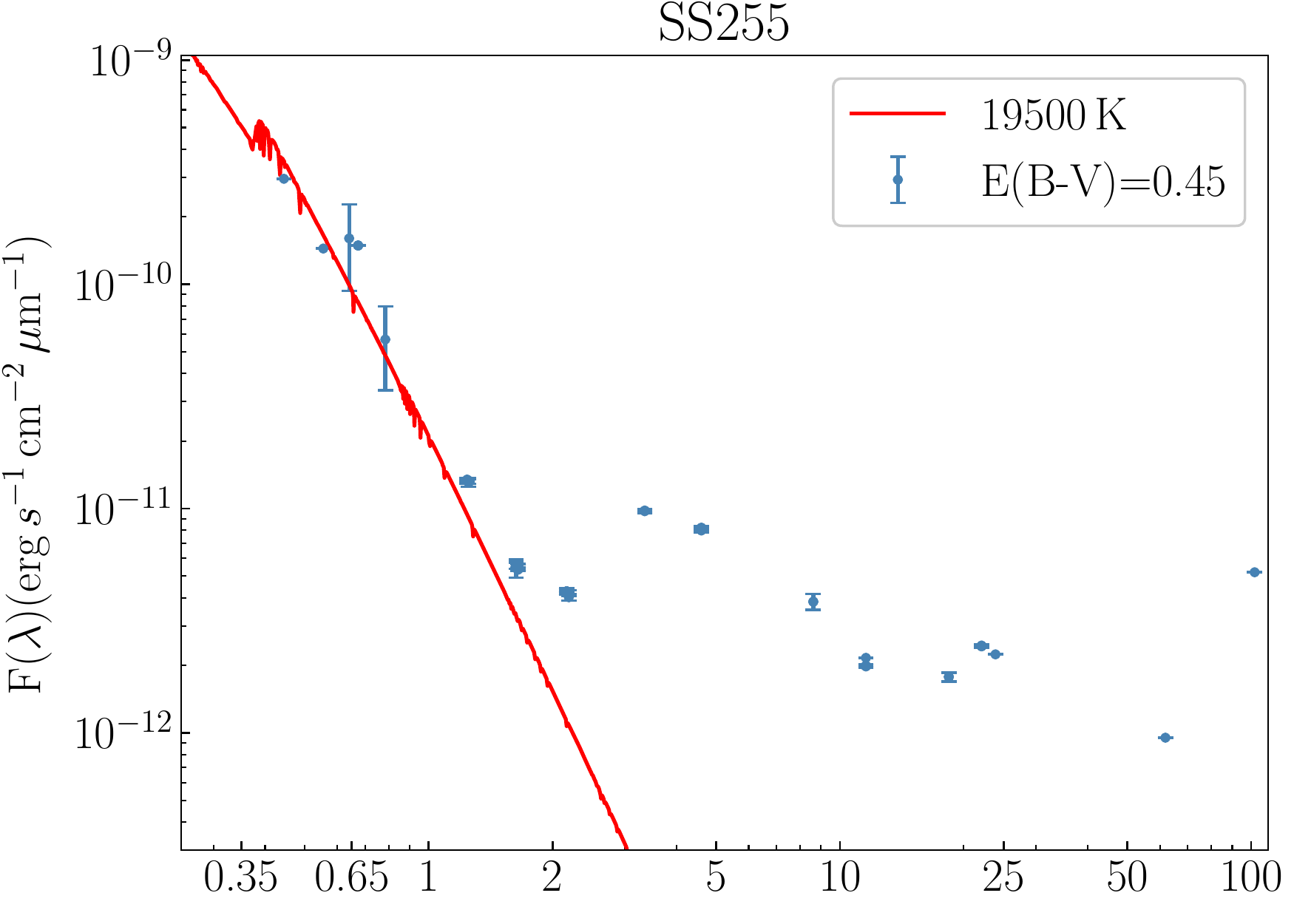}
\includegraphics[width=0.32\textwidth]{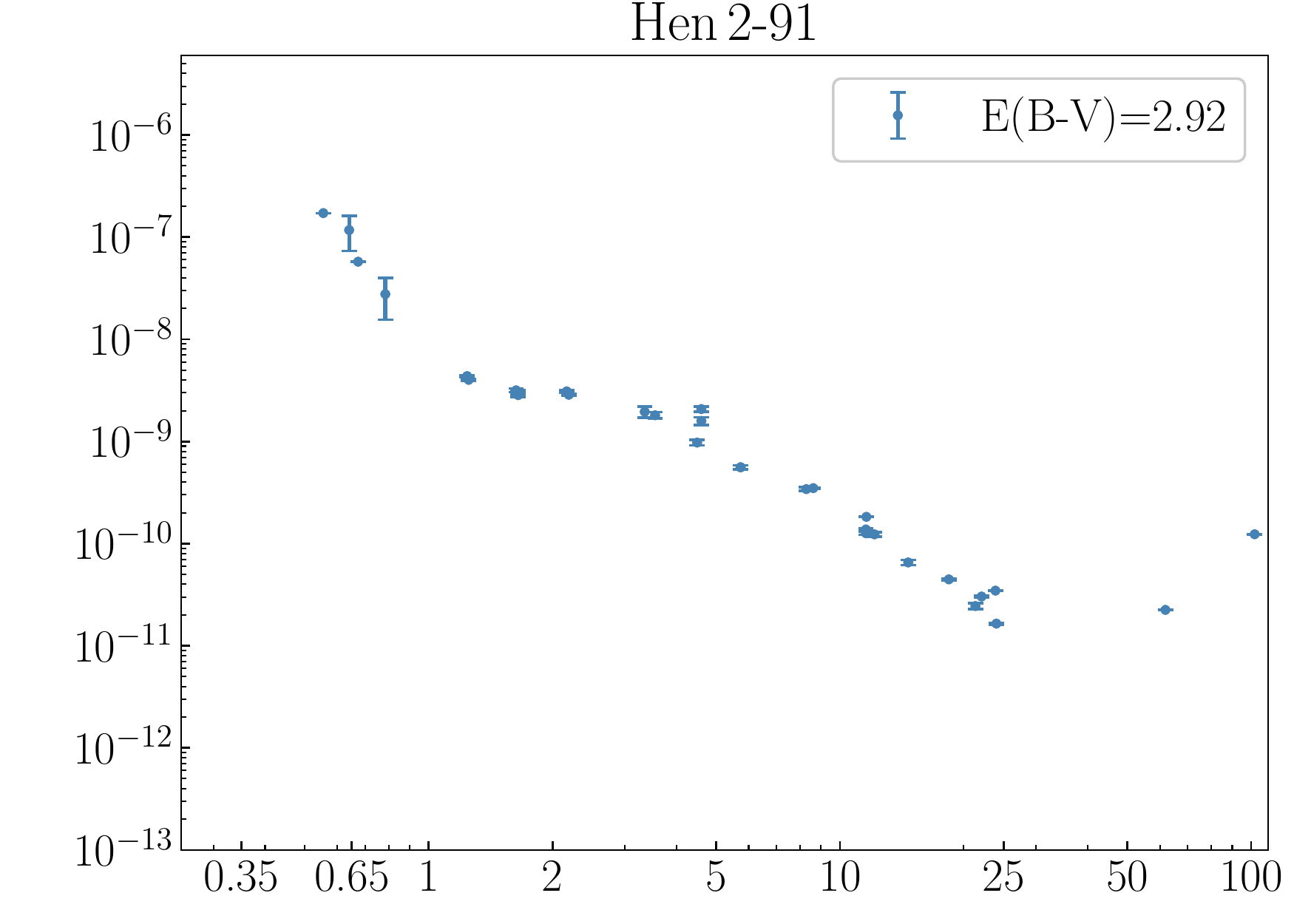}\\
\includegraphics[width=0.32\textwidth]{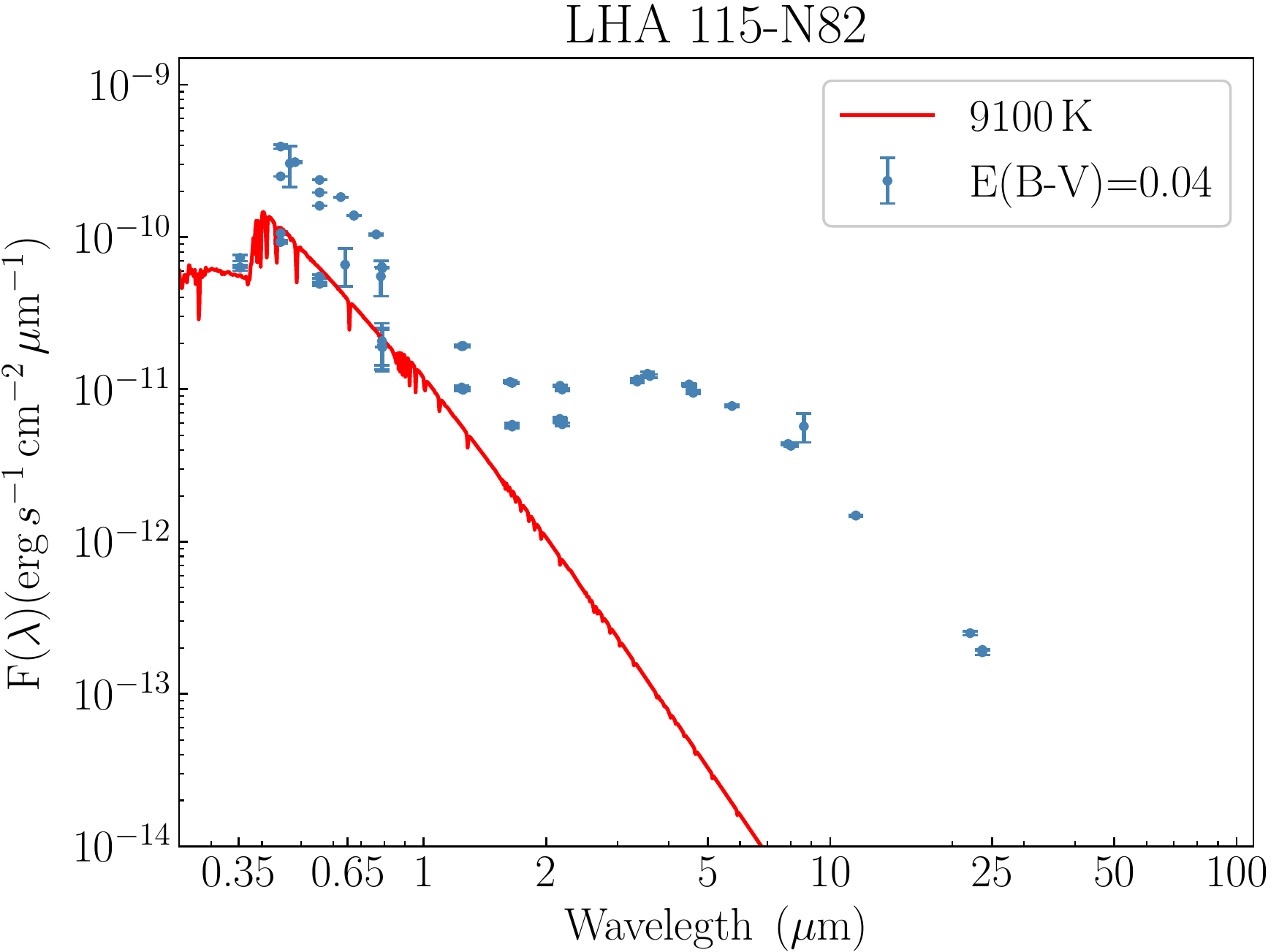}
\includegraphics[width=0.32\textwidth]{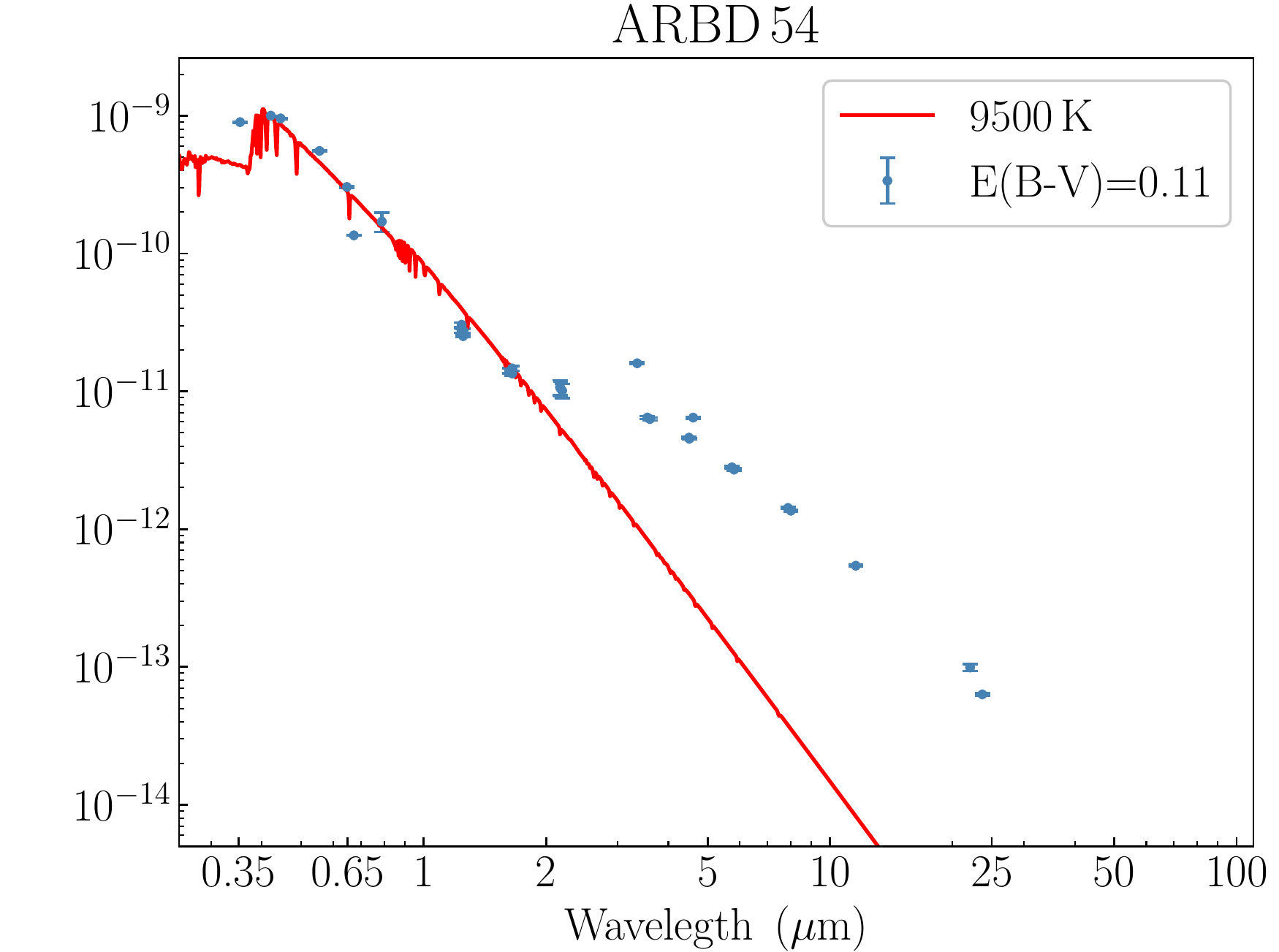}
\includegraphics[width=0.32\textwidth]{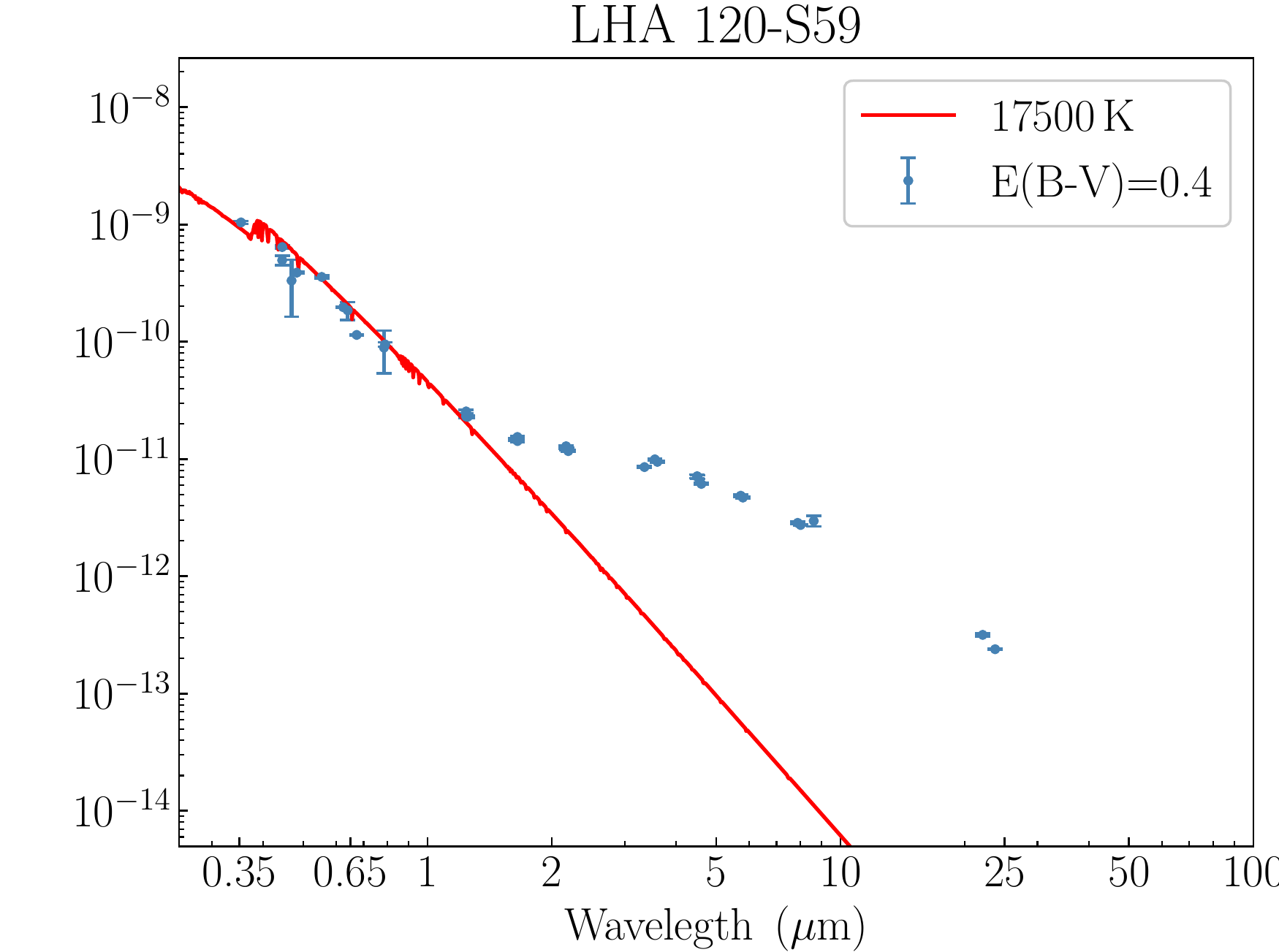} \\
\includegraphics[width=0.32\textwidth]{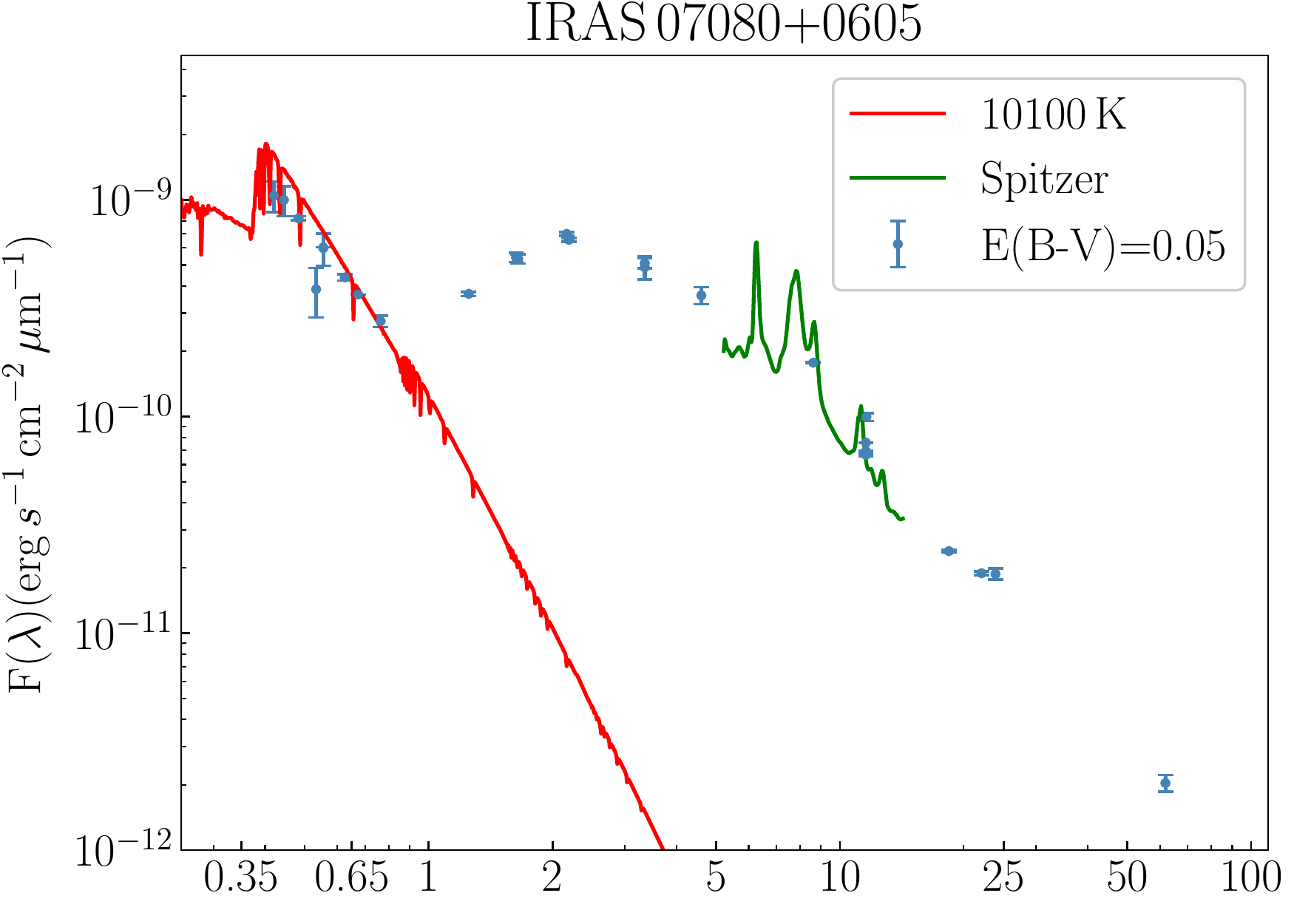}
\includegraphics[width=0.32\textwidth]{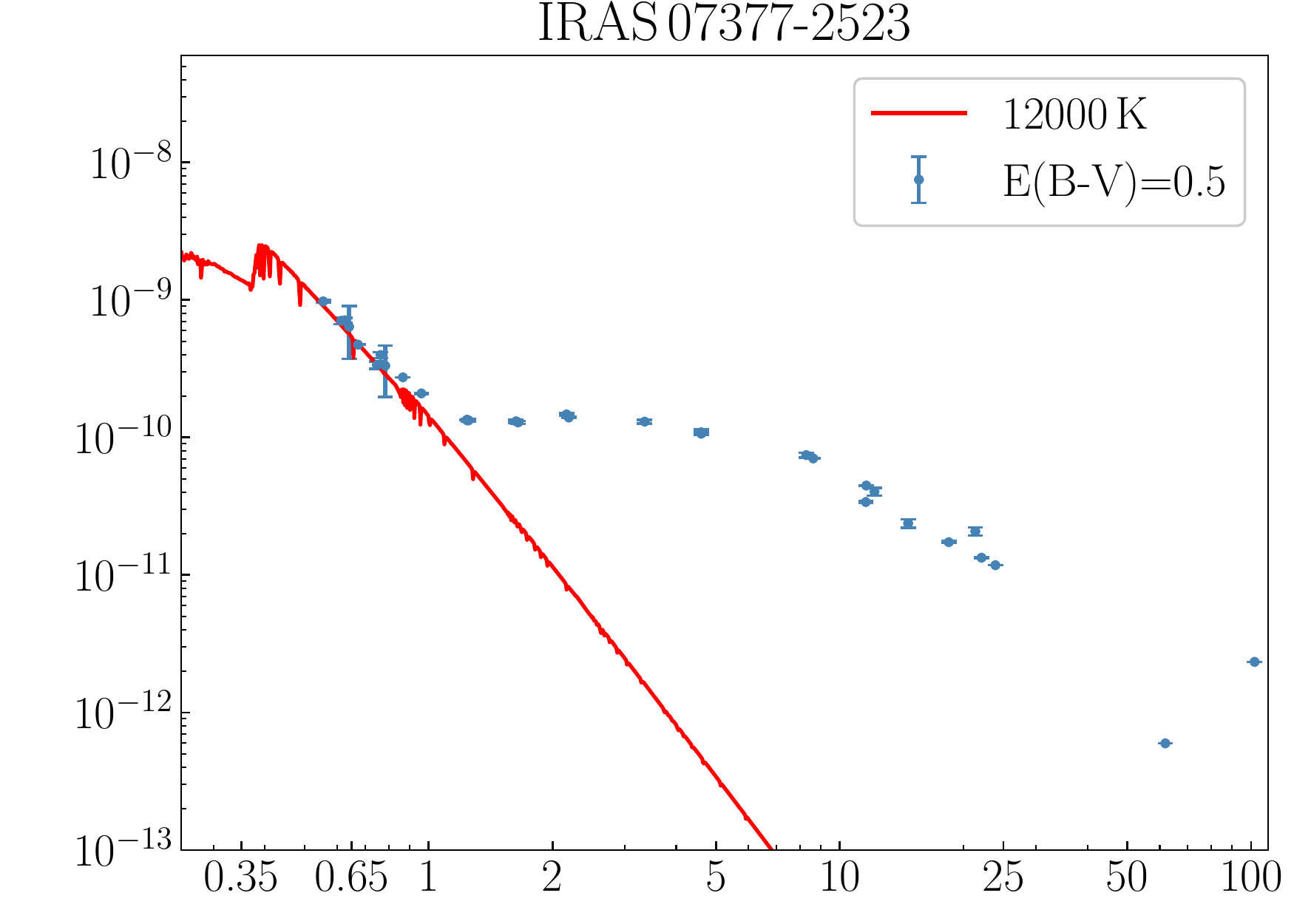}
\includegraphics[width=0.32\textwidth]{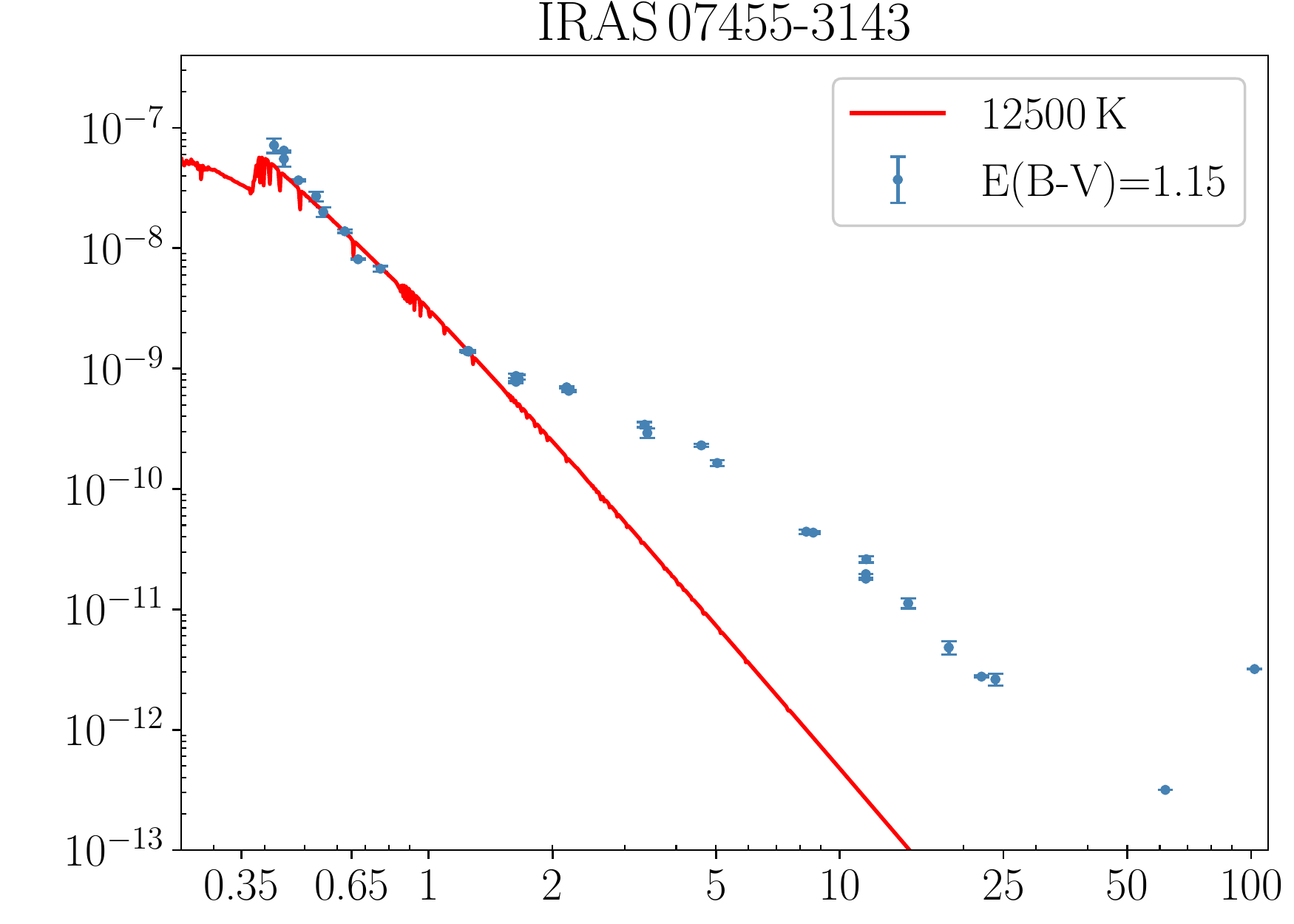}\\
\includegraphics[width=0.32\textwidth]{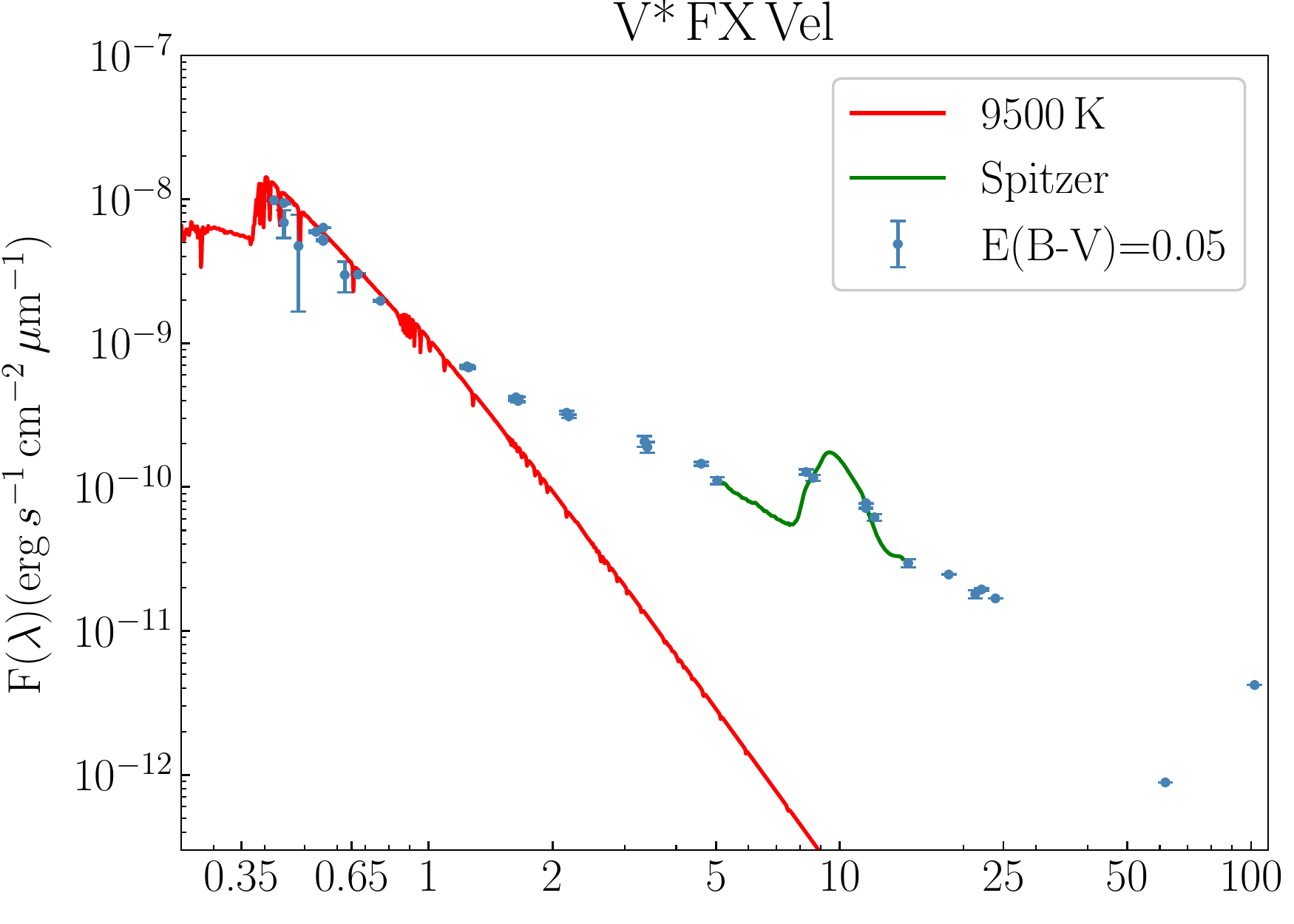}
\includegraphics[width=0.32\textwidth]{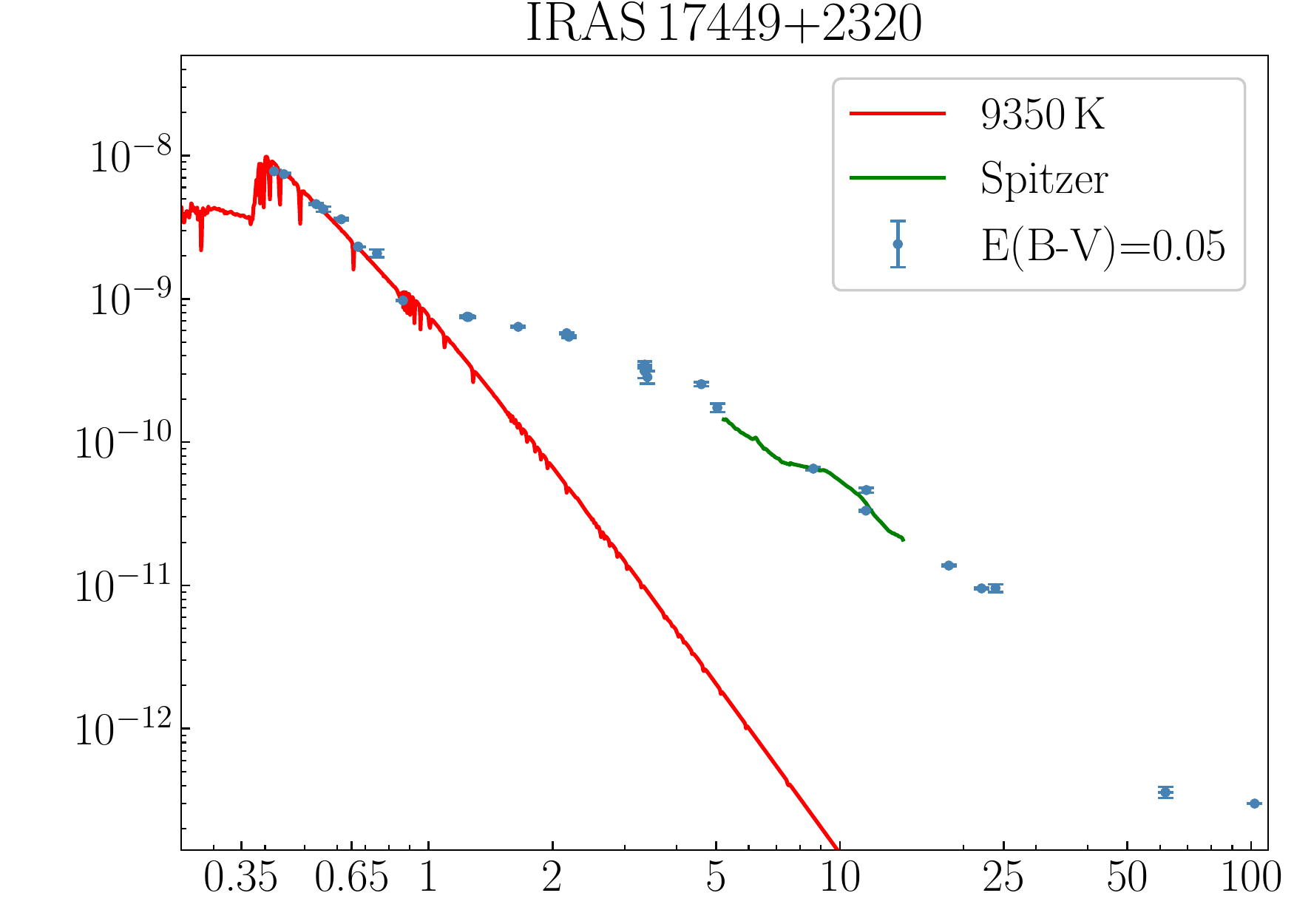}
\includegraphics[width=0.32\textwidth]{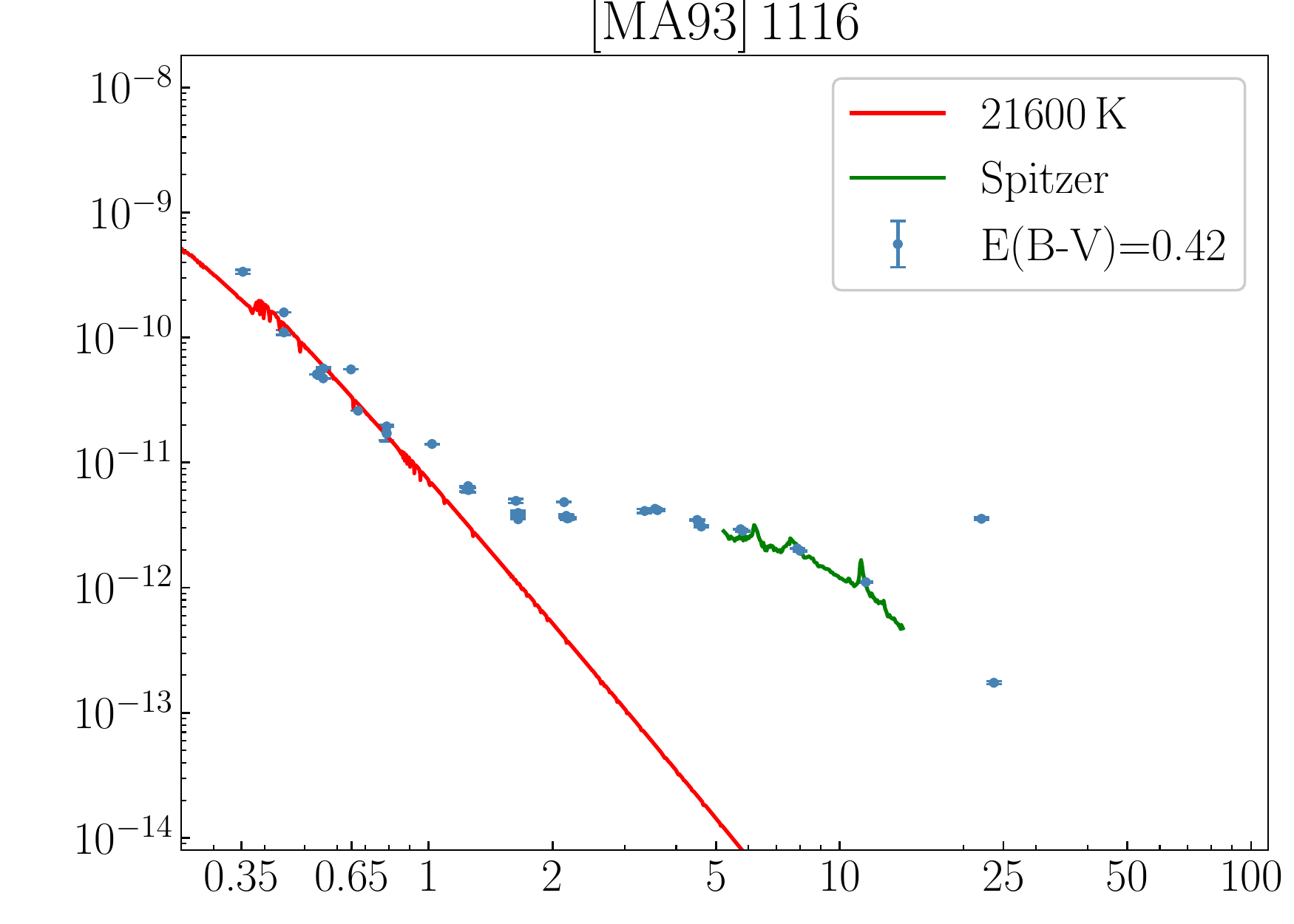}
\caption{Spectral energy distributions (SED) of our sample. Blue dots denote photometric data, green lines represent Spitzer spectra, and red lines represent Kurucz atmosphere models, considering the effective temperatures from Table~\ref{table:Physical-parameters} (except for Hen 2-91). The observed photometric data were obtained from \textit{\href{http://vizier.u-strasbg.fr/viz-bin/VizieR}{VizieR}} (except U-, B-, and V-band data from Table~\ref{table:objectsphotometry}) and dereddened using the interstellar color excess from Table~\ref{table:interstellar-extinction}.}
\label{fig:SEDs-model}
% SOURCE: /h
\end{figure*}

Our spectra also present forbidden emission lines (nebular lines) from high-ionized ions, like [O\,{\sc iii}], [Ar\,{\sc iii}], and [Ar\,{\sc iv}] with single-peaked profiles. However, as cited by \citet{Wisniewski_2007}, after the subtraction of the sky contribution, just a small residual remains, indicating that these lines are not from the object itself.
%\clearpage

%-----------------------------------------------------------------------

\section*{Other figures referenced in the text}
%\section{Period Analysis}
Figure \ref{f:app} shows the highest-power frequencies for IRAS~07080-0106 and V* FX Vel. Fig.\ \ref{fig:SEDs-model} shows the SEDs of the 12 stars of our sample.

\bsp        % typesetting comment
\label{lastpage}
\end{document}